\documentclass[journal]{IEEEtran}
\usepackage{tabularx}
\usepackage{makecell}
\usepackage{amsmath}
\usepackage{multirow}
\usepackage{graphicx}
\usepackage{amsfonts}
\usepackage{amsmath,epsfig}
\usepackage{stfloats}
\usepackage{amssymb}
\usepackage{url}
\usepackage{bm}
\usepackage{cite}
\usepackage{amsmath}
\usepackage{amssymb}
\usepackage{latexsym}
\usepackage{multirow}
\usepackage{epsfig}
\usepackage{graphics}
\usepackage{mathrsfs}
\usepackage{algorithmic}
\usepackage{algorithm}
\usepackage{subfigure}
\usepackage[all]{xy}
\usepackage{soul}

\usepackage{pifont}
\usepackage{bbding}
\usepackage{stmaryrd}
\usepackage{amssymb}
\usepackage{amsfonts}
\usepackage{epic}
\usepackage{stfloats}
\usepackage{latexsym}
\usepackage{epstopdf}
\usepackage{epic}
\usepackage{bm}
\usepackage{xcolor}
\usepackage{mathrsfs}
\usepackage{pifont}
\usepackage{bbding}
\usepackage{amsmath,epsfig}
\usepackage{stmaryrd}
\usepackage{amssymb}
\usepackage{amsfonts}
\usepackage{epic}
\usepackage{graphicx}
\usepackage{subfigure}

\usepackage{enumerate}

\usepackage{stfloats}
\usepackage{latexsym}
\usepackage{epstopdf}
\usepackage{epic}
\usepackage{multirow}
\usepackage{stfloats}
\usepackage{bm}
\usepackage{amsmath}
\usepackage{amssymb}
\usepackage{amsthm}

\usepackage{booktabs}
\usepackage{color}
\usepackage{cases}
\usepackage{array}
\usepackage{makecell}
\usepackage{times}

\graphicspath{{./figs/}}

\begin{document}
\title{{\color{black}Next Generation Advanced Transceiver \\Technologies for 6G and Beyond}}
\author{\IEEEauthorblockN{Changsheng You, Yunlong Cai, Yuanwei Liu, \emph{Fellow, IEEE}, Marco Di Renzo, \emph{Fellow, IEEE}, \\ Tolga M. Duman, \emph{Fellow, IEEE}, Aylin Yener, \emph{Fellow, IEEE},  A. Lee Swindlehurst, \emph{Fellow, IEEE}}
 \thanks{
 C. You is with the Department of Electronic and Electrical Engineering, Southern University of Science and Technology, Shenzhen, China (email: youcs@sustech.edu.cn).

 Y. Cai is with the College of Information Science and Electronic Engineering, Zhejiang University, Hangzhou 310027, China (email: ylcai@zju.edu.cn).

 Y. Liu is with the School of Electronic Engineering and Computer Science, Queen Mary University of London, E1 4NS, London, U.K. (email: yuanwei.liu@qmul.ac.uk).

 M. Di Renzo is with Universit\'{e} Paris-Saclay, CNRS, CentraleSup\'{e}lec, Laboratoire des Signaux et Syst\`{e}mes, 91192 Gif-sur-Yvette, France (email: marco.di-renzo@universite-paris-saclay.fr).

T. M. Duman is with the Department of Electrical and Electronics Engineering, Bilkent University, Ankara, 06800, Turkey (email: duman@ee.bilkent.edu.tr).
 
 A. Yener is with the Departments of Electrical and Computer Engineering, Computer Science and Engineering, and Integrated Systems Engineering,
The Ohio State University, OH, USA (email: yener@ece.osu.edu).

A. L. Swindlehurst is with the Center for Pervasive Communications and Computing, University of California, Irvine, CA 92697, USA (e-mail: swindle@uci.edu).

(\emph{Corresponding author: Yunlong Cai.})

}

}

\maketitle

\begin{abstract}
%no italics in abstracts
To accommodate new applications such as extended reality, fully autonomous vehicular networks and the metaverse, next generation wireless networks are going to be subject to much more stringent performance requirements than the fifth-generation (5G) in terms of data rates, reliability, latency, and connectivity. 
It is thus necessary to develop next generation advanced transceiver (NGAT) technologies for efficient signal transmission and reception. In this tutorial, we explore the evolution of NGAT from three different perspectives. Specifically, we first provide an overview of new-field NGAT technology, which shifts from conventional far-field channel models to new near-field channel models. Then, three new-form NGAT technologies and their design challenges are presented, including reconfigurable intelligent surfaces, flexible antennas, and holographic multi-input multi-output (MIMO) systems. Subsequently, we discuss recent advances in semantic-aware NGAT technologies, which can utilize new metrics for advanced transceiver designs. Finally, we point out other promising transceiver technologies for future research.
\end{abstract}


\section{Introduction}
While fifth-generation (5G) wireless networks are still under deployment globally, both academia and industry have been looking forward to the next frontier of \emph{sixth-generation} (6G) technologies, formally referred to by ITU-R as the ``IMT-2030 Framework.'' Among others, various research groups, standardization bodies, regulatory entities, and government agencies have launched initiatives to explore and define the future 6G.
6G envisions a transition from solely connected things to a more pronounced emphasis on \emph{connected intelligence}\cite{6gusage,you2021towards,9349624}. 
To accommodate emerging applications such as {\color{black} immersive extended reality (XR)}, autonomous vehicles, metaverse, holographic-type telepresence, and the tactile Internet, 6G needs to achieve significantly enhanced performance over 5G. 
% For instance, it has to offer terabits-per-second rates, hyper-reliable communications, near-zero latency, massive connectivity, etc.
\textcolor{black}{
% It is generally known that the main 5G services include: 1) enhanced mobile broadband (eMBB), which provides high data rates up to 1 Gigabit per second (Gbps) for mobile users; 2) ultra-reliable low-latency communication (URLLC), achieving a reliability of no less than 99.999\% with millisecond (ms)-level latency; and 3) massive machine-type communication (mMTC), which can simultaneously connect up to one million devices per square kilometer ($10^6$ devices per $\mathrm{km}^2$) in the Internet-of-Things (IoT) network.
% Compared to 5G, 6G is expected to significantly enhance network performance across all key metrics, including data rate, latency, energy efficiency, and reliability. Additionally, 6G will introduce a fundamental paradigm shift from merely connecting people and things to enabling \emph{connected intelligence}\cite{8808168,8869705,9349624}. This involves integrating communication, computing, control, sensing, and learning functions.
Specifically, the representative key performance indicators (KPIs) advocated for 6G are summarized as follows\cite{6gusage,you2021towards,VIMT,8808168,8869705,9349624,shafi20175g}:}
\begin{itemize}
\item \textbf{Peak data rate:} Up to 1 Tbps (terabits per second), which is 100-100 times that of 5G.

\item \textbf{Spectral efficiency:} 
Up to 100 bits/s/Hz, which is about 5 times that of 5G.

\item \textbf{Latency:} 
About  0.1$\sim$1 ms to support real-time applications like holographic communications and hyper-reliable low-latency communications (HRLLC).

\item \textbf{Network reliability:} 
About 1-$10^{-5}$$\sim$1-$10^{-7}$ to support critical applications such as remote surgery and autonomous vehicles.

\item \textbf{Energy efficiency:} 
Up to a 100 times increase in energy efficiency compared to 5G for achieving green communication networks.

\item \textbf{Positioning accuracy:} Achieve centimeter (cm) level accuracy to support high-accuracy applications.

\item \textbf{Connection density:} 
Up to $10^8$ devices/$\mathrm{km}^2$ to enable  extensive IoT deployments.

\item \textbf{Mobility support:} 
Offer seamless connectivity for high-mobility communications with velocity up to 1000 km/h, supporting high-speed trains and aerial vehicles.

\item \textbf{Intelligence level:} Achieve higher intelligence level than 5G, enabling a vast expansion of smart services and use cases. 

\item \textbf{AI service accuracy/efficiency:} Achieve over 90\% AI service accuracy/efficiency across various tasks to ensure high reliability in intelligent applications and human-machine interactions.

\end{itemize}

\begin{table}[t!]
	\caption{Main Abbreviations and their definitions}
	\label{table:abbreviation}
    \centering
    \begin{tabular}{ll}
    \Xhline{1.1pt}
    Abbreviation & Definition \\
    \Xhline{1.1pt}
    AI & Artificial Intelligence \\
 AO&    Alternating Optimization\\
BCD & Block Coordinated Descent \\
BS & Base Station \\
CS & Compressed Sensing \\
CSI & Channel State Information \\
DAS & Distributed Antenna System \\
DFT & Discrete Fourier Transform \\
DL & Deep Learning \\
DoF& Degree-of-Freedom\\
EM & Electromagnetic \\
ES & Energy Splitting \\
FA & Flexible Antenna \\
HARQ & Hybrid Automatic Repeat Request \\
HMIMO & Holographic Multi-Input Multi-Output \\
IB & Information Bottleneck \\
IRS & Intelligent Reflecting Surface\\
ISAC & Integrated Sensing and Communications \\
JSCC & Joint Source and Channel Coding \\
LoS&Line-of-Sight\\
LWA & Leaky-Wave Antenna \\
MIMO & Multi-Input Multi-Output \\
MISO&Multi-Input Single-Output\\
MMSE & Minimum Mean-Square Error \\
MS & Mode Switching \\
NGAT & Next Generation Advanced Transceiver \\
NOMA& Non Orthogonal Multiple Access\\
OFDM & Orthogonal Frequency Division Multiplexing \\
OMP & Orthogonal Matching Pursuit \\
RF& radio-frequency\\
RIS & Reconfigurable Intelligent Surface \\
RK & Randomized Kaczmarz \\
SC & Semantic Communication \\
SIMO&Single-Input Multi-Output\\
SNR & Signal-to-Noise Ratio \\
THz & Terahertz \\
TS & Time Switching \\
TTD & True Time Delay \\
ULA & Uniform Linear Array \\
UPA & Uniform Phased Array \\
USW & Uniform Spherical Wavefront \\
VAE & Variational Adversarial Auto-Encoder \\
VR & Visibility Region \\
    \Xhline{1.1pt}
    \end{tabular}
\end{table}

In 5G, three primary usage scenarios were identified based on user demands, including enhanced mobile broadband (eMBB), massive machine-type communications (mMTC), and ultra-reliable and low latency communications (URLLC)\cite{shafi20175g}. Entering the 6G era, these scenarios are expected to evolve towards their advanced versions, namely, immersive communications, massive communications, and HRLLC\cite{6gusage}.
Immersive communications encompass novel use cases such as {\color{black} immersive XR} and holographic communications, which demand more bandwidth than eMBB in 5G. Massive communications, an advanced version of mMTC, aim to support connectivity for a vast number of devices, even in a small area. Emerging use cases include e.g., dense sensor networks for industries, the metaverse, logistics, and transportation.
HRLLC extends URLLC by covering more specialized use cases with more stringent requirements on reliability and latency. Failure to meet these requirements may cause severe consequences for these applications.
Beyond these scenarios, there are three new additional use cases for 6G, including ubiquitous connectivity, integrated sensing and communications (ISAC), and integrated 
artificial intelligence (AI) and communication \cite{6gusage}. Ubiquitous connectivity seeks to enhance connectivity and bridge the digital divide, particularly in areas that are currently poorly covered or even uncovered, such as maritime and aircraft communications. 
%Typical use cases include the Internet of Things (IoT) and mobile broadband communication.
In addition, ISAC introduces new possibilities for end-users by offering sensing as a new network service, hence providing more options to the market and boosting vertical industries. Integrated AI and communications will support a wide range of intelligent applications, such as human-machine interactions, digital twin-based predictions and decision-making, which require AI-oriented communications and computing support.

The demanding performance requirements of 6G, however, cannot solely be achieved by existing 5G technologies such as massive multi-input multi-output (MIMO) and millimeter-wave (mmWave) communications \cite{shafi20175g,boccardi2014five},  calling for the development of revolutionary technologies and new network paradigms for future 6G networks. 
Among others, the design of next generation advanced transceivers (NGAT)  lays the foundations for efficient signal transmission and reception in 6G, which comprises a broad range of innovative designs in transceiver architecture, hardware, modulation and channel coding, waveforms, channel estimation, detection, interference management, etc. While existing transceiver designs  mainly target far-field communications in low-frequency bands, employ classical antenna manufacturing techniques, and follow bits-oriented design principles, NGAT has the ambitious goal to boost next generation network performance by exploring different design paradigms. In this paper, three new features of NGAT are elaborated:  \emph{new-field NGAT}, \emph{new-form NGAT}, and \emph{new-metric-aware NGAT}.
\begin{itemize}
\item {\bf New-field NGAT:} In conventional wireless systems, far-field communications with planar-wavefronts have been widely assumed. This is reasonable when the base stations (BSs), operating in relatively low-frequency bands, are equipped with a small or modest number of antennas, since the users are usually located in the far-field region. In these cases,  the BS-user distances are much larger than the Fraunhofer far-field (or Rayleigh) distance, which increases quadratically with the largest dimension of the array and  decreases linearly with the carrier wavelength  \cite{cui2022near,liu2023near,lu2023tutorial,you2023near,liu2024near}. However, this assumption will not be valid for future 6G systems for two main reasons. First, massive MIMO in 5G is evolving towards   \emph{extremely large-scale MIMO} (XL-MIMO) with an increase in the number of antennas by an order-of-magnitude, from several dozens to a few hundreds or even thousands. This leads to a much increased Fraunhofer distance, assuming that the inter-antenna distance is kept the same. Second, to be able to exploit increased  spectral resources, future wireless systems are migrating to higher frequency bands, such as mmWave and even Terahertz (THz) frequencies. Even without changing the array dimensions, the Fraunhofer far-field  distance is greatly enlarged at much higher frequencies. The above technology trends render users in future wireless systems more likely to be located in the \emph{near-field}, and the corresponding wireless channels are characterized by spherical wavefronts and spatial non-stationarity, leading to new opportunities and design challenges.

\item {\bf New-form NGAT:} With recent advances in antenna manufacturing technologies, a number of new-form antennas have been proposed to improve the channel capacity of future wireless systems. For example, unlike conventional transceiver designs for active beamforming and power/rate control at the BS, reconfigurable intelligent surfaces (RISs) (or equivalently intelligent reflecting surfaces (IRSs) working in reflection mode) have emerged as a new technology to sculpt the radio propagation environments by leveraging a massive number of low-cost metamaterial elements \cite{wu2019towards,wu2021intelligent,di2020smart,8796365,liu2021reconfigurable,swindlehurst2022channel}. By establishing virtual line-of-sight (LoS) links between the transmitter and receiver, RISs are able to bypass obstacles, hence greatly improving the communication performance in complex environments. Second, in conventional MIMO systems, the antennas are usually fixed once manufactured, and thus, MIMO systems may suffer considerable performance losses when the channels experience deep fading. Recently, flexible antennas (FAs),  also known as fluid antennas\cite{wong2022bruce} or movable antennas \cite{zhu2023movable}, have been developed to tackle this issue by dynamically adjusting the position of antennas within a designated space, and hence more efficiently exploit the spatial diversity and multiplexing gains. Third, unlike conventional MIMO with half-wavelength inter-antenna spacing, holographic MIMO (HMIMO) systems employ a spatially-continuous array/surface, consisting of a massive number of densely-spaced antennas \cite{huang2020holographic
,an2023tutorial}. Such a nearly spatially-continuous aperture enables HMIMO to achieve flexible control of the electromagnetic (EM) field, which generally provides more degrees-of-freedom (DoFs) for enhancing the communication performance.

\begin{table*}[!ht]
	\label{table:6g}
	\caption{Summary of representative overview, survey and tutorial Papers for 6G}
    \centering
    \setlength{\tabcolsep}{3pt} % Adjust column spacing
    \renewcommand{\arraystretch}{1.5} % Adjust row height
\begin{tabularx}{\textwidth}{|>{\centering\arraybackslash}p{2cm}
    |>{\raggedright\arraybackslash}X
    |>{\centering\arraybackslash}p{1.5cm}|
    }
    \Xhline{1.1pt}
    \textbf{Topics} &  \multicolumn{1}{c|}{\textbf{Major Contributions}} & \textbf{References} \\ \Xhline{1.1pt}
    \multirow{6}{*}{\begin{minipage}{2cm}
		\centering
		Applications,\\Trends \& Problems
	\end{minipage}} & Gives an overview of AI-empowered wireless networks. & \cite{8808168} \\ \cline{2-3}
	~ & Discusses  potential 6G technologies that enhance IoT applications. & \cite{9509294} \\ \cline{2-3}
    ~ & Provides an overview of the state-of-the-art and a vision for future 6G communication systems, covering key driving factors, use cases, requirements, architectures and enabling technologies. & \cite{9349624} \\ \cline{2-3}
    ~ & Provides a comprehensive vision of 6G wireless systems, identifying its drivers, service classes, enabling technologies, and research agenda. & \cite{8869705} \\ \Xhline{1.1pt}
    \multirow{2}{*}{New Spectrum} & Discusses the regulatory developments, potential applications in cognition and sensing, long-distance communication techniques, signal processing improvements, and new models for propagation and positioning for frequencies above 100GHz. &\cite{8732419} \\ \Xhline{1.1pt} 
	New Multiple Access & Introduces next generation multiple access schemes based on NOMA, as well as their 
	 requirements, candidate technologies, multi-antenna innovations, application scenarios, and mathematical tools. & \cite{2022liunoma} \\ \Xhline{1.1pt}
    \end{tabularx}
\end{table*}

\item {\bf New-metric-aware NGAT:}
Existing wireless systems are designed based on the classical framework of reliably communicating bit sequences \cite{Shannon_BSTJ}. 
Future 6G wireless networks are expected to provide seamless integration of the physical and cyber worlds, which demands an exponential growth in both bandwidth and complexity due to the transmission of massive amounts of data with diverse modalities such as text, images and video. Semantic communication (SC) has emerged as a promising paradigm to address this issue by effectively incorporating semantics into the communication system design \cite{guler_13_globalsip,semantic_index_assignment,guler_18_semantic_game,sagduyu2024will}. The implementation of SC requires semantic-aware transceiver designs that capitalize on the specific meaning that the messages represent \cite{9450827,emrecan_semantic_overview,NiuKai_ICM,lan2021semantic}. 
\end{itemize}

Compared to the existing overview, survey, and tutorial papers on 6G listed in Table~II, this paper offers a comprehensive overview of NGAT for 6G, including a systematic treatment of three types of NGAT, including near-field NGAT, new-form NGAT based on RISs, FAs, and HMIMO, as well as semantic-aware NGAT. For each technology, we introduce the corresponding new hardware architectures and system models, transceiver design challenges, as well as the current state-of-the-art, and identify open problems and challenges for future research.

The rest of the paper is organized as follows. Section \ref{sec_nf} provides an overview of near-field transceiver designs with a focus on system models and transceiver designs.
Section \ref{sec_nform} presents the new hardware architectures, system models, and design challenges for three new-form NGATs, namely, RISs, FAs, and HMIMO.
Section \ref{sec_nmetr} introduces semantic-aware transceiver designs for supporting semantics and goal-oriented communications.
Other 6G transceiver technologies are discussed in Section \ref{sec_extens}, and conclusions are given in  Section \ref{sec_con}.

\section{New Field: Near-field Transceivers} \label{sec_nf}

In this section, we first introduce the near-field region and its promising applications. Then, we provide a comprehensive overview for new near-field channel models and key transceiver design challenges in near-field communications.

\subsection{Near-field Region and Near-field Applications}
\subsubsection{Near-field Region}
%Let $D$ denote the XL-array aperture size, $\lambda$ denote the carrier wavelength, and $r$ denote the user$\to$XL-array distance. Then, 
{\color{black}Based on antenna theory, the EM field surrounding an XL-array can be divided into three regions based on the strongest LoS channel component between the transceivers, as illustrated in Fig.~\ref{1}.}
\begin{itemize}
\item { \bf Reactive near-field region}: When the user is located very close to the XL-array, ie., within the Fresnel distance \cite{lu2023tutorial,liu2023near}, it is said to be located in the reactive near-field region. 
%say,  in the reactive near-field region of the XL-array when their distance is very small within $r\le 0.62\sqrt{\frac{D^3}{\lambda}}$. 
In this case, the EM field is dominated by evanescent waves whose power decreases exponentially with distance. As such, evanescent waves generally store energy rather than propagate EM waves. 
\item  { \bf  Radiative near-field or Fresnel region}: When the user distance is smaller than the so-called Fraunhofer far-field or Rayleigh distance (to be defined later), it is located in the radiative near-field region where EM waves propagate. Due to the relative proximity of users in this region, the EM field must be characterized by spherical wavefronts that produce non-linear phase variations across a uniform linear array (ULA).
\item { \bf  Far-field or Fraunhofer region}:  When the user is located beyond the Rayleigh distance, the user is located in the far-field region, in which the spherical wavefronts can be approximated as planar waves with linear phase variations across a ULA. 
\end{itemize}

Since in practice the reactive near-field region is very small, in the following we mainly focus on the radiative near-field region, which hereafter is simply referred to as the near-field.  More specifically, depending on different criteria such as phase, amplitude, and power variations over the XL-array, the boundaries between the different types of EM fields can be defined as follows.

\begin{itemize}

	\item {\bf  Direction-dependent Rayleigh distance:} The classical Rayleigh or Fraunhofer distance distinguishes the far-field and near-field regions from the perspective of phase variations over an antenna array. Specifically, when the signal is incident on the broadside of a ULA, the Rayleigh distance specifies the distance at which the phase difference of the received signal at all array antennas is no larger than $\pi/8$. It can be obtained that for  XL-MIMO systems serving single-antenna users, the Rayleigh distance is $r_{\mathrm{Rayl}} = 2D^{2}/\lambda$, where $D$ is the size of the array aperture and $\lambda$ is the carrier wavelength \cite{cui2022near,selvan2017fraunhofer}. For example, consider an XL-MIMO  system where a BS equipped with an array with aperture $0.5$m serves several single-antenna users at $30$ GHz. In this case, the classical Rayleigh distance is about $50$m. For XL-MIMO systems serving multi-antenna users, the \emph{double-sided} Rayleigh distance can be obtained as $ 2(D_1+D_2)^{2}/\lambda$, where $D_1$ and $D_2$ are the aperture sizes of the BS and user arrays, respectively. However, the  Rayleigh distance does not take into account the effect on the phase variations of the angle of the user relative to the array broadside. To address this issue, a new \emph{direction-dependent} Rayleigh distance was introduced in \cite{lu2021communicating}. Specifically, for each direction of incidence, this new metric represents the shortest distance at which the phase difference of the received signals across the array aperture is no larger than $\pi/8$. For example, for XL-MIMO systems with single-antenna users, when the exact distance is approximated by a second-order Taylor approximation, the corresponding direction-dependent Rayleigh distance can be approximated as $r_{\mathrm{DDRayl}}(\theta) \approx 2D^{2}\sin^{2}\theta/\lambda$ \cite{lu2023tutorial}, which is jointly determined by the array aperture size $D$, carrier wavelength $\lambda$, and the incidence angle $\theta$. Thus, one can infer that the direction-dependent Rayleigh distance decreases with the signal angle $\theta$ and converges to the classical Rayleigh distance when $\theta = \pi/2$. 

\begin{figure}
	\centering
	\includegraphics[width=0.45\textwidth]{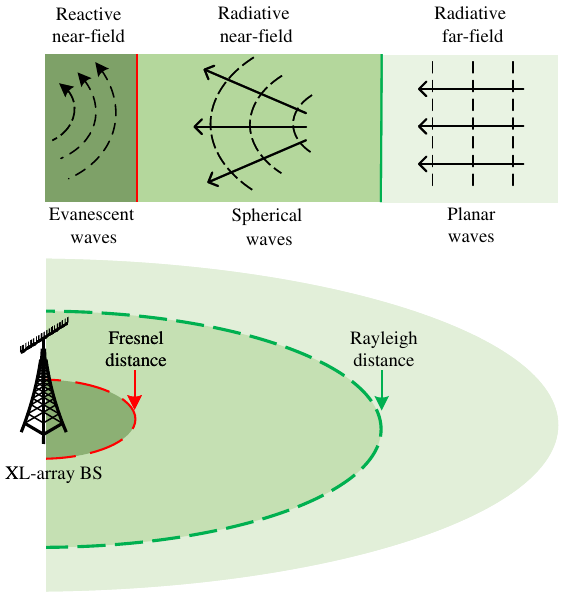}
	\caption{Illustration of XL-MIMO field regions.}
	\label{1}
\end{figure}

	\item { \bf  Uniform power distance:} The uniform-power distance is used to distinguish the far-field and near-field regions from the perspective of amplitude variations across the XL-array antennas. Specifically, the uniform-power distance defines the shortest distance where the ratio of the weakest and strongest received powers at the XL-array is greater than a specific threshold \cite{lu2021communicating}, which generally decreases with the incidence angle $\theta$. As such, for any given signal angle $\theta$, if the link distance is smaller (or larger) than the uniform-power distance, the amplitude variations across the XL-array is regarded as non-negligible (or negligible). In particular, for a ULA-based XL-MIMO system, the uniform power distance at the signal angle $\theta = \pi/2$  can be approximated as  $r_{\mathrm{UPD}}(\theta) \approx 1.2D$ when the threshold is set as $\cos^2{\pi/8}$ \cite{lu2021does}.
	
	\item {\bf Effective Rayleigh distance: }The effective Rayleigh distance distinguishes the far-field and near-field regions from the perspective of array gain, which directly affects the achievable rates of the users. Specifically, the effective Rayleigh distance, denoted by $r_{\mathrm{ERayl}}$, specifies the distance where the normalized coherence between the near-field channel and its far-field approximation is less than a certain threshold $\delta$ \cite{cui2021near}. For ULA-based XL-MIMO systems, the effective Rayleigh distance $r_{\mathrm{ERayl}}$ is given by $r_{\mathrm{ERayl}}^{(\mathrm{ULA})}(\phi)=\epsilon \frac{{2 D}^{2} \cos^{2} \theta}{\lambda}$, where $\epsilon = 0.367$ \cite{cui2021near}. Moreover, the Bj\"{o}rnson distance is also defined from the perspective of array gain, which is obtained by replacing the normalized coherence value in the effective Rayleigh distance by the normalized antenna array gain, which is the ratio of the gain of a given receive antenna array gain to the that of the largest gain of any antenna in the array \cite{bjornson2021primer}.
\end{itemize}

\begin{figure*}
	\centering
	\includegraphics[width=0.9\textwidth]{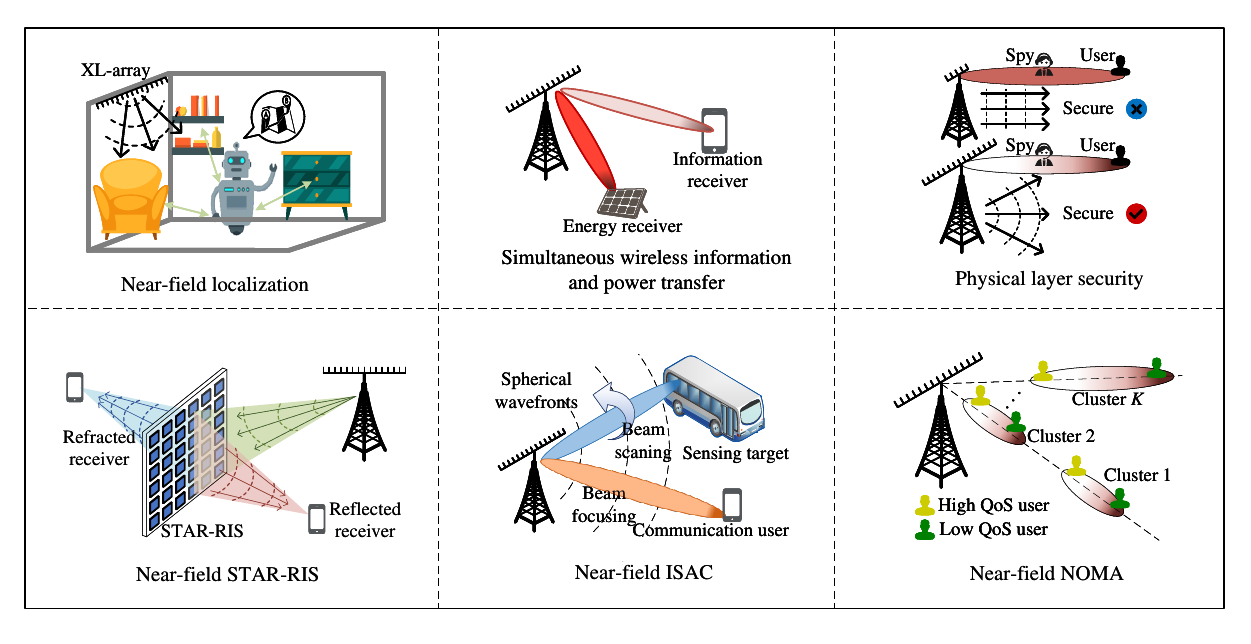}
	\caption{Illustration of near-field application scenarios.}
	\label{2}
\end{figure*}

\subsubsection{Near-field Applications}
Compared with massive MIMO, near-field XL-MIMO with spherical wavefronts provides  unique and appealing \emph{beam-focusing} function that allows the beam energy to be concentrated on a specific location/region rather than steering the beam energy towards a certain angle as in a far-field scenario \cite{zhang20236g}. This beam-focusing effect provides several advantages and enables new applications as illustrated in Fig.~\ref{2}. For example, it allows the XL-array BS to serve multiple users at the same angle but different distances with small (or even negligible) inter-user interference, hence enhancing the system spectrum efficiency and multi-user accessibility. Next, near-field beam-focusing can be leveraged to boost the energy-harvesting efficiency in wireless power transfer (WPT) systems since it can effectively reduce undesired energy leakage \cite{deutschmann2022location}, \cite{you2023near}. Moreover, the accuracy and resolution achieved by indoor and outdoor sensing/localization can be greatly improved by using XL-MIMO since it provides both user angle and distance information \cite{guerra2021near}, \cite{cong2023near}. Last but not least, near-field beam-focusing can be exploited for improving physical-layer security performance and protecting private information from eavesdropping \cite{zhang2023physical}.

\subsection{Near-field Channel Models}\label{nf_chan_mod}

{\color{black}In the following, we introduce near-field channel models that account for two unique channel characteristics, namely, near-field spherical wavefronts and spatial non-stationarity, as summarized in Table III.}

\subsubsection{Deterministic Channel Model for Spherical Wavefronts} In high-frequency bands such as mmWave and sub-THz, the transmitted signals are highly susceptible to severe path-loss and blockage. As such,  the near-field channel can in practice be modeled using deterministic methods, which generally include a possible LoS path and/or a few non-LoS (NLoS) paths. In the following, we first introduce the near-field array response vector and then present the deterministic LoS/multi-path channel model.

\paragraph{Near-field Array Response Vector}
Consider a BS equipped with an $N$-antenna ULA array. Denoting $r$ as the distance between the user and the center of the array, the near-field array response vector can be modeled in two different ways, depending on the user distance.

\begin{itemize}
	\item { \bf USW model:} When $r_{\mathrm{UPD}} \le r < r_{\mathrm{DDRayl}}$, the channel response vector can be modeled based on a uniform spherical wavefront (USW) \cite{lu2023tutorial}. Specifically, the amplitude variations across the antenna array are negligible and the phase variations are non-linear. The USW-based array response vector is thus given by \cite{cui2022channel}	
    \begin{equation}
		\mathbf{b}^{\mathrm{USW}}(r, \theta) = \left[e^{-j \frac{2\pi}{\lambda} (r_1-r)},\cdots ,e^{-j \frac{2\pi}{\lambda} (r_N-r)}\right]^{T}, \label{equ_1}
	\end{equation}
    where $r_n = \sqrt{r^2+2\delta_{n} dr\cos\theta + \delta_{n}^2 d^2}$ is the distance between the user and the $n$-th array antenna, $\delta_{n}=(2n-N-1)/2, n = 1,\cdots,N$, and $d$ denotes the antenna spacing.

	\item { \bf NUSW model:} When $r < \min \left\{ r_{\mathrm{UPD}}, r_{\mathrm{DDRayl}} \right\}$, the channel response exhibits a non-uniform spherical wavefront (NUSW) feature. In this case, the amplitude variations across the array are non-negligible, while the phase variations are still non-linear, and the array response vector is modeled as \cite{lu2021communicating}
	\begin{equation}
		\!\!\!\mathbf{b}^{\mathrm{NUSW}}(r,\theta) \!=\! \left[\frac{r}{r_1}e^{-j \frac{2\pi}{\lambda} (r_1-r)},\!\cdots\! ,\frac{r}{r_N}e^{-j \frac{2\pi}{\lambda} (r_n-r)}\right]^{T}.\!\!\! \label{equ_2}
	\end{equation}
Note that the NUSW model should be used when characterizing the asymptotic signal-to-noise ratio (SNR) for XL-MIMO systems when the number of antennas grows large ($N\to\infty$). 
\end{itemize}

\paragraph{Near-field LoS Channel}
Consider an XL single-input-multi-output (SIMO) system that comprises an LoS path only. The near-field LoS channel, denoted by $\mathbf{h} \in \mathbb{C}^{N \times 1}$, can be modeled as $\mathbf{h}_{\mathrm{LoS}} = \beta \mathbf{b}^{\rm SW}(r),$ where $\beta = \frac{\sqrt{\lambda}}{2 \pi r} e^{-j\frac{2 \pi}{\lambda}r}$ denotes the complex-valued channel gain at the center reference point of the array and $\rm{SW} \in \left\{ \mathrm{USW}, \mathrm{NUSW} \right\}$.

\paragraph{Near-field Multi-path Channel} For the more general multi-path channel case, the near-field channel can be modeled as \cite{cui2022channel,wei2021channel}
\begin{equation}
	\mathbf{h} = \mathbf{h}_{\mathrm{LoS}}(r, \theta) + \sum\limits_{\ell \in \mathcal{L}} \beta_{\ell} \mathbf{b}^{\rm SW}(r_{\ell}, \theta_{\ell}),\label{equ_3}           
\end{equation}
where $\beta_{\ell}$ denotes the complex channel gain of the $\ell$-th NLoS channel path, and $\mathbf{b}^{\rm SW}(r_{\ell},\theta_{\ell})$ denotes the near-field array response vector of the $\ell$-th path. The above near-field LoS and multi-path XL-SIMO channel models can be extended to  more general cases with multiple antennas at the users. Specifically, let $N_t$ and $N_r$ respectively denote the number of transmit and receive antennas. Then the general multi-path channel $\mathbf{H} \in \mathbb{C}^{N_r \times N_t}$ can be modeled as \cite{lu2023near,dong2022near}
\begin{equation}
	\mathbf{H} = \mathbf{H}_{\mathrm{LoS}} + \sum_{\ell \in \mathcal{L}} \beta_{\ell} \mathbf{b}_{r}^{\rm SW}(r_{\ell}, \theta_{\ell})[\mathbf{b}_{t}^{\rm SW}(r_{\ell}, \theta_{\ell})]^{H},\label{equ_4}
\end{equation}
where for the LoS path, we have  $\left[ \mathbf{H}_{\mathrm{LoS}} \right]_{n_r,n_t} = \beta \frac{r}{r_{n_r,n_t}} e^{-j \frac{2\pi}{\lambda}(r_{n_r,n_t}-r)}$ with $r_{n_r,n_t}$ representing the distance between receiver antenna $n_r$ and transmitter antenna $n_t$, $\mathbf{b}_{r}^{\rm SW}$ and $\mathbf{b}_{t}^{\rm SW}$ denote the near-field transmit and receive array response vectors with respect to (w.r.t.) the $\ell$-th path, respectively. 
\begin{table*}[!ht]
	\label{table:Channel Model}
	{\color{black}\caption{Summary of near-field channel models.}
    \centering
    \setlength{\tabcolsep}{3pt} % Adjust column spacing
    \renewcommand{\arraystretch}{2} % Adjust row height
\begin{tabularx}{\textwidth}{
	|>{\centering\arraybackslash}p{3.5cm}
    |>{\centering\arraybackslash}p{3.5cm}
	|>{\centering\arraybackslash}X
    |>{\centering\arraybackslash}p{2cm}|
    }
    \Xhline{1.4pt}
    \multicolumn{2}{|c|}{\textbf{Types}} &  \textbf{Channel Modeling}  \\ \Xhline{1.2pt}
    \multirow{4.5}{*}{Deterministic channel model} & \multirow{1.5}{*}{Near-field LoS channel} & LoS channel based on USW/NUSW, i.e., $\mathbf{h}_{\mathrm{LoS}} = \beta \mathbf{b}^{\rm SW}(r)$, $\rm{SW} \in \left\{ \mathrm{USW}, \mathrm{NUSW} \right\}$.  \\ \cline{2-3}
	~ & \multirow{1.5}{*}{Near-field multi-path channel} & Multi-path channel with LoS and NLoS components based on spherical wavefronts, i.e., $\mathbf{h} = \mathbf{h}_{\mathrm{LoS}}(r, \theta) + \sum_{\ell \in \mathcal{L}} \beta_{\ell} \mathbf{b}^{\rm SW}(r_{\ell}, \theta_{\ell})$. \\ \cline{2-3}
    ~ & \multirow{1.5}{*}{\makecell[c]{Spatial non-stationary channel}} & Near-field multi-path channel with spatial non-stationarity, i.e., $\mathbf{h} = \sum_{\ell \in \mathcal{L}} \beta_{\ell} \mathbf{b}^{\mathrm{\rm SW}}(r_{\ell},\theta_{\ell}) \odot \mathbf{p}(\Psi_{\ell})$, $\mathbf{p}(\Psi_{\ell}) = \left\{ 0,1 \right\}^{N \times 1}$. \\ \Xhline{1.2pt}
    \multirow{8}{*}{Stochastic channel model} & \multirow{4}{*}{\makecell[c]{Spatial correlation based \\near-field channel}} & 
	Channel responses are spatially correlated according to covariance matrix $\mathbf{R}$, and the NUSW feature is reflected in the large-scale fading
	coefficients, i.e., $\mathbf{h} = \sqrt{\boldsymbol{\xi}} \odot \left( \mathbf{R}^{1/2}  \hat{\mathbf{h}} \right)$. \\ \cline{3-3}
    ~ & ~ & Channel responses are spatially correlated according to covariance matrix $\mathbf{R}$, and the NUSW feature is reflected in the covariance matrix, i.e., $\mathbf{h} = \mathbf{R}^{1/2}  \hat{\mathbf{h}}$, where $[\mathbf{R}]_{m, n} = \int_{\ell \in \mathcal{L}} \frac{r^{2}(\ell) e^{-j \frac{2 \pi}{\lambda}\left(r_{m}(\ell)-r_{n}(\ell)\right)}}{r_{m}(\ell) r_{n}(\ell)} f(\ell) \mathrm{d} \ell$.    \\  \cline{2-3}
	~ & \multirow{5}{*}{\makecell[c]{Spatial non-stationary channel}}  & Channel responses are spatially correlated according to covariance matrix $\mathbf{R}$, and the spatial non-stationarity is reflected by a deterministic diagonal matrix $\mathrm{D}$, i.e., $\mathbf{h} = \sqrt{\boldsymbol{\xi}} \odot \left( \mathbf{R}_{\mathrm{VR}}^{1/2}  \hat{\mathbf{h}} \right)$, where $\mathbf{R}_{\mathrm{VR}}= \mathbf{D}^{1/2} \mathbf{R} \mathbf{D}^{1/2}$, $\mathbf{D} \in \left\{0, 1 \right\}^{N \times N}$. \\ \cline{3-3}
	~ & ~ & Channel responses are spatially correlated according to covariance matrix $\mathbf{R}$, and the spatial non-stationarity is reflected in the covariance matrix, i.e., $\mathbf{h} = \mathbf{R}^{1/2}  \hat{\mathbf{h}}$, where $[\mathbf{R}]_{m, n}= \int_{\ell \in \mathcal{L}} \mathbb{E}\left\{ \left[ \mathbf{p}(\Psi_{\ell}) \right]_{m} \left[ \mathbf{p}(\Psi_{\ell}) \right]_{n} \right\} \frac{r^{2}(\ell) e^{-j \frac{2 \pi}{\lambda}\left(r_{m}(\ell)-r_{n}(\ell)\right)}}{r_{m}(\ell) r_{n}(\ell)} f(\ell) \mathrm{d} \ell$. \\ \Xhline{1.2pt}
    \end{tabularx}}
\end{table*}
\subsubsection{Stochastic Channel Model for Spherical Wavefronts} In relatively low-frequency bands (e.g., sub-6G),  a stochastic model can be used to describe near-field channels in rich scattering environments \cite{bjornson2019massive}. Specifically, one can use the channel covariance matrix to characterize the correlation between different transmit and receive antenna pairs, which facilitates analyses using performance metrics such as signal-to-interference plus noise ratio (SINR) \cite{ali2019linear} and ergodic capacity \cite{li2015capacity}. In XL-MIMO systems with single-antenna users, different array antennas generally experience different large-scale fading due to the NUSW property. As such, the near-field stochastic channel   can be modeled as \cite{lu2023tutorial,feng2022mutual,may2020stochastic}
\begin{equation}
	\mathbf{h} = \sqrt{\boldsymbol{\xi}} \odot \left( \mathbf{R}^{1/2}  \hat{\mathbf{h}} \right),\label{equ_5}
\end{equation}
where {\color{black} the notation $\odot$ represents the Hadamard product}, $\hat{\mathbf{h}} \sim \mathcal{N}_{\mathbb{C}}\left(\mathbf{0},\mathbf{I}_N\right)$ denotes a circularly symmetric complex Gaussian random vector, $\mathbf{R} \in \mathbb{C}^{N \times N}$ is a symmetric positive semi-definite channel covariance matrix that takes into account the spatial channel correlation, $\boldsymbol{\xi} \in \mathbb{R}^{N \times 1}$ is the large-scale channel path-loss vector whose entries $\left[ \boldsymbol{\xi} \right]_{n}$ are determined by the link distance $r_{n}$.

Another method for stochastic modeling of the NUSW channel is characterizing the near-field covariance matrix accounting for spherical wavefronts. For example, the authors of  \cite{dong2022near} showed that the near-field spatial correlation matrix is determined by the power location spectrum (PLS), whose scatterer distribution is affected by both the angles and distances of environmental scatterers. More specifically, under the NUSW model, the $(m, n)$-th element of channel covariance matrix $\mathbf{R}$ can be expressed as
\begin{equation}
    [\mathbf{R}]_{m, n} = \int_{\ell \in \mathcal{L}} \frac{r^{2}(\ell) e^{-j \frac{2 \pi}{\lambda}\left(r_{m}(\ell)-r_{n}(\ell)\right)}}{r_{m}(\ell) r_{n}(\ell)} f(\ell) \mathrm{d} \ell,\label{equ_6}
\end{equation}
where $\mathcal{L}$ denotes the set of scatterers, $r(\ell)$ denotes the distance between scatterer $\ell$ and the array reference element, $r_{m}(\ell)$ denotes the distance between scatterer $r(\ell)$ and the $m$-th array element, and $f(\ell)$ represents the power location spectrum (PLS) of the scattering environment. 

%It can be observed  from \eqref{equ_6} that the near-field channel spatial correlation not only depends on the arrival angle of the scatterer, but also depends on the power spectrum.

\subsubsection{Near-field Spatial Non-stationarity}  
\begin{figure}
	\centering
	\includegraphics[width=0.35\textwidth]{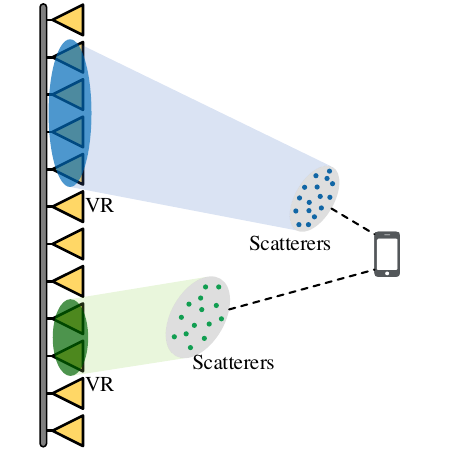}
	\caption{Illustration of VRs in near-field communication systems.}
	\label{3}
\end{figure}

{\color{black}Besides the spherical wavefronts, recent channel measurements have revealed that spatial non-stationarity is more pronounced in near-field scenarios compared to the far-field case \cite{han2023towards}.} This phenomenon generally arises from two aspects including 1) spherical wavefronts that produce different signal amplitudes at different array antennas, as modeled in NUSW, and 2) the near-field visibility region (VR) for which different segments of the XL-array may observe distinct radio propagation environments. Such VR information is particularly important in complex environments with obstacles and/or scatterers, since generally only a portion of the XL-array antennas are visible to the user depending on the geometric relationships among the XL-array, user, and obstacles, as illustrated in Fig.~\ref{3}. 

When the VR is considered for the deterministic channel model, the near-field channel vector can be modified as \cite{zhu2021bayesian, han2020channel, han2020deep}
\begin{equation}
	\mathbf{h} = \sum_{\ell \in \mathcal{L}} \beta_{\ell} \mathbf{b}^{\mathrm{\rm SW}}(r_{\ell},\theta_{\ell}) \odot \mathbf{p}(\Psi_{\ell}), \label{equ_7}
\end{equation}
where the set $\Psi_{\ell}$ denotes the VR of the user w.r.t. the XL-array on the $\ell$-th path, and $\mathbf{p}(\Psi_{\ell}) = \left\{ 0,1 \right\}^{N \times 1}$ is defined as the VR mask vector for the $\ell$-th path. If the user is visible (or invisible) to antenna $n$ for the $\ell$-th path,  we have $\left[ \mathbf{p}(\Psi_{\ell}) \right]_{n} = 1$ (or  $\left[ \mathbf{p}(\Psi_{\ell}) \right]_{n} = 0$).

For the stochastic channel model in low-frequency bands, the near-field channel vector accounting for the  VR information can be modeled as \cite{ali2019linear,yang2020uplink, rodrigues2020low, croisfelt2021accelerated}
\begin{equation}
	\mathbf{h} = \sqrt{\boldsymbol{\xi}} \odot \left( \mathbf{R}_{\mathrm{VR}}^{1/2}  \hat{\mathbf{h}} \right), \label{equ_8} 
\end{equation}
where {\color{black} the notation $\left( \cdot \right)^{1/2}$ represents the matrix square root}, $\mathbf{R}_{\mathrm{VR}}$ denotes the channel covariance matrix with VR, expressed as
\begin{equation}
	\mathbf{R}_{\mathrm{VR}} = \mathbf{D}^{1/2} \mathbf{R} \mathbf{D}^{1/2},\label{equ_9}
\end{equation}
where $\mathbf{D} \in \left\{0, 1 \right\}^{N \times N}$ is a diagonal matrix indicating whether or not each antenna is visible to the user. In addition, the authors of \cite{Z2023near} further obtained a near-field spatial correlation matrix that takes VR into account, modifying the spatial correlation matrix in \eqref{equ_6} as
    \begin{align}
		[\mathbf{R}]_{m, n}= \int_{\ell \in \mathcal{L}} & \mathbb{E}\left\{ \left[ \mathbf{p}(\Psi_{\ell}) \right]_{m} \left[ \mathbf{p}(\Psi_{\ell}) \right]_{n} \right\} \\
		& \times  \frac{r^{2}(\ell) e^{-j \frac{2 \pi}{\lambda}\left(r_{m}(\ell)-r_{n}(\ell)\right)}}{r_{m}(\ell) r_{n}(\ell)} f(\ell) \mathrm{d} \ell, \label{equ_10}
	\end{align}
where $\left[ \mathbf{p}(\Psi_{\ell}) \right]_{m}$ indicates the visibility of path $\ell$ to the $n$-th array element.

In the above model, we focus on the VR under the assumption that the environmental scatterers are always visible to the users. However, in complex environments, the scatterers may only be partially visible to the users \cite{li2015capacity}. For this case, a two-tier VR model as proposed in \cite{han2023towards} can be adopted, where the scatterers are partially visible to the XL-array and the user, potentially in different ways \cite{han2023towards,amiri2021distributed, guerra2022clustered}. 

\subsection{Transceiver Design Challenges}
{\color{black}In this section, we discuss  new transceiver design challenges for near-field XL-MIMO channel models, including near-field beamforming design, beam training, and channel estimation.}

\subsubsection{Near-field Beamforming} {\color{black}We first discuss analog/digital beamforming design for XL-MIMO systems assuming spherical wavefronts and present a system performance analysis of the beam pattern, SNR, and DoFs.} Then, we introduce efficient hybrid beamforming designs for XL-MIMO systems that balance the trade-off between the performance and hardware/energy cost for both narrow-band and wide-band systems.

\paragraph{Analog/Digital Beamforming}
For XL-MIMO systems, analog beamforming with only phase shifters is a cost-efficient method for maximizing the rate performance by exploiting the appealing near-field beam-focusing effect. This not only enhances the received power at targeted users but also reduces undesired interference to others. {\color{black}Specifically, under the USW model in \eqref{equ_1},
it was shown in  \cite{wu2023location} that when the number of antennas is sufficiently large, the near-field array response vectors exhibit asymptotic orthogonality in the angular and distance domains.} Using this beam-focusing effect, the authors proposed a new location division multiple access (LDMA) scheme for more effectively utilizing the available spatial resources, where users at the same angle can be simultaneously served with small (even negligible) interference, as long as they have sufficiently different distances to the BS. However, in practice, an XL-array is composed of only a limited number of antennas, and thus the orthogonality between different near-field array response vectors will not be exactly achieved. To this end, the beam-width and beam-depth of the near-field beam pattern, which reflect the degree of orthogonality of the near-field array response vectors in the angular and distance domains, respectively, are used to characterize the achievable beam-focusing region. More specifically, the near-field beam-depth defined in \cite{kosasih2023finite} specifies the distance interval over which the beam pattern is larger than a certain threshold. Based on the 3-dB focusing depth, the authors analyzed the achievable beam-focusing area for antennas of different geometrical shapes. As a further study, the authors of \cite{ding2023resolution} revealed the conditions for imperfect near-field beamforming resolution and demonstrated that imperfect resolution enables near-field beams to be reused for serving multiple users. {\color{black} In addition, \cite{li2022analytical,xie2023near} showed that the beam-focusing region constitutes an ellipse and the beam-focusing pattern is jointly determined by system parameters such as carrier frequency, array aperture, and user location.} In addition to the correlation between near-field channels, the correlation between near- and far-field channels was studied in \cite{zhang2023mixed}, \cite{zhang2023swipt} for characterizing inter-user interference in mixed near- and far-field scenarios. It was shown that when a far-field oriented beam is used at a near-field observation location, an \emph{energy-spread} effect occurs in which the energy of a far-field beam steered towards a specific angle is spread over an angular region.

{\color{black} For near-field communications in USW/NUSW-based LoS channels, it is convenient to apply analog beamforming for characterizing the SNR performance.} Specifically, the authors of \cite{lu2021does} revealed that under the near-field NUSW model, the asymptotic SNR for maximum ratio transmission (MRT)-based beamforming is determined by the angular span formed by the two line segments connecting the user and the two ends of the XL-array. For XL-MIMO architectures where the antennas are arranged in a modular manner, \cite{li2022near} characterized the single-user SNR scaling law and showed that as the number of modules increases, the SNR gradually tends to a constant value that depends on the module separation and the user's projected distance to the normal of the XL-MIMO array. In \cite{zhi2023performance} an SNR expression was derived for the case of polarization mismatch, demonstrating that the array can only capture $1/3$ of the power when the number of antennas is large. In addition, the authors of \cite{zhou2024sparse} exploited sparse arrays for enabling near-field communications and showed that the grating lobes of coprime arrays generally have smaller power than linear sparse arrays.

\begin{figure*}[ht] %hb代表放在文章底部，%ht为放在文章顶部
	\centering
	\includegraphics[width=6.2in]{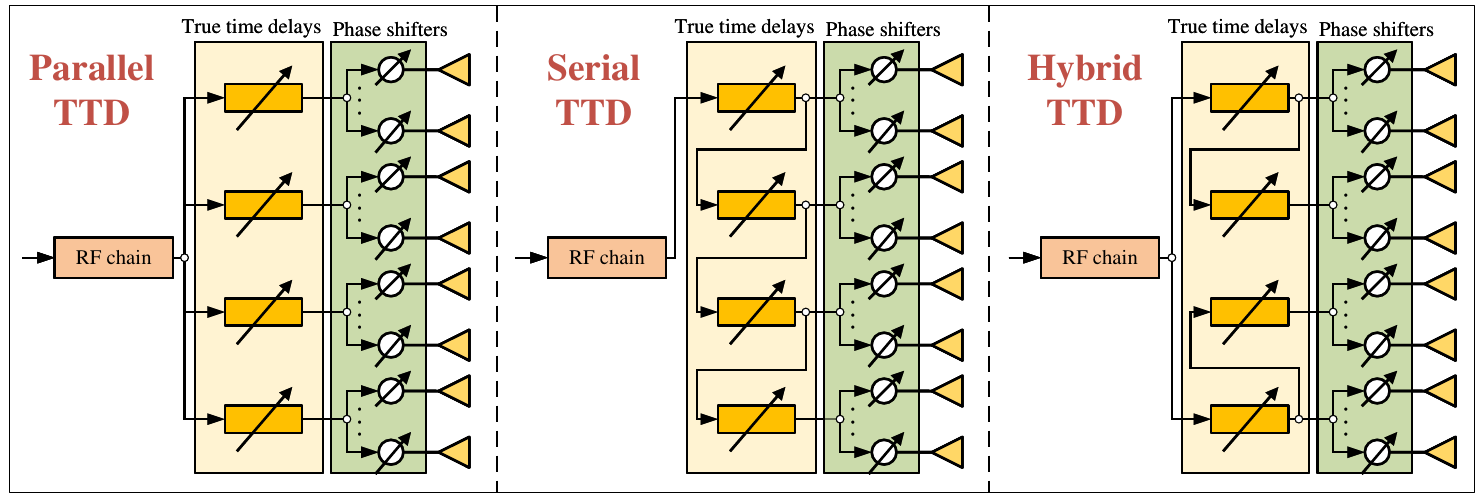}
    \caption{{\color{black} Different configurations of TTD architectures for near-field communication systems \cite{wang2023ttd}.  }}
    \label{4}
\end{figure*}

Although analog beamforming may be energy efficient, it is generally unable to perfectly mitigate the inter-user interference in multi-user systems, motivating the need for efficient yet low-complexity digital beamforming designs for near-field communications \cite{wang2024tutorial}. For example, the authors of \cite{croisfelt2021accelerated,xie2023performance}  proposed an accelerated randomized Kaczmarz (RK) algorithms to achieve performance close to that of regularized zero-forcing (RZF). The RK approach can solve linear equations in a cyclic and iterative manner, thus avoiding the high computational complexity of direct matrix inversion. The RK algorithm was further improved in  \cite{xu2023low}, where a sampling-without-replacement randomized Kaczmarz (SwoR-RK) algorithm was proposed for downlink precoder design in XL-MIMO systems, which was shown to enjoy faster convergence than the RK algorithm. An alternative was proposed in \cite{sun2021low,wang2020expectation} based on a low-complexity expectation propagation (EP) detector, where polynomial expansion is used to approximate the matrix inversion and reduce the computational complexity in each EP iteration. To further reduce the complexity of the matrix inversion operation, \cite{amiri2019message} described a low-complexity variational message passing (VMP) receiver whose computational complexity scales only linearly with the number of array antennas and users.

\paragraph{Hybrid Beamforming}

To reduce the hardware cost of digital beamforming, efficient hybrid beamforming designs need to be developed for near-field communication systems to reduce the number of required radio-frequency (RF) chains. For narrow-band XL-MIMO systems, a near-optimal design for hybrid beamforming can be obtained by alternately optimizing the analog and digital beamformers to minimize the gap between the hybrid beamformer and the optimal fully-digital beamformer. However, this approach is generally computationally demanding. To tackle this issue, a low-complexity yet effective approach is to perform the hybrid beamforming in two phases. {\color{black}First, a high-dimensional analog beamformer is devised for maximizing the received power at the target user by leveraging the energy-focusing effect under the spherical wavefront model.} Then, a low-dimensional digital beamformer is obtained to mitigate the inter-user interference. Although this two-phase approach was proposed for conventional far-field communications, it is particularly useful for the near-field case since the analog beamformer can generate focused beams toward specific locations/regions. 

{\color{black}For wide-band XL-MIMO systems, phase-shifter-based analog beam-focusing may result in the so-called \emph{beam-split} effect under the spherical wavefront model \cite{cui2021near}.} Since the phase shifters are frequency-flat devices, the beam wavefronts generated at widely separated frequencies will fail to focus the beam energy at the target location, which results in degraded SNR. Unlike the far-field case where the beam-squint effect causes the beams to be propagated towards different angles, in near-field scenarios the beam-split effect causes beams at different frequencies to be focused at different angles and distances \cite{cui2021near}. {\color{black}Efficient algorithms have been proposed to handle this effect in the far-field case, e.g. \cite{chen2020hybrid}, but such methods are not suitable for near-field spherical wavefronts.} A possible solution is to employ hardware-based methods to generate frequency-dependent beams that eliminate the beam-split effect. For example, true time delays (TTDs) can be used instead of phase shifts to compensate for the signal propagation delays between antennas, and can effectively eliminate the beam-split effect \cite{dai2022delay, zhai2020thzprism, rotman2016true}. {\color{black} In the existing literature, three different TTD architectures for near-field wide-band communication systems have been proposed, as illustrated in Fig.~\ref{4}.} For the \emph{parallel} configuration, the time delays of the TTD outputs are directly fed to the phase-shift network and are thus decoupled \cite{cui2022near,wang2023beamfocusing}. The authors of \cite{wang2023beamfocusing} proposed both a fully- and partially-connected TTD-based hybrid beamforming architecture for the near-field case, and showed that the sub-connected architecture exhibits higher energy efficiency and imposes fewer hardware limitations on the TTDs and system bandwidth. However, due to the large array aperture, parallel TTDs are usually constrained by the maximum possible delay of the TTD devices. To address this issue, the authors of \cite{wang2023ttd} proposed a serial TTD configuration that can effectively bypass the maximum delay limit by accumulating delays across multiple TTDs. However, compared to phase shifters, the cost and power consumption of TTDs are relatively high, which hinders their practical implementation. To balance the tradeoff between performance and cost, an alternative \emph{hybrid} TTD configuration was proposed in \cite{wang2023ttd} to avoid the excessive use of TTDs. With this architecture, although only a limited number of TTDs are inserted between the RF chains and phase shifters, near-optimal communication performance can be achieved. 

\begin{figure*}[ht] %hb代表放在文章底部，%ht为放在文章顶部
	\centering
	\includegraphics[width=5.8 in]{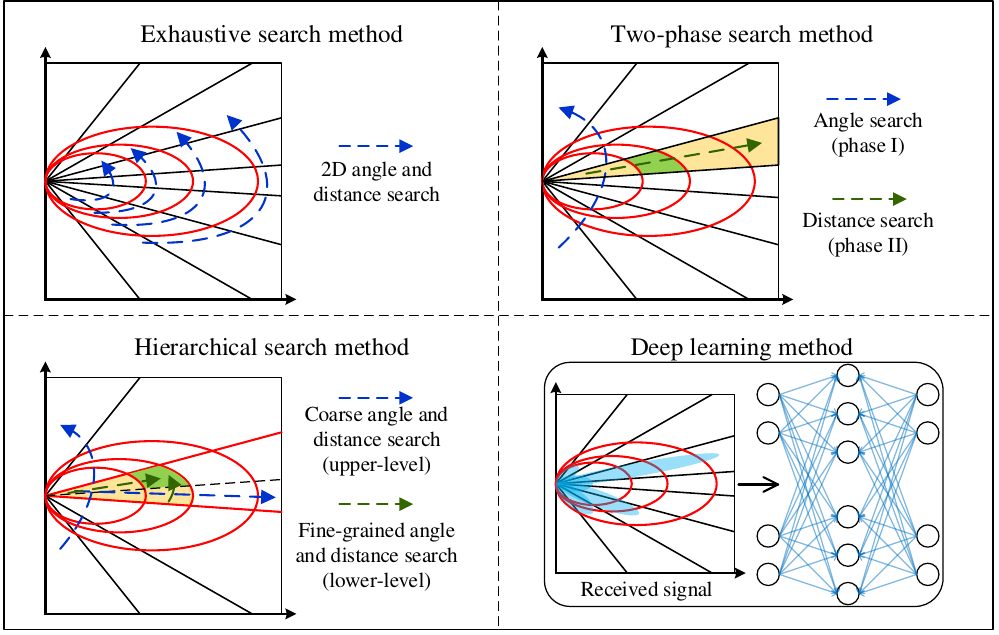}
    \caption{Illustration of different near-field beam training methods.}
    \label{5}
\end{figure*}

\subsubsection{Near-field Beam Training}
{\color{black}To achieve the potential XL-MIMO beam-focusing gain, channel state information (CSI) is indispensable, although obtaining CSI is more challenging for spherical rather than planar wavefront models.} At high-frequencies, beam training is an effective approach for establishing an initial high-SNR link between the BS and users without explicit CSI. The purpose of beam training is to search for the best beam from a dedicated codebook consisting of multiple candidate beam codewords. Therefore, the choice of the codebook directly determines the beam training efficiency. In far-field communications, discrete Fourier transform (DFT)-based angle domain codebooks are widely used. However, when a DFT-based codebook is adopted for near-field beam training, the user may receive high signal energy for multiple codewords due to the energy-spread effect. Therefore, the design of new codebooks based on both the angle and distance domains are needed for near-field beam training. Widely used near-field codebooks include Cartesian domain \cite{wei2022codebook}, \cite{han2020channel}, polar domain \cite{cui2022channel}, and slope domain codebooks \cite{shi2023chirp}. The Cartesian domain codebook is obtained by designing beam-focusing solutions for uniformly sampled points on a Cartesian grid, leading to a codebook with very high dimension and training overhead. The polar domain codebook is the one most commonly used for near-field beam training, employing a uniformly sampled angular grid and a non-uniformly grid for distance where the interval between grid points gradually increases, since larger distances result in smaller phase variations across the array \cite{cui2022channel}. The key idea for the slope-intercept domain codebook originates from linear frequency modulation signals in continuous wave radar, and the codewords are obtained by projecting each near-field steering vector onto one point in the slope-intercept domain \cite{shi2023chirp}.

{\color{black}Under the spherical wavefront channel model, various beam training schemes based on the above codebooks have been proposed to improve beam training accuracy and efficiency, as illustrated in Fig.~\ref{5}.} For narrow-band XL-MIMO systems in high-frequency bands, the LoS path is typically much more dominant than NLoS paths. For this case, the most straightforward near-field beam training method is to perform an exhaustive search over all polar domain codewords \cite{wei2022codebook}. However, this approach incurs a very high overhead, leaving less time for data transmission. To reduce the training overhead of exhaustive search, the authors of \cite{zhang2022fast} proposed a two-phase near-field beam training method that first estimates the candidate user angles using a far-field DFT-based codebook, and then identifies the user distance via the polar-domain codebook given the candidate user angles. This method was further improved in \cite{wu2023near}, where the authors proposed to directly employ the DFT codebook to jointly estimate the user angle and range by leveraging the key observation that the user angle is approximately located at the middle of a defined angular support and the user distance can be inferred from the width of the beam's angular support. In addition, \cite{hu2023design} proposed a three-dimensional (3D) near-field beam training method in which a two-dimensional (2D) angle-domain far-field beam training is conducted first, followed by a one-dimensional (1D) distance-domain beam training. Unlike the two-phase beam training method in \cite{zhang2022fast} that first estimates the user angle, \cite{wang2023near} proposed to first estimate the user distance based on a codebook containing omnidirectional beams. Subsequently, the optimal beam angle is determined via an angle domain search.

\begin{table*}[!htp] \small
	\caption{Comparison of beam training overhead of different methods.}
	\centering
	\renewcommand{\arraystretch}{1.2}
	\setlength{\arrayrulewidth}{0.9pt}
	\begin{tabular}{|c|c|c|}   \hline
		\textbf{Beam training method} 					&  	\textbf{Key ideal}         &  	\makecell{\textbf{Order of beam training overhead}\\ $U$: number of angular domain codewords\\ $S$: number of distance domain codewords} 	\\   \hline 
		Exhaustive search  \cite{cui2022channel}    &   Search all polar-domain codewords    &	$\mathbb{O}(US)$  \\   \hline 
		Two-phase search  \cite{zhang2022fast,wu2023near,hu2023design}      &    \makecell{First search in the angle domain, \\ and then search in the distance domain} 		       &	 $\mathbb{O}(U+S)$	  \\   \hline    
		Hierarchical search  \cite{lu2023hierarchical,chen2023hierarchical,wu2023twonfh}  &     \makecell{   Search from the lowest-resolution near-field \\ codebook  to the highest-resolution \\ codebook layer by layer}	  &	 $\mathbb{O}(\log U + \log S)$	     \\   \hline   
		DL method \cite{liu2022deep,jiang2023near}        &	Search using DL methods        &	 --   \\  \hline      
		\end{tabular}				
\end{table*}

Although the two-phase beam training methods greatly reduce the training overhead compared to an exhaustive search, in practice the required training overhead is still large. To address this issue, efficient hierarchical near-field beam training methods have recently been proposed  \cite{he2015suboptimal}. For example, the authors of \cite{chen2023hierarchical} proposed an adaptive near-field hierarchical codebook with gradually reduced beam coverage in the angle and distance domains, where lower-layer codebooks are designed to provide coverage for the Fresnel region, while upper-layer codebooks are devised to generate a desired beam pattern towards the target locations using beam rotation and relocation techniques. 
Leveraging a similar idea, \cite{lu2023hierarchical} presented a 2D hierarchical near-field beam training method based on hybrid beamforming by solving a multi-resolution optimization problem that minimizes the error between the ideal beam pattern in the near-field hierarchical codebook and that achieved by practical codewords based on hybrid beamforming. In \cite{shi2023chirp}, the authors proposed a novel spatial-chirp beam-assisted near-field hierarchical codebook in which the near-field channel model is represented by a joint angle and slope intercept. Additionally, \cite{wu2023twonfh} integrated hierarchical beam training into the aforementioned two-phase near-field beam training process and proposed a new two-stage hierarchical codebook design for accelerating near-field beam training. In this approach, a coarse user angle is first estimated by activating only a portion of the XL-array and applying conventional far-field hierarchical beam training. Then, a finer-grained user angle and distance are resolved by designing a dedicated near-field hierarchical codebook. In addition, near-field multi-beam methods have also been studied recently to reduce the beam training overhead of existing single-beam training methods. It is worth noting that the conventional sub-array based multi-beam training methods designed based on far-field channel models cannot be directly applied in the near-field scenario, since different sub-arrays may observe different user angles and there exist coverage holes in the angular domain. To address these issues, recent works \cite{zhou2024near2,zhou2024near} have proposed to design the near-field multi-beam codebook and beam training by sparsely activating a portion of antennas to form a sparse linear array, hence generating multiple beams simultaneously by effective exploiting the near-field grating-lobs. Moreover,deep learning (DL) methods have also recently been proposed to significantly reduce the training overhead by learning the mapping between the optimal beam codeword and the received pilot signals. For instance, \cite{liu2022deep} proposed to independently predict the optimal beam angle and distance using neural networks whose inputs are the received signal powers generated by a set of wide far-field beams. The beam training accuracy of this method was further improved in  \cite{jiang2023near} by jointly predicting both the optimal beam angle and distance using DL methods.  

For wide-band XL-array systems, the beam-split phenomenon may affect beam training performance since variations in frequency will defocus the beams at the target location. The authors of \cite{liu2022deep} showed that TTD devices can be used for actively controlling beam-split in near-field beam training.  In addition, \cite{cui2021near} designed an efficient ``rainbow'' beam training method for wide-band XL-MIMO systems that simultaneously generates multiple beams pointing to all the sampled angles at the same distance, while the XL-MIMO array controls the beam sweep at different ranges for determining the best beam distance. 

\begin{table*}[!ht]
\caption{Summary of main design challenges for near-field communications.}
\label{table:1}
    \centering
    \setlength{\tabcolsep}{3pt} % Adjust column spacing
    \renewcommand{\arraystretch}{1.5} % Adjust row height
\begin{tabularx}{\textwidth}{
    |>{\centering\arraybackslash}X
    |>{\centering\arraybackslash}X
    |>{\centering\arraybackslash}p{4cm}|
    }
        \Xhline{1.1pt}
        \textbf{Design Challenges} & \textbf{Solutions} & \textbf{References} \\ \hline
        \multirow{3}{*}{Near-field Beamforming} & Analog Beamforming & \cite{wu2023location,kosasih2023finite,ding2023resolution,li2022analytical,xie2023near,zhang2023mixed,lu2021does,li2022near,zhi2023performance} \\ \cline{2-3}
        ~ & Digital Beamforming & \cite{wang2024tutorial,croisfelt2021accelerated,xie2023performance,xu2023low,sun2021low,wang2020expectation,amiri2019message} \\ \cline{2-3}
        ~ & Hybrid Beamforming & \cite{cui2021near,chen2020hybrid,dai2022delay,zhai2020thzprism,rotman2016true,cui2022near,wang2023beamfocusing,wang2023ttd} \\ \Xhline{1.1pt}
        \multirow{5}{*}{Near-field Beam Training} & Exhaustive Search & \cite{wei2022codebook} \\ \cline{2-3}
        ~ & Two-phase Search & \cite{zhang2022fast,wu2023near,hu2023design} \\ \cline{2-3}
        ~ & Hierarchical Search & \cite{he2015suboptimal,chen2023hierarchical,lu2023hierarchical,shi2023chirp,wu2023twonfh} \\ \cline{2-3}
        ~ & DL-based Beam Training & \cite{liu2022deep,jiang2023near} \\ \cline{2-3}
        ~ & Wideband Beam Training & \cite{liu2022deep,cui2021near} \\ \Xhline{1.1pt}
        \multirow{4}{*}{Near-field Channel Estimation} & CS-based Estimation & \cite{cui2022channel,guo2023compressed,jiang2023near,lu2023near} \\ \cline{2-3}
        ~ & DL-based Estimation & \cite{lei2023channel,cui2023near,elbir2023near} \\ \cline{2-3}
        ~ & Hybrid-field Estimation & \cite{wei2021channel,hu2022hybrid,yang2023practical,tarboush2024cross,chen2021hybrid,yu2023adaptive} \\ \cline{2-3}
        ~ & VR-aware Estimation & \cite{han2020deep,tian2023low,han2020channel,liu2023location, iimori2022joint, iimori2022grant} \\ \Xhline{1.1pt}
    \end{tabularx}
\end{table*}
\subsubsection{Near-field Channel Estimation}

Directly estimating the CSI based on pilot signals is also possible for near-field scenarios with XL-MIMO arrays, although clearly more channel parameters must be estimated per pilot. Substantial research efforts have recently been made to reduce the channel estimation overhead by exploiting channel sparsity. {\color{black}However, conventional angle-domain sparsity does not hold for XL-MIMO systems due to the near-field spherical wavefronts.} It was shown in \cite{cui2022channel} that near-field channels typically exhibit sparsity in the polar domain, a fact that can be exploited for reducing the channel estimation overhead via compressive sensing (CS) techniques. For narrow-band systems, \cite{cui2022channel} proposed on-grid polar domain simultaneous orthogonal matching pursuit (P-SOMP) and off-grid iterative weighted channel estimation methods, both of which were demonstrated to achieve superior channel estimation accuracy compared to existing angle-domain counterparts. Additionally, \cite{guo2023compressed} developed a near-field channel estimation method for planar antenna arrays that solves a complex three-dimensional CS task using a triple parametric decomposition. The authors of \cite{jiang2023near} designed an efficient sparse Bayesian learning algorithm for near-field channel estimation based on CS using off-grid sampling in the angle and distance domains. {\color{black} For the mixed LoS/NLoS near-field channel model in \eqref{equ_3}, the authors of \cite{lu2023near} proposed a hierarchical parameter estimation and orthogonal matching pursuit (OMP) algorithm to estimate both the LoS and NLoS path components.} Apart from CS-based methods, DL methods have also proven to be effective for near-field channel estimation. For example, \cite{lei2023channel} devised a multiple residual dense network (RDN) to estimate the near-field channels by exploiting polar-domain sparsity. Atrous spatial pyramid pooling RDN and a polar-domain multi-scale RDN channel estimation schemes were proposed to further improve the accuracy. For wide-band systems, a bilinear pattern detection method was developed in  \cite{cui2023near} to estimate near-field wide-band channels, which was inspired by the classical SOMP technique. In addition, \cite{elbir2023near} devised a near-field wide-band OMP approach based on a beam-squint-aware dictionary.

For hybrid-field channels that include both near- and far-field components,  channel estimation generally is more challenging due to the coupling among the near- and far-field channel paths. To tackle this issue, \cite{wei2021channel} proposed to leverage  OMP techniques to estimate the hybrid-field channel, while \cite{hu2022hybrid} derived an efficient algorithm with low channel estimation overhead using the support detection orthogonal matching pursuit (SD-OMP) algorithm. However, these methods require prior information on the number of far-field and near-field channel paths, which may not be available in practice. This issue was addressed in \cite{yang2023practical}, which presented an efficient method for finding the number of far-field and near-field channel paths before employing the hybrid OMP method. For wide-band hybrid-field channels, \cite{tarboush2024cross} proposed a reduced dictionary method to determine the appropriate channel model for each user, followed by a cross-field channel estimation algorithm for XL-MIMO THz systems. The authors of \cite{chen2021hybrid,yu2023adaptive} employed fixed point networks to estimate the near-field channel parameters.

{\color{black}To account for near-field spatial non-stationarity, VR-aware channel estimation methods have also been recently proposed for XL-MIMO systems \cite{han2020deep}, \cite{tian2023low}, \cite{han2020channel}, \cite{liu2023location, iimori2022joint, iimori2022grant}.} 
 For example, in \cite{han2020channel}, subarray-wise and scatterer-wise channel estimation schemes were developed for estimating the spatially non-stationary XL-MIMO channel. In the subarray-wise method, the visible scatterers for each subarray can be posited based on a refined OMP algorithm. The scatterer-wise method utilizes the array gain to localize the scatterer while simultaneously determining its mapping to the subarrays. In \cite{tian2023low}, the authors proposed a low-overhead channel reconstruction scheme that is able to simultaneously locate the user and identify its VR. A model-driven DL-based downlink channel reconstruction scheme was proposed in \cite{han2020deep} using a powerful neural network referred to as ``You Only Look Once'' to detect the angles and delays of the channel paths and identify the VRs of the scatterers. In \cite{liu2023location}, the authors established a VR model for XL-MIMO systems that relates a user's VR to its location given the VR information of other users. In addition, \cite{iimori2022grant,iimori2022joint} studied channel estimation methods for stochastic models with VR. In \cite{iimori2022grant},  a novel bilinear inference-based joint activity and channel estimation (JACE) algorithm was proposed to jointly estimate the associated channel coefficients, user activity patterns and VR. In \cite{iimori2022joint}, the authors further proposed an iterative JACE algorithm and utilized an EM-based auto-parameterization method to estimate instantaneous sub-array activity factors, which can enhance the practicality of the proposed algorithm.

In Table \ref{table:1}, we summarize the design issues for near-field communications along  with solutions developed to date.

\subsection{Future Directions}

\subsubsection{Hybrid-field Communications}

In the exiting literature, most works have considered cases where all environmental scatterers are located either in the near- or far-field of the XL-array \cite{chen2023beam}. However, for scenarios where the Rayleigh distance is not large, it may happen that some scatterers closer to the BS reside in the near-field region, while others are located in the far-field, hence leading to hybrid-field multi-path channels and introducing new challenges. Although a few recent works have investigated hybrid-field channel modeling and estimation methods, more research is still needed for this new area. For example, it would be interesting to characterize the spatial non-stationarity of hybrid-field channels accounting for VRs. Moreover, in wide-band systems, the beam-split effect becomes more complicated in hybrid-field channels, thus calling for the development of new TTD-based beamforming designs. In addition, for multi-cell systems, inter-cell interference in hybrid channels becomes more involved and new interference management schemes should be devised.

\subsubsection{DL for Near-field Communications}

Compared with far-field planar wavefronts, near-field spherical wavefronts differ in the nonlinear phase variations they produce across a ULA.
The transceiver design issues arising from such non-linear phase variations may be tackled using DL methods. Although several works have developed different DL frameworks for designing efficient near-field beam training and channel estimation approaches (see, e.g., \cite{lu2023tutorial,liu2023near}), more in-depth and comprehensive studies are still needed for near-field communications in more complicated scenarios. For example, in mixed- or hybrid-field channels, end-to-end DL methods can be developed for efficient CSI acquisition and beamforming design, thus avoiding separate approaches for each phase. In addition, the challenging design of coordinated beamforming among multiple BSs for heterogeneous wireless networks with near-field and far-field users can be tackled by developing efficient AI techniques for robust networking control. 

\section{New Form: Antenna Transceiver Design}\label{sec_nform}

Here we introduce three new-form NGAT technologies for enhancing the performance of conventional wireless communication systems, including RISs, FAs, and HMIMO. 

\subsection{Reconfigurable Intelligent Surfaces: From Metals to Metamaterials}
RISs (or equivalently IRSs) have emerged as a promising technology for reconfiguring the radio propagation environment to enhance energy and spectral efficiency \cite{wu2019towards,di2020smart,wu2021intelligent,liu2021reconfigurable,wu2023intelligent,8796365,10380596}. Generally speaking, RISs are digitally controllable metasurfaces equipped with a large number of reflecting/refracting elements.
The reflection/refraction properties of the RIS elements can be dynamically adjusted to shape the propagation behavior of electromagnetic waves, hence allowing for real-time and dynamic control of wireless links to enhance ubiquitous communication performance \cite{you2021enabling,kang2021irs} and enabling various applications such as edge computing and edge learning \cite{zhou2020delay,mao2023roar}.
RISs are generally divided into three categories: nearly-passive, active, and simultaneous transmitting and reflecting RIS (STAR-RIS). In the following, we provide an overview of these three types of RISs.
% \footnote{A more recent RIS architecture referred to as beyond diagonal RIS (BD-RIS) has been proposed, in which different RIS elements are connected via controllable impedance to provide further DoFs for improving the performance of RIS-assisted systems \cite{9913356}.}.

\subsubsection{Nearly-passive RISs}
Unlike conventional active transceivers equipped with power-consuming RF chains, nearly-passive RISs are composed of a large number of low-cost reflecting/refracting elements, whose properties are tuned by a smart controller that consumes low power  \cite{wu2019intelligent,huang2019reconfigurable}. Thanks to low-power control circuits, these RIS elements can individually adjust the phase shifts and/or amplitudes of incident signals. 
% for signal power enhancement and/or interference mitigation. 
The reflected/refracted signals can be combined constructively to enhance the signal power of a desired link or destructively to eliminate undesired interference.

\begin{figure*}
\centering
\includegraphics[width=0.9\textwidth]{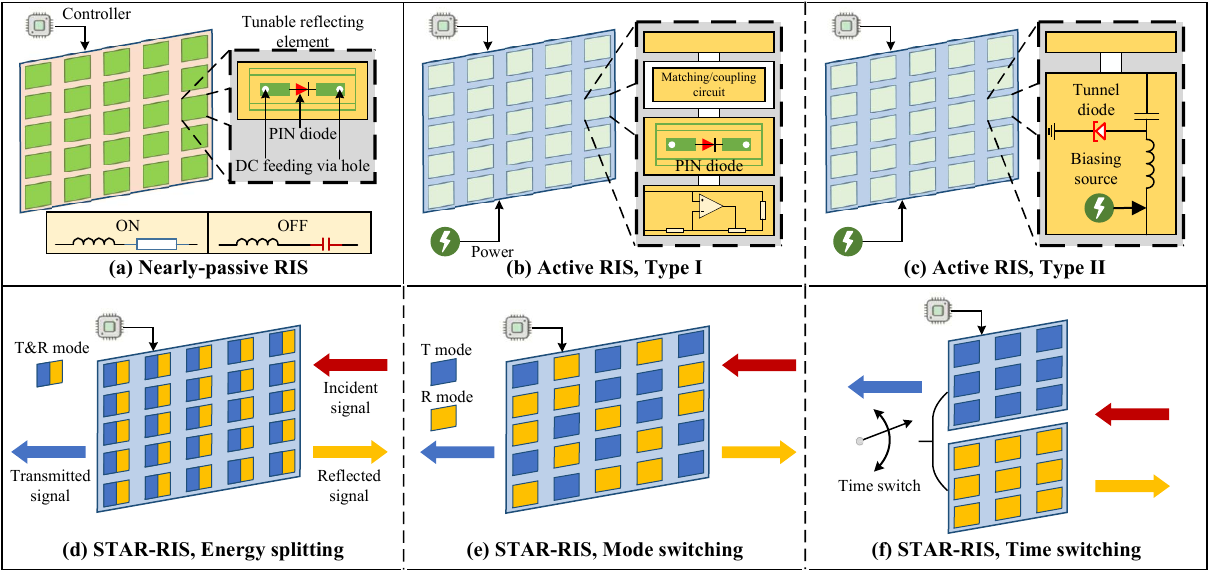}
\caption{Architectures for different types of RIS.}
\label{ris_fig}
\end{figure*}

\paragraph{Hardware Architecture and System Model}
\textcolor{black}{Generally speaking, 
there are three main approaches for controlling the signal reflections/refractions from nearly-passive RIS \cite{nayeri2018reflectarray}: 1) mechanical actuation, 2) functional materials such as liquid crystals and graphene, and 3) electronic devices such as positive-intrinsic-negative (PIN) diodes, varactor diodes, field-effect transistors (FETs), or micro-electromechanical system (MEMS) switches. Among these, the first approach has recently gained considerable popularity in practical implementations due to its rapid response, minimal reflection/refraction loss, and relatively low energy consumption and hardware cost. For example, as shown in Fig.~\ref{ris_fig}(a), applying different biasing voltages through a direct-current (DC) feed to the PIN diode can toggle it between its ON and OFF states, thus enabling the element to introduce a phase-shift difference of $\pi$ in the incident signal.}
Since reflection- and refraction-based RISs follow similar models and involve similar signal processing techniques, we mainly consider reflection-based RISs in the rest of this subsection. 

\begin{figure}
\centering
\includegraphics[width=0.45\textwidth]{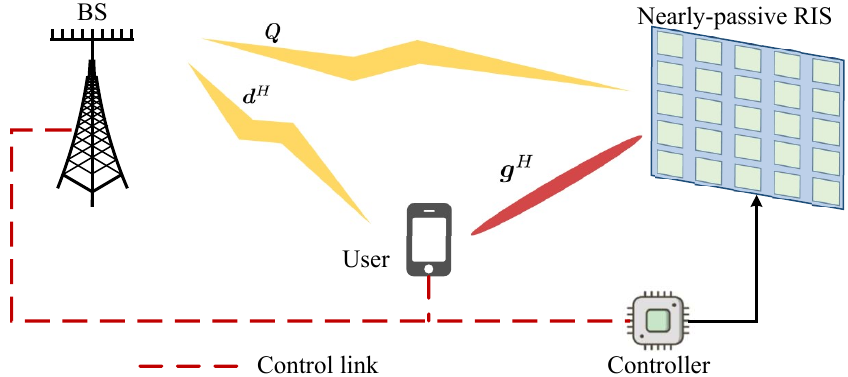}
\caption{Illustration of a nearly-passive RIS-assisted MISO system.}
\label{ris_sys}
\end{figure}

We introduce the model for the downlink of a nearly-passive RIS-assisted downlink multiple-input-single-output (MISO) system such as that shown in Fig.~\ref{ris_sys}, where the RIS is equipped with $M$ elements and the BS has $N$ antennas.
%We consider the challenging scenario where the BS-user direct channel is blocked by obstacles.
Let $\boldsymbol{g}^H\in \mathbb{C}^{1\times M}$, $\boldsymbol{Q}\in \mathbb{C}^{M\times N}$, and $\boldsymbol{d}^H\in \mathbb{C}^{1\times N}$ denote the RIS$\to$user, BS$\to$RIS, and BS$\to$user channels, respectively. The RIS reflection response is denoted by $\boldsymbol{\nu} = \left[\alpha_1 e^{j\theta_1}, \ldots,\alpha_m e^{j\theta_m},\ldots,\alpha_M e^{j\theta_M}\right]$, where $\alpha_m \in [0,1]$ and $\theta_m \in [0,2\pi )$ are the reflection amplitude and phase shift of the $m$-th element, respectively.
Based on the above setup, the effective channel from the BS to the user is given by  \cite{wu2021intelligent}
%received signal $y\in \mathbb{C}$ at the user via the reflecting link of RIS can be modeled as \cite{wu2021intelligent}
% \begin{align}\label{sig_mo_pass}
%    \boldsymbol{h}^H  &= \boldsymbol{g}^H 
%    \boldsymbol{V}
%     \boldsymbol{Q}  +\boldsymbol{d}^H,
% \end{align} 
% where $\boldsymbol{V}\triangleq \operatorname{diag}\left(\boldsymbol{\nu}\right)$ denotes the diagonal reflection matrix and  $\boldsymbol{h}^H_r\triangleq \boldsymbol{g}^H \boldsymbol{V}\boldsymbol{Q} $ is referred to as the cascaded BS-RIS-user channel. 
\begin{align}\label{sig_mo_pass}
   \boldsymbol{h}^H  &= \boldsymbol{g}^H 
   \operatorname{diag}\left(\boldsymbol{\nu}\right)
    \boldsymbol{Q}  +\boldsymbol{d}^H,
\end{align} 
where $\boldsymbol{h}^H_r\triangleq \boldsymbol{g}^H \operatorname{diag}\left(\boldsymbol{\nu}\right)\boldsymbol{Q} $ is referred to as the cascaded BS-RIS-user channel. 
{\color{black}
The promising applications of nearly-passive RIS  have motivated extensive worldwide endeavors on prototyping, measurements, and experiments \cite{wu2023intelligent}. The performance gains of nearly-passive RIS have been demonstrated in practical wireless systems operating at frequencies ranging from sub-6 GHz to mmWave and even THz bands\cite{9020088,9551980,9749219,10373827,9999288}. Interested readers may refer to \cite{wu2023intelligent} for more details on the state-of-the-art implementation of nearly-passive RIS.
}

\paragraph{Performance Gains}
The key performance advantages of nearly-passive RISs are mainly due to two properties. First, RISs can greatly extend the coverage of wireless systems by establishing virtual LoS links between the BS and users. This property is particularly appealing when the direct BS-user links experience blockages. Second, RISs are able to achieve a {squared} SNR scaling order due to the beamforming and aperture gains, which is higher than those of conventional active relays and MIMO systems. Specifically, \cite{wu2019intelligent} showed that for an RIS-assisted single-input-single-output (SISO) system with a sufficiently large number of reflecting elements $M$, the received signal power at the user increases quadratically with $M$. In other words, it is possible to scale down the transmit power of the BS by $1/M^2$  without compromising the received SNR.
The above analysis assumes a far-field channel model and is not accurate for $M\rightarrow \infty$ since it violates the energy conservation law. Thus,  near-field channel modeling needs to be considered for extremely large-scale RIS systems, for which the total received signal power will saturate when $M\to\infty$.

\paragraph{Transceiver Design Issues} 
To reap the substantial passive beamforming gains offered by RISs, it is indispensable to acquire accurate CSI for the links between the RIS and its associated BS/users. However, this task is challenging since a nearly-passive RIS lacks RF chains for transmission/reception of pilot signals, and the large number of RIS elements means a very large number of channel coefficients must be estimated. To address these issues, efficient cascaded RIS channel estimation methods have been proposed in the literature (see e.g., \cite{wu2023intelligent,zheng2022survey,swindlehurst2022channel} and references therein) to directly estimate the cascaded BS-RIS-user channel instead of the individual BS-RIS and RIS-user channels. Research efforts have primarily focused on how to reduce the prohibitively high CSI estimation overhead. Typical solutions include RIS element grouping \cite{yang2020intelligent,you2020channel}, exploiting shared BS-RIS links \cite{wang2020channel,zheng2020intelligent,10028753}, channel sparsity \cite{swindlehurst2022channel,zhou2022channel}.
These cascaded RIS channel estimation methods require significant modifications to existing CSI estimation/training protocols, and require coordination between the BS and RISs. Moreover, the RIS CSI is estimated based on pilot signals received at the users, which may not be feasible in practice for portable devices \cite{wu2023intelligent}. 
To address these issues, an alternative approach is to design the RIS response using received signal power measurements at the users, which are easier to acquire in existing cellular networks, e.g.,  using the Reference Signal Received Power, or RSRP. This approach also eliminates the need for pilot signals dedicated for explicit RIS-related CSI estimation.

For high-frequency bands that typically suffer from severe path-loss and shadowing, codebook-based beam training is an efficient approach to establish high-SNR links between the BS and users by exploiting channel sparsity. The major challenge lies in reducing the beam training overhead. Several new and efficient beam training methods have been recently proposed in the literature for both far- and near-field scenarios using hierarchical RIS codebooks \cite{wang2023hierarchical,wu2023twonfh,alexandropoulos2022near}, multi-beam training \cite{you2020fast}, and machine learning (ML)-based methods \cite{10100676}. On the other hand, for low-frequency bands, blind channel estimation \cite{ren2022configuring} and channel recovery methods \cite{yan2023channel,sun2023user} have been proposed for acquiring CSI based on received signal powers. For example, \cite{ren2022configuring} presented a method based on the conditional sample mean to design the RIS reflection coefficients. In addition, \cite{yan2023channel} proposed to first estimate the autocorrelation matrix of the RIS-cascaded channel based on received signal power prior to designing the RIS coefficients. The computational complexity of this method was further reduced in \cite{ren2022configuring} by designing a neural network-based channel estimation scheme using power measurements at the users.

{\color{black}
To facilitate the implementation of RISs in future wireless systems, several practical issues need to be addressed. First, RIS/transceiver hardware constraints should be taken into account for nearly-passive RIS beamforming designs. For example, the common assumption of continuously adjustable RIS phase shifts/amplitudes is not valid in practice, as RIS control will likely be based on the use of a finite number of discrete phase shifts, thus resulting in combinatorial RIS beamforming optimizations. To tackle this challenge, 
various optimization methods have been proposed to obtain high-quality suboptimal RIS designs using, e.g., the branch-and-bound method \cite{wu2019beamforming,di2020hybrid}, the relax-and-quantize technique \cite{you2020channel,wu2019beamforming}, element-wise block coordinated descent (BCD) \cite{xiu2021secrecy}, and penalty-based methods \cite{zhao2021exploiting}.
Moreover, as revealed in \cite{abeywickrama2020intelligent}, another practical hardware constraint is that the reflection amplitude of each RIS element is coupled with its phase shift. This issue can be properly addressed using, e.g., element-wise BCD methods that account for the phase-dependent amplitude for both narrow-band \cite{zeng2023resource,abeywickrama2020intelligent} and wide-band systems \cite{li2021intelligent}.} 

{\color{black}Second, various transceiver/RIS hardware impairments (such as RIS phase noise, transmitter/receiver RF impairments, quantization errors, amplifier non-linearity) may degrade the performance of RIS-assisted systems. For instance, in \cite{khel2021effects}, the RIS phase noise was shown to significantly degrade the achievable ergodic capacity. In \cite{badiu2019communication}, it was shown that RIS-assisted systems can still achieve the squared SNR scaling law and the linear diversity order scaling as the number of RIS elements grows, although the achievable rate is degraded due to the phase error. The authors of \cite{zhou2020spectral} showed that the performance degradation in the high-SNR regime is mainly attributed to the transceiver's hardware impairments rather than the RIS phase noise, since the RIS beamforming simultaneously affects the desired signal and imperfection-induced noise.  Channel impairments due to wet or dirty RIS surfaces may also affect communication performance in practice, although this issue has not been well studied and thus deserves future investigation.} 

{\color{black}Under the above hardware imperfections as well as channel estimation errors, robust RIS beamforming designs have been developed to achieve reliable communications \cite{9293148,9316283,9973349}. For example, the authors of \cite{9293148} modeled the estimation error as Gaussian noise and proposed a low-complexity algorithm to optimize the RIS phase shifts subject to outage probability constraints. To reduce the computational complexity for optimizing the RIS phase shifts under estimation error, \cite{9316283} proposed a novel penalty dual decomposition-based algorithm for an RIS-assisted MISO system. In \cite{9973349}, an RIS-aided massive MIMO system subject to imperfect aggregated CSI is considered, and a gradient ascent method was proposed to maximize the minimum user rate based only on statistical CSI.}
{\color{black} Third, an RIS installed by one operator may cause interference to users of another operator in an adjacent channel, since RIS elements, although designed for use at a particular frequency, reflect wireless signals over a wider frequency band with different amplitudes and phase shifts. This can produce undesired pilot contamination even without the presence of interfering signals or intra-band pilot reuse.
To address this issue, \cite{10504275} proposed to exploit orthogonal RIS configurations during uplink pilot transmissions, which effectively mitigates the inter-operator pilot contamination. However, studies on the effect of inter-operator interference due to RIS are still limited to theoretical analyses, and deserve more comprehensive investigation.
}

\subsubsection{Active RISs}
\textcolor{black}{Nearly-passive RIS wireless systems  face  practical performance limitations due to the significant  reflection-induced product-distance path loss \cite{9206044,9433568,9119122}. This issue can be mitigated by either increasing the number of RIS elements or by placing the RIS in close proximity to the transceivers. However, these approaches may encounter practical hurdles due to the RIS size and constraints on the RIS deployment.
To address these concerns, \emph{active} RISs have recently been proposed to facilitate simultaneous signal reflection and amplification, thus providing a means for mitigating the severe product-distance path loss compared to nearly-passive RIS systems \cite{long2021active,you2021wireless,zhang2022active,kang2023activemag}.}

\paragraph{Hardware Architecture and System Model}

Generally speaking, active RISs can be implemented by two types of hardware architectures. The first uses the cascaded amplifying and phase-shifting circuit shown in Fig.~\ref{ris_fig}(b), which is the combination of an operational amplifier (op-amp) and a phase shifter \cite{kishor2011amplifying}. The weak incident signal is amplified by the op-amp prior to phase-shift adjustment by a PIN- or varactor diode circuit.
The second active RIS implementation can be realized using load modulation circuits based on components like tunnel diodes \cite{amato2018tunneling} and complementary metal-oxide-semiconductor (CMOS) devices \cite{landsberg2017low}, as shown in Fig.~\ref{ris_fig}(c). {\color{black}
Discussion of an active RIS prototype that operates at $3.5$ GHz frequency band and has  $64$ elements for signal amplification can be found in \cite{10001687}.} 
Taking the tunnel diode \cite{amato2018tunneling} as an example, application of a proper bias voltage allows it to operate in a region exhibiting negative differential resistance, thereby establishing an active load impedance (or equivalently a negative resistance) for the reflecting element. Consequently, the amplitude of the reflected signals is increased by converting DC power into RF power, while the phase shift is adjusted by tuning the circuit capacitance. While the hardware and energy cost of an active RIS is larger than for nearly-passive RISs, it is still considerably smaller than for conventional full-duplex (FD) relays \cite{kang2023activemag}. 

For an active RIS, let $x_m \in \mathbb{C}$ denote the incident signal on the $m$-th active RIS element, where $m \in \{1,\ldots, M\}$ and  $M$ is the number of RIS elements. With reflection-based amplifiers supported by a power supply, the reflected and amplified signal at the $m$-th active RIS element can be modeled as
\begin{align}\label{sig_mo_act}
    y_m = \alpha_m e^{j\theta_m} \left(x_m + z_m\right), 
\end{align} 
where $\alpha_m \ge 1$ denotes the gain of the $m$-th element due to the integrated reflection-type amplifier, and $\theta_m \in [0,2\pi )$ denotes the reflection phase shift. A key difference with active RISs is that they introduce non-negligible amplification noise, denoted by $z_m \in \mathbb{C}$  in (\ref{sig_mo_act}).

\paragraph{Performance Gains}
The key metric for demonstrating the performance gains of active-RIS aided wireless systems lies in understanding how the received SNR scales with an asymptotically large number of reflecting elements \cite{li2022active}. 
First, consider the scenario where there is a maximum power constraint for all active RIS elements. For this case, the authors of \cite{long2021active}  showed that the received SNR experiences a linear increase with $M$ when $M$ is sufficiently large. This phenomenon is expected because the signal power at the receiver grows quadratically with $M$, owing to the RIS beamforming gain. At the same time, the amplification noise power increases linearly with $M$, thus resulting in an SNR scaling order of $M$. However, an active-RIS-aided systems attains a smaller SNR scaling order ($\mathcal{O}(M)$) than its nearly-passive RIS counterpart (i.e., $\mathcal{O}(M^2)$), it outperforms the nearly-passive RIS in terms of rate  when the number of reflecting elements is moderate and/or its amplification power is sufficiently large \cite{long2021active}. This is due to the power amplification gain offered by active RISs, and the limited beamforming gain of nearly-passive RISs for values of $M$ that are not very large. In another scenario where there is a fixed per-element power at each active-RIS element for double-active-RIS systems, the total power budget scales linearly with the number of active elements \cite{kang2023double}. For this case, it was shown that for double-active-RIS systems, the received SNR scaling order for the far-field case is $\mathcal{O}(M^2)$, owing to the additional amplification power scaling of $\mathcal{O}(M)$. On the other hand, if the total amplification power is fixed, the double-active-RIS systems attains a linear capacity scaling order only, which is the same as the single-active-RIS case \cite{li2023double}.

\paragraph{Transceiver Design Challenges}
A key challenge in the design of active-RIS is accounting for the non-negligible RIS amplification noise. An active-RIS-aided system must balance the objective of  maximizing the received signal power and minimizing the active-RIS correlated noise at the receiver. This interplay gives rise to non-convex constraints on the reflection coefficients, rendering the optimization problems more challenging. On the other hand, the amplification factors of active-RIS elements are more flexible, resulting in more DoFs compared to nearly-passive RISs for which the reflection amplitudes are typically fixed at unity. 

Various methods have been proposed to optimize the active-RIS communication performance from different perspectives. For example, \cite{long2021active} proposed an alternating optimization (AO) algorithm to maximize the SNR for an active-RIS-assisted SIMO system. Equipping each RIS element with a dedicated amplifier may not be cost-effective due to the high hardware and energy cost of active-RIS elements, and several active-RIS hardware architectures have been proposed, such as hybrid and sub-connected active RISs \cite{nguyen2022hybrid,liu2021active,kang2023active}.
Specifically, \cite{nguyen2022hybrid} proposed a hybrid active RIS architecture in which a few active RIS elements are included among conventional nearly-passive RIS elements. Instead of a fully-connected active RIS, a sub-connected active RIS architecture was proposed in \cite{liu2021active} to reduce the energy consumption and hardware complexity, where active-RIS elements are grouped into multiple sub-surfaces, each supported by one power amplifier, thus offering a flexible tradeoff between minimizing the power consumption and maximizing the communication performance. The authors of \cite{kang2023active} further proposed to deploy both nearly-passive and active RIS elements to enable cooperative beamforming. 

Since active RISs operate in FD mode, self-interference can degrade system performance. Note that the self-interference of active RISs is generally different from that of conventional FD relays. Specifically, due to the long processing delay at relays, their self-interference originates from different symbols transmitted in adjacent time slots and is usually modeled as colored Gaussian noise, which can be mitigated by conventional beamforming methods \cite{lioliou2010self}. However, due to the near-instantaneous behavior of active RISs, the incident and reflected signals carry the same symbol in the same time slot. Due to the non-ideal inter-element isolation in active RISs, part of  reflected signals may be received again by the active RIS, thus resulting in self-interference feedback \cite{zhang2022active}. In such cases, the active-RIS self-interference may not be considered as Gaussian noise, which renders conventional self-interference suppression approaches inapplicable. To address this issue, \cite{zhang2022active} modeled the self-interference feedback and proposed alternating direction method of multipliers (ADMM) and sequential unconstrained AO minimization techniques to suppress the active-RIS self-interference. 

\subsubsection{STAR-RISs}
Reflection-based RISs face the fundamental limitation that both the source and destination have to be located on the same side (i.e., the reflection half-space) of the RIS, which limits the ability of RISs to achieve full-space coverage. To overcome this limitation, a new type of simultaneously transmitting and reflecting RIS (STAR-RIS) was proposed in \cite{xu2021star}.
STAR-RISs enable both signal reflection and transmission/refraction, such that signals incident on the surface can be redirected on both sides of the STAR-RIS \cite{10133841}. As such, a full-space smart radio environment can be realized by STAR-RISs with $360^{\circ}$ coverage. This type of RIS is also referred to as an intelligent omni-surface (IOS) \cite{zhang2020beyond}.

\paragraph{Hardware Architecture and System Model}
\label{starris_mode}
To describe the STAR-RIS model, let $x_m \in \mathbb{C}$ denote the incident signal on the $m$-th RIS element, where $m \in \{1,\ldots, M\}$ and $M$ is the number of STAR-RIS elements. The signals reflected and transmitted by the $m$-th element can be modeled as $r_m = x_m \alpha^r_m e^{j \theta^r_m} $ and $t_m = x_m \alpha^t_m e^{j \theta^t_m} $, respectively, where $\alpha^r_m \in [0,1]$ (or $\alpha^t_m \in [0,1]$) and $\theta^r_m \in [0, 2\pi)$  (or $\theta^t_m \in [0, 2\pi)$) respectively denote the amplitude and phase shift of the $m$-th element's reflection (or transmission) coefficients.
Following the energy conservation law and assuming no thermal loss, the incident signal power should equal the sum of the reflected and transmitted signal power, i.e., $\left|x_m\right|^2 = \left|r_m\right|^2+\left|t_m\right|^2$, thus yielding $\left|\alpha^r_m\right|^2+\left|\alpha^t_m\right|^2 = 1$.
By properly adjusting the amplitude coefficients for signal reflection and transmission, a STAR-RIS element can operate in full reflection mode (i.e., $\alpha^r_m = 1, \alpha^t_m = 0$), full transmission mode (i.e., $\alpha^r_m = 0, \alpha^t_m = 1$), or simultaneous transmission and reflection mode (i.e., $\alpha^r_m \in [0,1], \alpha^t_m \in [0,1]$). Based on the above operating modes, there are three possible operating protocols for STAR-RISs, namely energy splitting (ES), mode switching (MS), and time switching (TS)\cite{mu2021simultaneously}.

\begin{itemize}
\item \textbf{Energy splitting:}
In ES, all STAR-RIS elements operate in simultaneous transmission and reflection mode, where the power of the signals incident on each element
is split into transmitted and reflected signals, as shown in Fig.~\ref{ris_fig}(d).
For the ES case, the optimization of both the transmission and reflection coefficients for each element provides a higher number of DoFs for system performance optimization. However, the larger number of design variables also introduces a higher design complexity and more demanding overhead in the information exchange between the BS and STAR-RIS. 
\item \textbf{Mode switching:} 
In MS, the elements of the STAR-RIS are divided into two groups: one consists of elements operating in transmission mode, while the other consists of elements operating in reflection mode, as shown in Fig.~\ref{ris_fig}(e). In this case, we have $\alpha^r_m\in \{0,1\}$, $\alpha^t_m\in \{0,1\}$, and $\left|\alpha^r_m\right|^2+\left|\alpha^t_m\right|^2 = 1$. An MS-based STAR-RIS can be conceptualized as comprising a conventional reflection-based RIS and a transmission-based RIS, both with reduced sizes. The selection of the mode for each element, along with the corresponding transmission and reflection phase shift coefficients, are jointly optimized. However, MS generally cannot achieve the same passive beamforming gain as ES, since only a subset of the RIS elements are selected for transmission and reflection. 
\item \textbf{Time switching:} 
In contrast to the ES and MS modes, the TS protocol periodically switches all elements between the transmission and reflection mode in different time slots, as shown in Fig.~\ref{ris_fig}(f). Let $0\leq \rho_r \leq 1$ and $0\leq \rho_t \leq 1$ denote the percentage of time allocated for reflection and transmission, respectively, where $\rho_r + \rho_t = 1$. Unlike the ES and MS protocols, the use TS decouples the design of the transmission and reflection coefficients, which simplifies the RIS transmission and reflection coefficient optimization. However, the TS mode requires stringent time synchronization.
\end{itemize}
{\color{black}
The authors of \cite{9895224} have built an IOS-aided wireless communication prototype that validates its ability to achieve full-space coverage.
}

\paragraph{Performance Gains}
The key advantage of STAR-RIS is extending the one-sided coverage of conventional RIS aided wireless networks, thanks to its ability to serve users on both its sides.
Motivated by the above, STAR-RIS designs have recently been intensely investigated for various scenarios.
For example, \cite{zhang2020beyond} considered an easy-to-implement phase shift model for STAR-RIS/IOS where the same phase shift for each element is applied to both the transmission and reflection links. In addition, \cite{xu2021star} considered a more general model for phase shift and amplitude control in which the STAR-RIS transmission and reflection responses are decoupled, and showed that significant coverage extension and the full diversity order can be achieved. The coverage maximization problem was investigated in \cite{wu2021coverage} by jointly optimizing the  BS resource allocation and the STAR-RIS transmission and reflecting coefficients.

\paragraph{Transceiver Design Challenges}

To reap the performance gains of STAR-RISs, accurate CSI acquisition is of paramount importance. However, the channel estimation schemes for nearly-passive RISs cannot be directly applied since STAR-RISs require CSI for both the reflection and transmission links, the channel estimation depends on which protocol (ES, MS, TS) that is adopted as discussed in Section~\ref{starris_mode}, and both the training transmission and reflection patterns must be jointly designed during training to minimize the estimation error. To address the above issues, \cite{wu2021channel} proposed efficient channel estimation algorithms for ES- and TS-based STAR-RIS. For the TS-based case, the cascaded channels on the transmit and reflection sides are estimated separately. For the ES case, a more realistic phase-shift model was considered in which the transmission-and-reflection coefficients are coupled. To solve this latter problem, an optimal solution is first obtained under the ideal independent transmission-and-reflection phase-shift model, based on which a sub-optimal solution satisfying the practical phase-shift model is obtained.
For MS-based STAR-RIS-assisted non-orthogonal-multiple-access (NOMA) systems, two distinct transmission protocols were proposed in \cite{wu2023two} to efficiently obtain STAR-RIS CSI for maximizing the average achievable sum-rate. First, for the case with LoS dominant channels, the long-term STAR-RIS transmission and reflection coefficients were optimized based on statistical CSI, while the short-term power allocation at the BS was designed based on the estimated fading channels of all users. Second, for rich scattering environments, the STAR-RIS surface is first partitioned based on statistical CSI, and then the BS estimates the instantaneous CSI of each sub-surface, based on which the power allocation and STAR-RIS phase-shifts are optimized.

Most of the existing works on STAR-RIS assume that the phase shifts of the transmission and reflection coefficients are independent, which requires complex RIS hardware designs in practice. To evaluate the effect of coupled transmission and reflection phase shifts, \cite{liu2022simultaneously} studied the power consumption minimization problem for a STAR-RIS-assisted multi-user SISO system under the coupled phase-shift model. It was shown in  \cite{10133914} that the performance degradation due to the coupled transmission and reflection phase shifts is mild for multi-user MISO THz systems. Other work in \cite{9774942} investigated beamforming design for a multi-user SISO system taking EM theory and the law of energy conservation. into account. In addition, an AO-based approach was proposed in \cite{9751144} to maximize the weighted-sum-rate of a multi-user MISO system, where the transmission and reflection coefficients were alternately found with the other being fixed.
However, the optimality of the proposed designs in \cite{9774942,9751144} cannot be guaranteed. To address this issue, a general optimization framework for STAR-RIS-assisted systems was proposed in \cite{9935266}, which was shown to converge to the Karush-Kuhn-Tucker (KKT) optimal solution under some mild conditions. In this approach, the amplitudes and phase-shifts were optimized alternately in closed form via the penalty dual decomposition (PDD) approach.

STAR-RISs can exploit their full-space coverage to enhance the performance of NOMA networks via joint optimization of the active beamforming/resource allocation at the BS, as well as the transmission and reflection coefficients at the STAR-RIS \cite{wu2021coverage,9786807}. For example, \cite{wu2021coverage} characterized the coverage of  STAR-RIS-assisted networks to showcase their effectiveness in enhancing NOMA network performance. However, the required joint optimization is more challenging than for orthogonal-multiple-access (OMA) systems due to the coupled optimization variables. To address this issue, \cite{9740451} proposed a three-step approach to maximize the sum-rate of STAR-RIS-assisted multi-carrier NOMA networks, where the decoding orders, beamforming-coefficient vectors, and power allocation were optimized by semidefinite programming, convex upper bound approximations, and geometric programming. Moreover, the authors of \cite{9863732} proposed a two-layer iterative framework for maximizing the achievable sum-rate of a STAR-RIS-assisted NOMA system by decoupling the decoding order optimization problem. The more practical scenario with coupled transmission and reflection phase shifts further complicates the optimization of STAR-RIS for NOMA systems and thus deserves further investigation.

\subsubsection{Future Directions}

The existing literature on RIS has mostly considered single-RIS systems or multi-RIS systems with only a single reflection at each RIS. More in-depth research efforts are needed to study how to extend existing designs to more general and more complicated scenarios with multi-reflection links between multiple RISs, (see e.g., \cite{mei2022intelligent,you2020wireless,zheng2021double,zhang2023multi}). Moreover, existing RIS research has mainly focused on static or quasi-static wireless channels, which are not applicable in high-mobility scenarios characterized by fast time-variations, such as vehicle-to-everything (V2X) communications. Rapidly time-varying channels significantly complicate passive beamforming design, CSI acquisition, and RIS deployment and deserve further investigation \cite{9661068,10028753,you2022deploy,10316541}.

{\color{black}
Furthermore, most existing works on RIS assume that each RIS element operates in isolation from other elements without considering inter-element connections or effects (i.e., mutual coupling). Under such an idealized model, the associated RIS reflection/refraction matrix is diagonal.
To fully exploit the performance gain of RISs, a more recent RIS architecture, referred to as beyond diagonal RIS (BD-RIS), has been proposed in \cite{9913356} to exploit interconnections among different RIS elements. Specifically, the RIS elements are connected to each other via controllable impedances to provide further DoFs for controlling the propagation environment, and lead to models in which the reflection/refraction matrix is not limited to being diagonal\cite{10316535}.
However, the non-diagonal reflection/refraction matrix results in more system variables to optimize for achieving high-gain RIS beamforming and accurate channel estimation, which leads to increased complexity.
}

\subsection{Flexible Antennas: From Static to Movable}

In contrast to the conventional MIMO systems with fixed-position antennas (FPAs), flexible antennas (FAs), also known as \emph{fluid antennas} \cite{wong2020fluid} or \emph{movable antennas} \cite{zhu2023movable}, introduce a dynamic structure in which some or all of the antennas can adjust their locations within a designated space. Some FA designs allow near-continuous movement among possible locations (or ``ports'' \cite{wong2020fluid}) with very small inter-antenna spacing much less than half-wavelength. This type of reconfigurable design provides wireless systems with increased potential for diversity and multiplexing gains. 

\subsubsection{Hardware Architecture and System Model}
We first introduce two hardware architectures for implementing FAs: fluid antennas and movable antennas. 

\paragraph{Fluid Antennas}

\begin{figure*}
\centering
\includegraphics[width=0.85\textwidth]{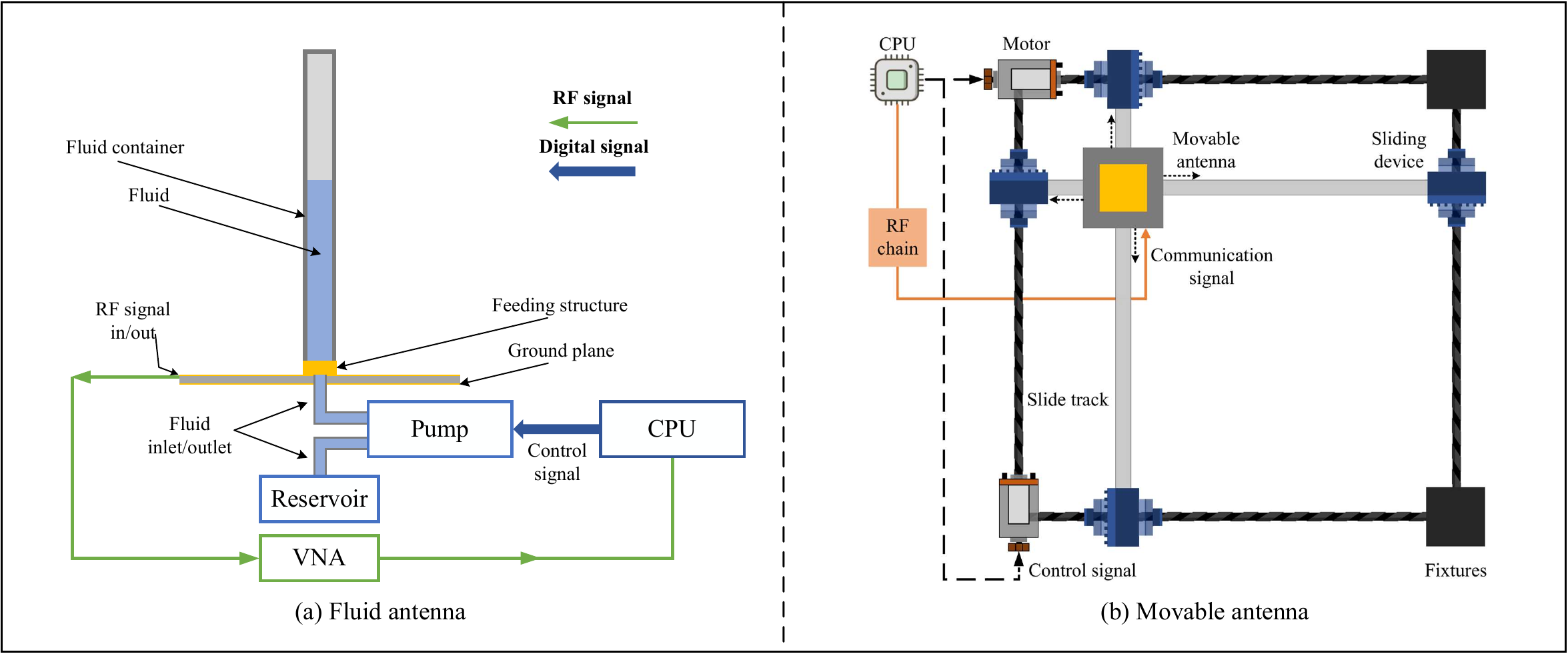}
\caption{Typical FA hardware architectures.}
\label{FA_fig}
\end{figure*}

Conventional antennas are usually composed of solid materials, such as metal wires, plates, and dielectric slabs. The predetermined geometries of these antennas generally limit their performance in dynamic wireless environments. 
However, an antenna that is ``fluid'' has no predefined form and thus can be molded into any desired shape to adapt to the wireless environment. As shown in Fig.~\ref{FA_fig}(a), a fluid antenna system typically consists of four key components \cite{wong2022bruce}: 1) a feed network for delivering/collecting the RF signal with a vector network analyzer (VNA) for signal measurement, 2) a movable fluid serving as the radiating element, 3) a fluid container and a reservoir for holding the fluid, and 4) a pump and central processing unit (CPU) for adjusting the antenna parameters. Based on the above architecture, the two most relevant types of fluid antennas are i) frequency-agile monopole fluid antennas, where the amount of the radiating fluid can be adjusted to control the resonant frequency \cite{borda2019low,singh2019multistate}, which is particularly suitable for applications where the transceiver must flexibly switch the operating frequency over a wide band, and ii) fluid antennas that change their radiation pattern by adjusting the position of the radiating fluid, e.g., using surface wave technology \cite{shen2020beam,shen2021reconfigurable}.

\paragraph{Movable Antennas}

Movable antenna systems share some similarities with distributed antenna system (DAS), where remote antenna units are distributed geographically within wireless networks \cite{castanheira2010distributed}. As shown in Fig.~\ref{FA_fig}(b), a movable antenna system typically consists of four main components: 1) a communication module resembling that of conventional fixed-position antenna systems, 2) the movable antenna connecting to the RF chain via a flexible cable to enable antenna movement, 3) an antenna positioning module such as a mechanical slide or rotation shaft driven by step motors, and 4) a CPU to facilitate signal processing and the control of the antenna positions. Despite their similarities, movable antenna systems stand apart from DAS in terms of application scenarios and system setups. In DAS, antennas are typically positioned far apart, resulting in independent channels, and their performance is mainly determined by the large-scale channel path loss. In contrast, movable antenna systems relocate the antennas within a confined region, typically with dimensions on the order of several to tens of wavelengths. This arrangement is more practical and requires shorter connecting cables compared to DAS. As such, the channels for multiple movable antennas exhibit inherent spatial correlation and thus the communication performance is predominantly determined by small-scale fading within the target region.
{\color{black} The authors of \cite{122111} have built a FA prototype for multiple access and experimentally evaluated the system performance in 5G mmWave bands, verifying the potential of FA to reduce outage probability and increase multiplexing gain.}

\paragraph{Channel Model}

\begin{figure}
\centering
\includegraphics[width=0.45\textwidth]{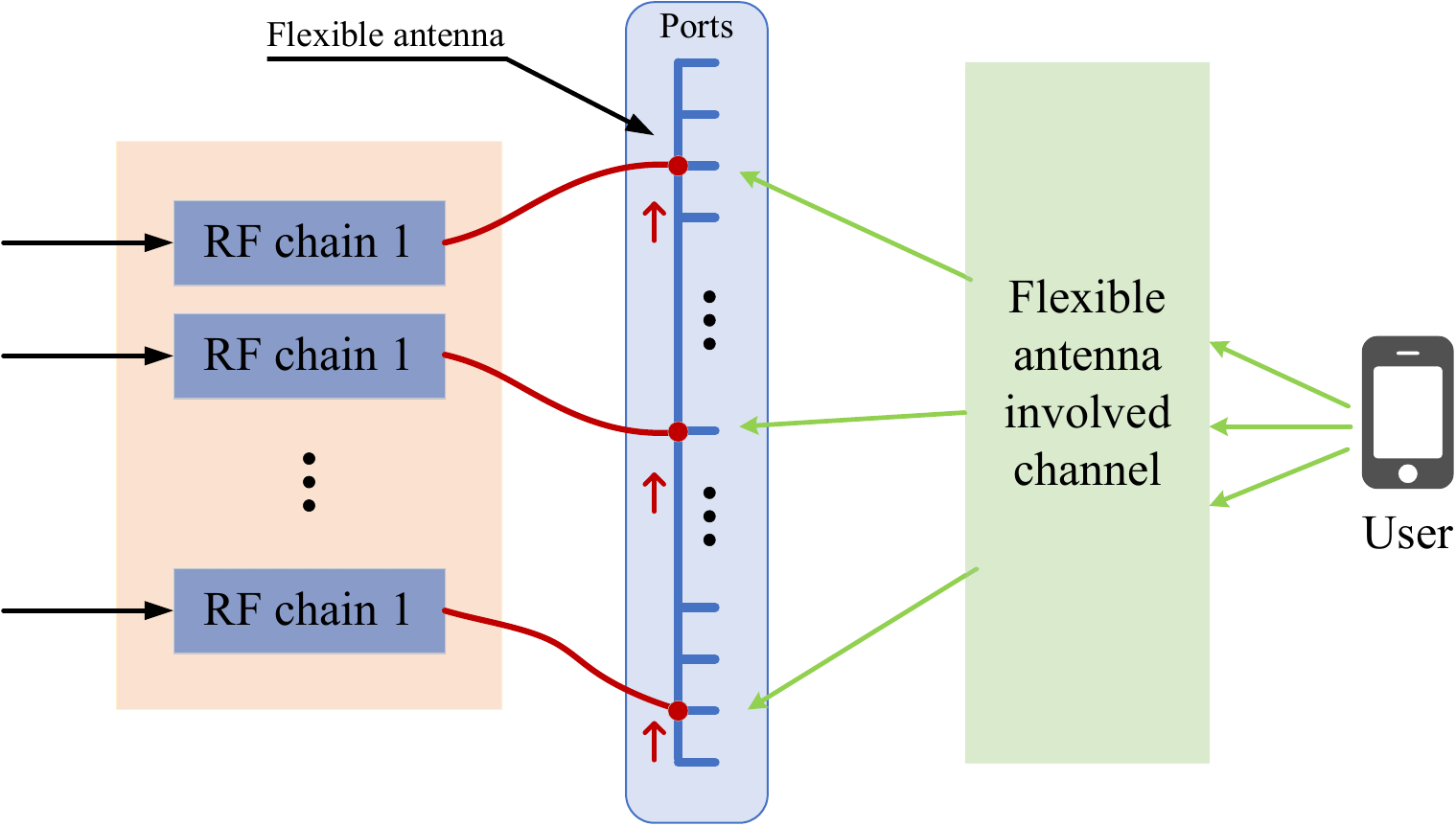}
\caption{Illustration of a movable antenna MISO system.}
\label{fa_sys}
\end{figure}

Fig.~\ref{fa_sys} illustrates an FA-enabled uplink MISO system in which a FAs (e.g., fluid or movable antennas) are deployed at the BS, which  consists of $N$ candidate locations (i.e., ports) and $M$ FA elements ($M \leq N$), each connected to an RF chain.
The location of each antenna can be switched to one of the $N$ candidate locations/ports. 
Let $\boldsymbol{h}\in\mathbb{C}^{N\times 1}$ denote the channel from the user to $N$ candidate locations/ports at the BS. 
To characterize the mobility of the FAs, let $\boldsymbol{S}\in\mathbb{R}^{M\times N}$ denote a switching matrix, where each entry of $\left[\boldsymbol{S}\right]_{m,n} \in \{0,1\}$ indicates whether the $m$-th antenna is located at the $n$-th port. This implies that $\|\left[\boldsymbol{S}\right]_{m,:}\| = 1,\forall m \in\{1,\ldots,M\}$, $\|\left[\boldsymbol{S}\right]_{:,n}\| = 1,\forall n \in\{1,\ldots,N\}$, and $\boldsymbol{S} \boldsymbol{S}^H = \mathbf{I}_M$, which are constraints imposed to ensure that $M$ of $N$ ports are selected at a given time instant. Based on the above, the signal vector $\boldsymbol{y}\in \mathbb{C}^{M\times1}$ received at the BS is given by
\begin{align}
    \boldsymbol{y} = \boldsymbol{S}\boldsymbol{h}x +\boldsymbol{z},
\end{align}
where $x \in \mathbb{C}$ denotes the signal transmitted by the user and $\boldsymbol{z} \in \mathbb{C}^{M\times1}$ is noise.

\subsubsection{Performance Gains} 
Compared with the conventional FPA case, FA systems provide appealing diversity and multiplexing gains, as elaborated below.

\paragraph{Diversity Gain}
For the SISO case, FA systems can significantly increase the received signal power at the users by exploiting spatial diversity via antenna position optimization or port selection, especially for a large  receiving aperture. In particular, for a given channel within a region of interest, the communication performance may be compromised if multiple FPAs are situated in positions with deep fading channels. In such scenarios, even with antenna selection (AS) or MRC at the receiver, the communication performance cannot be guaranteed. 
In contrast, the FAs can be moved to  positions with better channel quality to increase the received signal power. Thus, FA systems can more flexibly exploit the available spatial diversity by continuously moving the antennas to achieve maximum channel power gain \cite{zhu2023movableNull}.
The authors of \cite{wong2020fluid} demonstrated that by switching the antenna ports along a line, FA systems can significantly reduce the outage probability compared to conventional FPA systems and may even outperform multi-antenna systems with MRC.
For movement in a 2D space, \cite{zhu2023modeling} proposed a new field-response model leveraging the amplitude, phase, and angle of arrival/angle of departure (AoA/AoD) information of a multiple-path channel and demonstrated that the FA system can achieve greater performance gains as the number of channel paths increases, owing to the more pronounced small-scale fading effects in the spatial domain.

\paragraph{Multiplexing Gain}
The capacity of MIMO systems generally depends on the singular values of the channel matrix.
In contrast to conventional FPA systems in which the channel matrix is fixed given the transmitter/receiver locations, FA systems are able to flexibly alter the positions/ports of the transmit/receive antennas and thus enable the reconfiguration of the channels to achieve higher capacity. For example, \cite{ma2023mimo} characterized the capacity of FA-enabled point-to-point MIMO systems under geometric multi-path channel models for both low- and high-SNR regimes. On the other hand, \cite{new2023information} considered rich scattering environments and characterized the capacity of FA-enabled MIMO systems. The optimal diversity and multiplexing tradeoff was also obtained to showcase the performance enhancement over FPA systems. In low-SNR scenarios, single-stream beamforming is optimal where all transmit power is allocated to the strongest eigen-channel. Hence, the positions of the movable antennas can be optimized to maximize the maximum singular value of the channel matrix. For high-SNRs, the transmit power is allocated to multiple eigen-channels based on the water-filling principle. Consequently, optimization of the antenna positions must balance the distribution of the singular values to maximize the capacity.

\subsubsection{Transceiver Design Challenges} In the following, we discuss the main transceiver  design issues  for FA systems.
\paragraph{Channel Estimation}
To fully leverage the performance gains in the spatial domain, it is crucial for FA systems to obtain accurate CSI at the transmitter/receiver. {\color{black}In general, the FAs should be moved to measure the channel response from each point of the transmit/receive region. While an exhaustive search of all possible antenna positions could be employed, this is impractical for larger transmit/receive regions.
To reduce the channel estimation overhead, various methods have been recently discussed in \cite{skouroumounis2022fluid,zhu2023modeling,wang2023estimation,ma2023compressed,zhang2023successive}. For example,
\cite{skouroumounis2022fluid} proposed a sequential linear minimum mean square error (MMSE) channel estimator in which multiple FAs simultaneously move to equally spaced ports for channel measurements, reducing the training overhead by exploiting the spatial correlation between measured and unmeasured ports.
In \cite{zhu2023modeling}, a field-response estimation method was presented where AoAs/AoDs and channel gains are estimated to reconstruct the channel map between the transmitter and receiver. The required estimation overhead is only contingent on the number of channel paths, instead of the size of the transmitting/receiving region as in the case of an exhaustive search. Under the assumption of perfect knowledge of the AoAs/AoDs of all channel paths, \cite{wang2023estimation} proposed to estimate the channel gains with low overhead using least-squares (LS). Exploiting channel sparsity in the angular domain, \cite{ma2023compressed} proposed a OMP-based approach to extract the sparse multipath channel parameters. The above FA channel estimation methods rely on assumptions such as spatially slow variations \cite{skouroumounis2022fluid}, prior knowledge of AoAs/AoDs \cite{wang2023estimation}, and angular domain sparsity\cite{ma2023compressed,10497534}. However, in cases where these assumptions do not hold, model mismatch may lead to significant performance loss. To address this issue, \cite{zhang2023successive} proposed a successive Bayesian algorithm to find a nonparametric estimate of the FA channel. In particular, the FA channels are modeled statistically and the channel uncertainty is successively eliminated using empirical kernel-based sampling and regression with a few carefully selected ports. Simulation results show that the proposed scheme in \cite{zhang2023successive} can achieve 50\% training overhead reduction compared with benchmark approaches.}

\paragraph{Position Optimization/Port Selection}
{\color{black}
The performance gains of FA over conventional FPA systems hinge on the antenna positioning or port selection. However, determining the optimal antenna positions that yield the best communication performance is very challenging due to the high-dimensionality and non-linearity of the channel response as a function of the candidate FA positions. Assuming CSI is available, \cite{ma2023mimo} alternately optimized the transmitter/receiver FA positions using successive convex approximation (SCA) for a MIMO FA system, while \cite{zhu2023modeling} adopted the gradient descent method and \cite{zhu2024performance} developed a parallel greedy ascent  algorithm. 
For multi-user MISO FA systems, \cite{10354003} proposed an alternating multi-directional descent (MDD) optimization framework to optimize the BS beamforming, 
the positions of user-side MAs and the user transmit powers, in order to significantly reduce the total transmit power compared to conventional FPA systems employing antenna selection.
For cases without CSI, an ML-based framework was proposed in \cite{zhu2023movable},  where the positions were optimized leveraging decisions provided by a deep neural network (DNN) and reinforcement learning. Low-complexity port selection schemes have been proposed for SISO FA systems \cite{chai2022port} and multi-user MISO FA systems \cite{waqar2023deep} with linearly-movable ports, where DL was used to infer the SINR at all available ports given only a small number of channel measurements at each user, based on which the best port location for enhancing the communication performance was selected. Moreover, efficient optimization methods were proposed in \cite{hu2023movable} for minimizing transmit power in multi-user FA systems.
} {
\color{black}
It is worth noting that FA-enabled systems generally entail significant time overhead for antenna movement/port switching. Pump-free electrowetting techniques can be used to continuously traverse the fluid radiator at a speed up to 10 mm/s \cite{9977471} for enabling fluid antenna systems, while the speed of the antenna motion is determined by the driving step motors \cite{zhu2023movable}.
}

\subsubsection{Future Directions}
%Despite the above promising solutions to the transceiver design issues of FAs, 
Research on FA-enabled systems is still in its infancy and needs further investigation. This includes the effect of hardware imperfections various FA-based architectures on, including imprecise positioning of the FAs. 
{\color{black}
The feedback signaling overhead (e.g., the necessary CSI) may be large in practice due to closely-spaced candidate locations (i.e., a large number of ports). However, most of the existing works have neglected the effects of the antenna-motion/port-switching time and feedback signaling on the performance of practical systems. This issue deserves in-depth investigations for different system setups considering different channel models. 
}
It is also of interest to investigate the interplay between FAs and RISs, e.g., flexible-element-enabled RIS and RIS-assisted FA-enabled systems, both of which provide additional flexibility for enhancing system performance.

\begin{figure*}
	\includegraphics[width=0.7\textwidth]{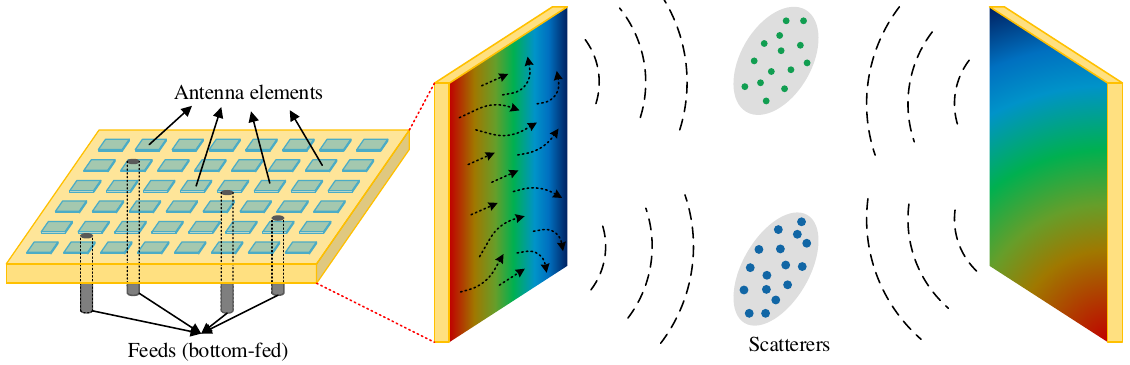}
	\centering
	\caption{Illustration of a communication system with HMIMO surfaces.}
	\label{fig:HMIMO}
\end{figure*}

\subsection{Holographic MIMO: From Discrete to Continuous}
HMIMO refers to the use of large, ultra-thin, \emph{spatially-continuous} arrays/surfaces consisting of a massive number of densely-spaced antennas/meta-atoms. Such arrays/surfaces are capable of recording, manipulating, and shaping the amplitudes and phases of EM waves with great flexibility and efficiency, thus enabling a magnitude of holographic-type applications (e.g., holographic imaging). Compared to conventional massive MIMO or RIS, HMIMO enjoys several new features and advantages. First, the nearly spatially-continuous aperture allows HMIMO to achieve unprecedented EM wave/field control with high flexibility. Thus, signal processing for HMIMO can be performed not only in the digital domain, but also in the EM domain to provide more finely tuned control for improving wireless communication efficiency. Second, similar to XL-MIMO, the extremely large apertures of HMIMO means the users will more likely be located in the near-field where {\color{black}beam-focusing} effect and higher spatial multiplexing gains can be exploited. Third, the use of metamaterials for HMIMO can greatly reduce its energy and hardware cost, hence achieving higher energy efficiency than conventional MIMO and RIS systems. 
{\color{black}
The promising applications of HMIMO have motivated several prototypes, measurements, and experiments. The first holographic antenna was manufactured in \cite{checcacci1970holographic}, demonstrating the possibility of visible microwave images. A dual circularly polarized HMIMO surface was fabricated and tested in the Ku band, validating the ability of the HMIMO design to obtain a given broadside radiation pattern \cite{pereda2018experimental}.}

\subsubsection{Hardware Architecture}
 
\begin{figure*}
\centering
\includegraphics[width=\textwidth]{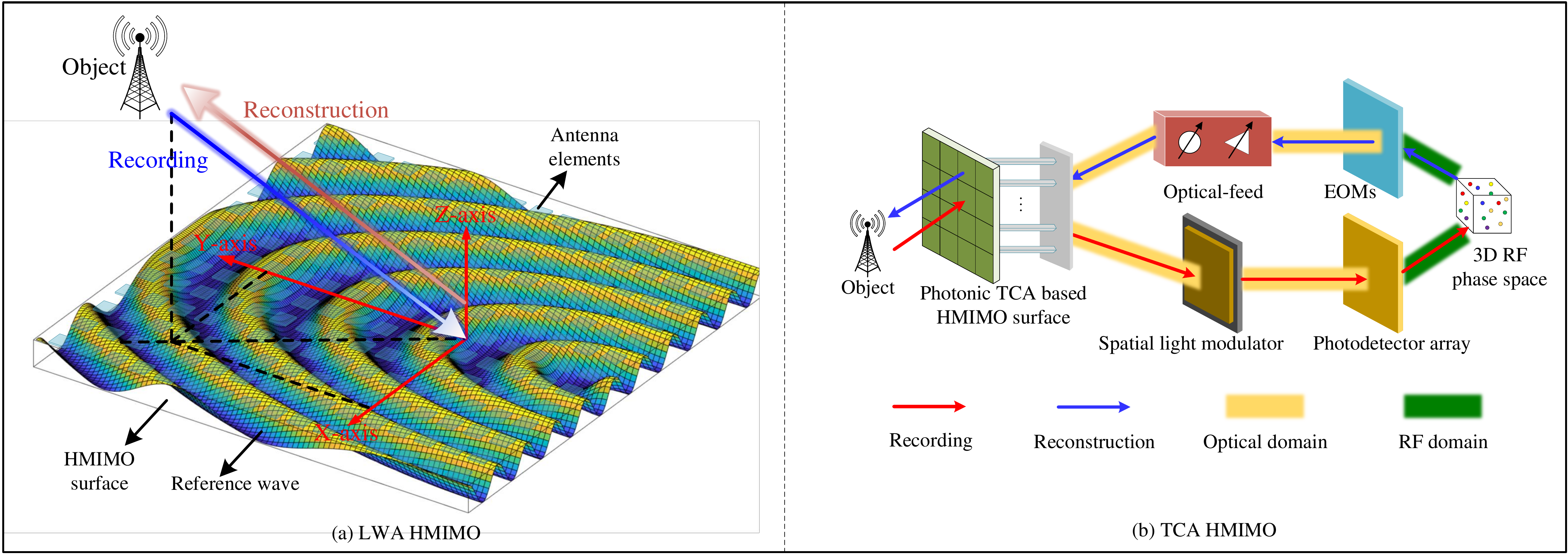}
\caption{{\color{black}Illustration of LWA and TCA HMIMO \cite{Gong_Vinieratou_Ji_Huang_Alexandropoulos_Wei_Zhang_Debbah_Poor_Yuen_2022}.}}
\label{LWA and TCA HMIMO}
\end{figure*}

{\color{black}There are two main categories of HMIMO implementations for achieving EM holography, including leaky-wave antenna (LWA) HMIMO \cite{araghi2021holographic} and photopic tightly coupled antenna array (TCA) HMIMO \cite{zong20196g}, as illustrated in Fig.~\ref{LWA and TCA HMIMO}.} Specifically, TCA-based HMIMO can transform RF signals into optical beams using electro-optic modulators and uni-traveling-carrier photo-detectors, thus realizing RF-optical mapping for implementing holography. On the other hand, LWA-based HMIMO realizes holography in the EM domain. Since signal processing for TCA-based HMIMO is carried out in the optical domain, in the following we primarily focus on LWA-based HMIMO.  

In LWA-based holography, the EM source is used for generating reference waves, while the EM antenna aperture or HMIMO surface is responsible for recording the hologram superimposed by the reference and object waves {\color{black}(see Fig. \ref{LWA and TCA HMIMO}(a))}. As such, the EM source and HMIMO surface essentially resemble a laser and photographic film in optical holography. Next, to reconstruct the object's EM wave after recording, one or more feeds are used for generating reference waves and illuminating the hologram on a carefully designed HMIMO surface. As illustrated in Fig.~{\ref{fig:HMIMO}}, HMIMO surfaces are primarily composed of three components: feeds, substrates, and antenna elements. The feeds are used for generating reference waves, and the substrate serves as a waveguide to funnel the reference waves to the antenna elements for radiation.

\subsubsection{System Model}
 
Compared with conventional massive MIMO, channel modeling for spatially-continuous HMIMO surfaces is more complicated due to the non-negligible spatial correlation and mutual coupling arising from the narrow inter-antenna spacing. The more densely the HMIMO antennas are packed, the stronger the spatial correlation and mutual coupling, and the more the channel rank is reduced relative to the channel dimension \cite{an2023tutorial}. The increased mutual couling also reduces the antenna radiation efficiency \cite{Balanis_1982}. For large-aperture HMIMO surfaces, near-field spherical wave propagation needs to be considered. However, for HMIMO, 
it is generally difficult to establish a self-contained and accurate channel model accounting for all the above channel characteristics. In the following, we introduce  HMIMO channel properties for discrete uniform planar arrays (UPAs) and spatially continuous planar surfaces.

\paragraph{UPA-based HMIMO Channel Model} For discrete UPA-based HMIMO, a deterministic channel model can be formulated using methods similar to those for near-field channels as described in Section~\ref{nf_chan_mod} and thus are omitted here. On the other hand, for stochastic channel modeling, the spatial correlation matrix of HMIMO in the near-field field can be expressed as
\begin{align}
	\mathbf{R}_{\rm SC}&=\mathbb{E}\left\{\mathbf{h} \mathbf{h}^{H}\right\}
	=\beta \iint f(\varphi, \theta) \mathbf{a}(\varphi, \theta) \mathbf{a}^{H}(\varphi, \theta) {\rm d} \theta {\rm d} \varphi,
\end{align}
where $ \varphi $ and $ \theta $ are the azimuth and elevation angle, $ \mathbf{a}(\varphi, \theta) $ represents the array response vector, $ \beta $ denotes the average channel gain, and $ f(\varphi, \theta) $ is the normalized {\it spatial scattering function} with $ \iint f(\varphi, \theta) {\rm d} \theta {\rm d} \varphi=1 $, accounting for the angular multipath distribution and the directivity gain of the antennas \cite{demir2022channel}.
%The authors of \cite{demir2022channel} further derived a tight approximation for the spatial correlation matrix and provided a subspace based channel estimation method.
To include the impact of the mutual coupling, the channel correlation matrix $ \mathbf{R}_{\rm MC}$ can be modeled by introducing the coupling effect into the array response as follows:
\begin{align}
	\mathbf{R}_{\rm MC}
		\! &=\! \beta\iint  f(\varphi, \theta) {\mathbf{C}^{\mathrm{T}} \mathbf{a}(\varphi, \theta) \mathbf{a}(\varphi, \theta)^{H} \mathbf{C}^* {\rm d} \varphi {\rm d} \theta } \nonumber \\
		\! &=\! \mathbf{C}^{\mathrm{T}} \mathbf{R}_{\rm SC} \mathbf{C}^*, \label{HMIMO:MuCoup}	
\end{align}
where element $ C_{m,n} $ of $ \mathbf{C} $ represents the mutual coupling coefficient related to the radiation pattern distortion between the $ m $-th and $ n $-th antennas \cite{sun2022characteristics}. The matrix $ \mathbf{C} $ simplifies to an identity matrix when there is no mutual coupling.
The authors of \cite{Williams_2022} leveraged circuit theory to describe an HMIMO channel model that accounts for the mutual coupling over both the air and waveguides for dynamic metasurface antennas (DMAs). In particular, the channel is modeled as a multi-port network with antennas that act as tunable admittances due to the reconfigurability of the metasurface, and the mutual coupling is characterized by the admittance matrix due to the interaction between the currents and different ports.

\paragraph{CPS-based Channel Model} 
To characterize the nearly-continuous aperture of an HMIMO surface, the tensor Green's function can be used to model the HMIMO LoS channel \cite{wei2023tri}. Mathematically, the tensor Green's function between a source point $ \mathbf{s} $ and receive point $ \mathbf{r} $ is given by
\begin{equation}
	{\small
	\begin{aligned}\label{tensor_green}
		{\mathbf{G}}\left(\mathbf{r}, \mathbf{s}\right) \triangleq& 
		\left(1+\frac{\imath}{k_0 r}-\frac{1}{k_0^2 r^2}\right) {\mathbf{I}} g\left(\mathbf{r}, \mathbf{s}\right) \\
		& +\left(\frac{3}{k_0^2 r^2}-\frac{3 \imath}{k_0 r}-1\right) {\mathbf{d}} {\mathbf{d}}^{T} g\left(\mathbf{r}, \mathbf{s}\right),
	\end{aligned}}
\end{equation}
where $ {\mathbf{I}} $ denotes the identity matrix, $ {\mathbf{d}} = \frac{{\mathbf{r}} -{\mathbf{s}}}{\Vert{\mathbf{r}} -{\mathbf{s}} \Vert}  $ represents the direction between the source and receiver, and $ g\left(\mathbf{r}, \mathbf{s}\right) \triangleq \frac{e^{\imath k_0|\mathbf{r}-\mathbf{s}|}}{4 \pi\left|\mathbf{r}-\mathbf{s}\right|} $ denotes the scalar Green's function. Based on \eqref{tensor_green} and Maxwell's equations, a bridge between the receiving and transmitting antennas was established in  \cite{wei2023tri}, which paves the way for a physically consistent channel model from the EM perspective. Mathematically, the channel between the $ n $-th transmitter and $ m $-th receiver is given by
\begin{equation}
	\boldsymbol{H}_{m,n}=\frac{\eta}{2 \lambda} \int_{S_R} \int_{S_T} \boldsymbol{G}\left(\boldsymbol{r}_m, \boldsymbol{s}_n\right) \mathrm{d} \boldsymbol{s}_n \mathrm{d} \boldsymbol{r}_m,
\end{equation}
where $ \eta $ represents the impedance of free-space, and $ S_R $ and $ S_T $ respectively denote the receive and transmit antenna apertures. This model can be used for mitigating the inter-user interference by proposing efficient precoding schemes for two HMIMO surfaces in parallel \cite{wei2023tri} or in an arbitrary arrangement \cite{gong2023holographic}.

For NLoS channels, the  Fourier plane-wave series expansion can be used to model the HMIMO channel \cite{Gong_Vinieratou_Ji_Huang_Alexandropoulos_Wei_Zhang_Debbah_Poor_Yuen_2022}. 
Specifically, when there is a unit impulse at the source point $ {\mathbf{s}} $,  the channel response at the receive point $ {\mathbf{r}} $ is given by \cite{pizzo2022fourier, pizzo2020spatially}
\begin{equation}
	\begin{aligned}\label{Fourier}
		& h(\mathbf{r}, \mathbf{s})=\frac{1}{(2 \pi)^2} \iiiint_{\mathcal{R}(\kappa) \times \mathcal{R}(\kappa)} a_r\left(k_x, k_y, \mathbf{r}\right) \\
		& \quad \times H_a\left(k_x, k_y, \kappa_x, \kappa_y\right) a_s\left(\kappa_x, \kappa_y, \mathbf{s}\right) d k_x d k_y d \kappa_x d \kappa_y,
	\end{aligned}
\end{equation}
where $ a_s\left(\kappa_x, \kappa_y, \mathbf{s}\right)\triangleq e^{-\jmath \mathbf{\kappa}^{\mathrm{T}} \mathbf{s}}=e^{-\jmath\left(\kappa_x s_x+\kappa_y s_y+\gamma\left(\kappa_x, \kappa_y\right) s_z\right)} $ with {\small$ \gamma\left(\kappa_x, \kappa_y\right) \triangleq \sqrt{\kappa^2-\kappa_x^2-\kappa_y^2}$} denotes the transmit response, 
$ a_r\left(k_x, k_y, \mathbf{r}\right)\triangleq e^{-\jmath \mathbf{k}^{\mathrm{T}} \mathbf{s}}=e^{-\jmath\left(k_x s_x+k_y s_y+\gamma\left(k_x, k_y\right) s_z\right)} $ with {\small$ \gamma\left(k_x, k_y\right) \triangleq \sqrt{k^2-k_x^2-k_y^2}$} represents the receive response,  
$ {\bf{\kappa}}=\left[\kappa_x,\kappa_y, \kappa_z\right]^T $  and $ \mathbf{k}=\left[k_x, k_y, k_z\right]^T $ respectively denote the transmit and receive wave vector,
{\small $ \mathcal{R}(\kappa) = \left\{\left(k_x, k_y\right) \in \mathbb{R}^2: k_x^2+k_y^2 \leq k^2\right\} $} denotes the integration region, and
$ H_a\left(k_x, k_y, \kappa_x, \kappa_y\right) $ represents the {\it wavenumber domain channel response}, which is related to the NLoS paths in the scattering environment.
By uniformly sampling the integration region to approximate the wavenumber angular response,  we can take the Fourier transformation to obtain \eqref{Fourier}.
This model is also applicable to the near-field scenario accounting for spherical wave propagation characteristics \cite{pizzo2022fourier}. Such Fourier plane-wave series channel models for HMIMO can be exploited to characterize the ergodic capacity for HMIMO communication systems \cite{pizzo2022fourier}. 
 
 \begin{table*}[!ht]
	\caption{Summary of Main Design Issues for New-form NGATs}
	\label{table:2}
    \centering
    \setlength{\tabcolsep}{3pt} % Adjust column spacing
    \renewcommand{\arraystretch}{1.5} % Adjust row height
\begin{tabularx}{\textwidth}{
    |>{\centering\arraybackslash}p{1.5cm}
    >{\centering\arraybackslash}p{2.3cm}
    |>{\centering\arraybackslash}X
    |>{\centering\arraybackslash}p{5cm}|
    }
        \Xhline{1.1pt}
        \multicolumn{2}{|c|}{\textbf{New-form NGAT}} & \textbf{Design Issues} & \textbf{References} \\ \Xhline{1.1pt}
        \multicolumn{1}{|c|}{\multirow{10}{*}{RISs}} & \multirow{4}{*}  {Nearly-passive RISs}  	
        & CSI Acquisition & \cite{wu2023intelligent,zheng2022survey,swindlehurst2022channel,yang2020intelligent,you2020channel,wang2020channel,zheng2020intelligent,10028753,zhou2022channel,9293148,9316283,9973349,wang2023hierarchical,wu2023twonfh,alexandropoulos2022near,you2020fast,10100676,ren2022configuring,yan2023channel,sun2023user} \\ \cline{3-4}
        \multicolumn{1}{|c|}{} & ~ & Beamforming Design & \cite{wu2019beamforming,di2020hybrid,you2020channel,xiu2021secrecy,10028753,zhao2021exploiting,abeywickrama2020intelligent,li2021intelligent} \\ \cline{3-4}
        \multicolumn{1}{|c|}{} & ~ & Hardware Impairments & \cite{khel2021effects,badiu2019communication,zhou2020spectral} \\ \cline{3-4}
        \multicolumn{1}{|c|}{} & ~ & Multiple RISs & \cite{mei2022intelligent,you2020wireless,zheng2021double} \\ \cline{2-4}
        \multicolumn{1}{|c|}{} & \multirow{3}{*}{Active RISs} & Beamforming Design & \cite{long2021active} \\ \cline{3-4}
        \multicolumn{1}{|c|}{} & ~ & High Hardware and Energy cost & \cite{nguyen2022hybrid,liu2021active,kang2023active} \\ \cline{3-4}
        \multicolumn{1}{|c|}{} & ~ & Self-interference & \cite{zhang2022active} \\ \cline{2-4}
        \multicolumn{1}{|c|}{}  & \multirow{3}{*}{STAR-RISs} & CSI Acquisition & \cite{wu2021channel,wu2023two} \\ \cline{3-4}
        \multicolumn{1}{|c|}{} & ~ & Transmission and Reflection Coupling & \cite{9774942,9751144,9935266} \\ \cline{3-4}
        \multicolumn{1}{|c|}{} & ~ & Enhance NOMA Networks & \cite{9740451,9863732} \\ \Xhline{1.1pt}
        \multicolumn{2}{|c|}{\multirow{2}{*}{FAs}} & Channel Estimation & \cite{skouroumounis2022fluid,wang2023estimation,ma2023compressed,zhang2023successive} \\ \cline{3-4}
        ~ & ~ & Position Optimization/Port Selection & \cite{ma2023mimo,zhu2023modeling,zhu2023movable,chai2022port,waqar2023deep} \\ \Xhline{1.1pt}
        \multicolumn{2}{|c|}{\multirow{3}{*}{HMIMO}} & Mutual Coupling & \cite{sun2022characteristics,Williams_2022,wang2022electromagnetic,bjornson2024towards,Marzetta_2019, An_Yuen_Huang_Debbah_Poor_Hanzo_2023, Han_Yin_Marzetta_2022} \\ \cline{3-4}
        ~ & ~ & EM Information Theory & \cite{sanguinetti2022wavenumber,zhang2023pattern,pizzo2022nyquist} \\ \cline{3-4}
        ~ & ~ & Channel Estimation & \cite{demir2022channel,wan2021terahertz,ghermezcheshmeh2023parametric,d2023dft,yu2023bayes} \\ \Xhline{1.1pt}
    \end{tabularx}
\end{table*}
 
To characterize the  mutual coupling for HMIMO surfaces, an EM-compliant channel model based on Fourier plane-wave series expansion theory was proposed in \cite{wang2022electromagnetic}. In this work, Hannan's efficiency, which reflects the maximum efficiency of an antenna in a dense array \cite{Kildal_Vosoogh_Maci_2016}, was introduced to characterize the mutual coupling. Hannan's efficiency can be expressed as $ \frac{\pi d_{x}d_{y}}{\lambda^2} $, which is proportional to the area allocated to the antenna. Obviously, when the antenna spacing is much smaller than a half wavelength, the radiation efficiency will decrease sharply, which significantly reduces the system capacity. The single-user EM-compliant channel model was extended in \cite{9779586} to the multi-user case, where the users are independently distributed in space.

\subsubsection{Transceiver Design Challenges} Several principal design challenges for HMIMO are discussed below.

\paragraph{Mutual Coupling} The effect of mutual coupling on HMIMO  performance has recently been studied in \cite{sun2022characteristics,Williams_2022,wang2022electromagnetic,bjornson2024towards}. It was originally shown in \cite{Balanis_1982} that when the antennas are densely arranged with sub-wavelength spacing much smaller than a half-wavelength, there is a significant reduction in the radiation efficiency. This thus results in a considerable capacity loss for HMIMO  systems \cite{Kildal_Vosoogh_Maci_2016}. The authors of \cite{bjornson2024towards} revealed that when the antenna spacing is narrowed to $ \lambda/4 $ or $ \lambda/6 $, the number of spatial DoFs respectively reduces to $1/2$ and $1/3$ of that of a collocated array with half-wavelength spacing given the same number of antennas. This indicates that it is generally not possible to increase the number of spatial DoFs by reducing the antenna spacing.
%and we should compensate the reduced element spacing by increasing the antenna number.
It was shown in \cite{wang2022electromagnetic} that when the antenna spacing is $ \lambda/8 $ at both the BS and user arrays (with no mutual coupling assumed at the user), HMIMO can achieve about three times the capacity than a corresponding collocated array, or only two times when Hannan's efficiency is reduced to $ 80\% $.
 
On the other hand,  mutual coupling may also provide some benefits for HMIMO  systems \cite{Marzetta_2019, An_Yuen_Huang_Debbah_Poor_Hanzo_2023, Han_Yin_Marzetta_2022}. For example, \cite{Marzetta_2019} showed that mutual coupling in HMIMO can be exploited to achieve {\it{super-directivity}} for attaining higher antenna array gains, which is particularly useful for wireless communication and WPT systems.
% both data transmission and wireless power transfer (WPT), which is termed as
Specifically, it was shown that when the antenna spacing approaches zero, the power transfer efficiency can be enhanced by a factor equal to the number of antennas compared to the scenario with no mutual coupling. This can be explained by the fact that a spectrum of plane waves characterized by super-wavenumber components with wavenumber greater than $ k $ are generated in the direction towards the user. Then, in the transverse direction, these waves exhibit evanescent behavior which decays exponentially fast, hence carrying solely reactive power and leading to the super-directivity in the transmit direction. For  HMIMO systems, this property implies that the array gain scales with the square of the number of antennas, which is beneficial for extending communication coverage without significantly increasing the transmit power. However, in practical implementations, resistive losses tend to significantly reduce the super-directive effect.

\paragraph{EM Information Theory}
While Shannon's information theory has been widely used in the design of modern communication systems, it is a mathematical abstraction that overlooks the specifics of the physical medium other than its statistical properties, i.e, the conditional probability distribution acting on the channel input. The emergence of HMIMO makes it possible to perform signal processing in the EM-domain instead of the digital domain only, hence giving rise to a new analytical framework referred to as {\it EM information theory} (EMIT)  \cite{di2024electromagnetic}. EMIT generally encompasses circuit theory, Maxwell's EM theory, and Shannon's information theory, and thus is envisioned to achieve improved channel capacity based on interdisciplinary techniques.
%Although the investigation of EMIT is still in its infancy, it has already raised numerous design issues, such as transmit pattern design and EM wave sampling.

From the EM perspective, the transmit current distribution can be regarded as the superposition of basic radiation transmit patterns: $ j(\mathbf{s})=\sum_{m=1}^M x_m \phi_m(\mathbf{s}), $ 
where $ \phi_m(\mathbf{s}) $ and $ M $ respectively denote the $ m $-th transmit pattern and the number of patterns.
How to design the patterns or \emph{basis functions} $ \phi_m(\mathbf{s}) $ is crucial to optimizing the HMIMO communication performance. To address this issue, 
\cite{sanguinetti2022wavenumber} first obtain the optimal transmit pattern for a single-user HMIMO system assuming a linear array aperture and far-field channels. Then, they proposed a simple suboptimal wavenumber-division multiplexing (WDM) scheme for the transmit pattern based on Fourier basis functions. However, due to the bounded channel response, the WDM scheme inevitably introduces interference among different communication modes.
To further improve the performance, a more general pattern-division multiplexing (PDM) scheme was proposed in \cite{zhang2023pattern}.

For practical implementations, the continuous EM waves must be sampled to facilitate digital signal processing. Therefore, effective spatial domain sampling schemes have recently been proposed to retain the maximum EM mutual information with the minimum number of samples. 
For example, an optimal Nyquist sampling and reconstruction scheme for the EM field was studied in \cite{pizzo2022nyquist} for arbitrary scattering environments. This work suggested regular hexagonal spatial domain Nyquist sampling, showing that sampling on quadrilateral grids with half-wavelength sides produced an increase of $13.4$ in redundancy compared with spatial domain Nyquist sampling. 

\paragraph{Channel Estimation}
Efficient and accurate CSI acquisition is essential for achieving the gains promised by HMIMO. When the spatial correlation across the array is known, MMSE-based channel estimation can achieve excellent performance. However, for HMIMO systems with nearly innumerable antennas, it is challenging to acquire the resulting high-dimensional spatial correlation matrix, and the resulting matrix inversion would require prohibitively high computational complexity. 

\begin{figure*} [t]
	\begin{centering}
		\includegraphics[width=0.650\textwidth]{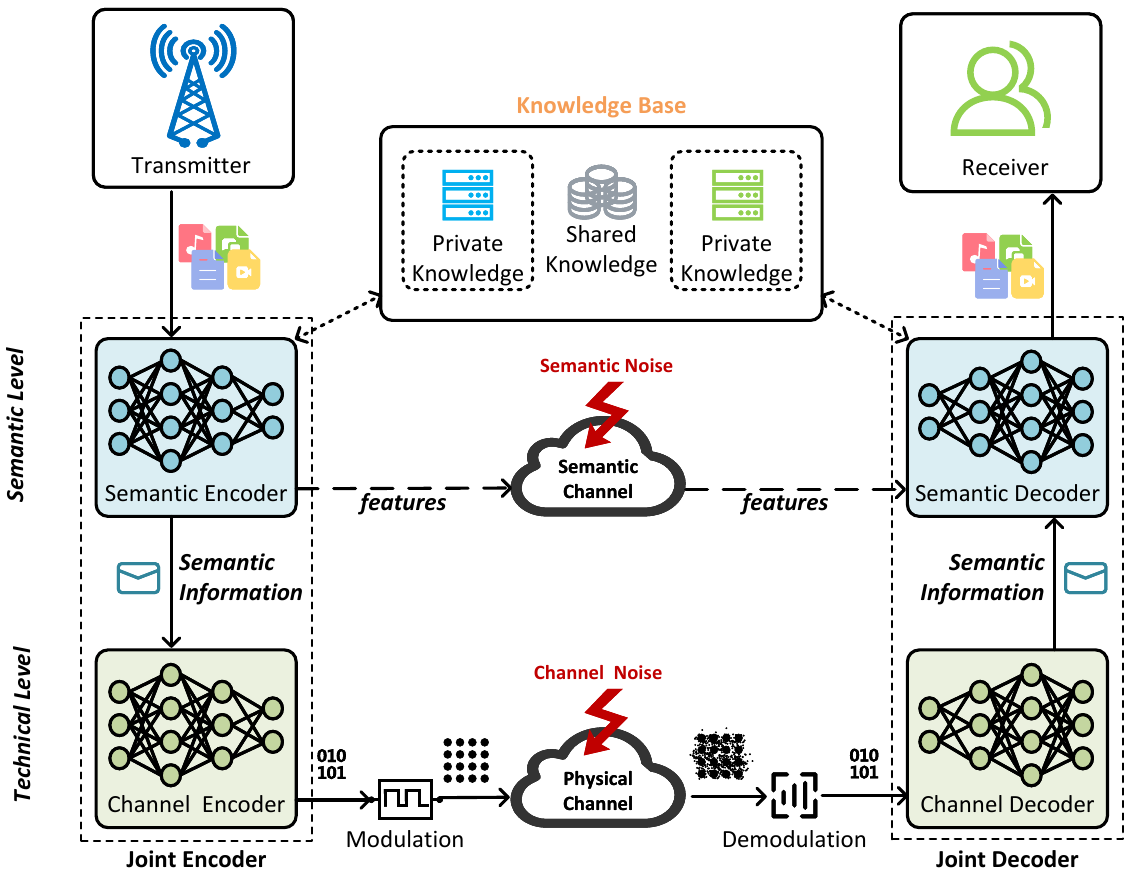}
		\par\end{centering}
	\caption{The architecture of a typical SC system.}
	\label{FrameArchitect1}
\end{figure*}	

To tackle the above challenges, several research efforts have recently been made to reduce the complexity of MMSE channel estimation for HMIMO
\cite{demir2022channel,wan2021terahertz,ghermezcheshmeh2023parametric}. 
For example, a subspace-based channel estimator for HMIMO was proposed in \cite{demir2022channel}, where the rank deficiency of the HMIMO spatial correlation matrix was exploited to obtain a low-dimensional MMSE estimator. Although this method outperforms the simple least-squares (LS) estimator which does not exploit any prior channel knowledge, it still incurs an excessive pilot overhead. To reduce the number of required pilot symbols, \cite{wan2021terahertz} used the dual sparsity of THz-HMIMO channels in both the angular and delay domains and proposed a CS-based channel estimation method to reduce the estimation overhead.
Moreover, a parametric channel estimator for LoS HMIMO channels was proposed in \cite{ghermezcheshmeh2023parametric}, which leveraged the structure of the radiated antenna pattern to estimate the channel parameters, including the user range, elevation, and azimuth AoDs. 
Its advantage is that the training overhead and computational complexity do not scale with the number of antennas.

While the aforementioned subspace- and CS-based HMIMO channel estimators achieve higher estimation accuracy than the simple LS method, they suffer a certain loss in  accuracy when compared to the MMSE approach. To fill the gap, \cite{d2023dft} proposed a low-complexity DFT-based HMIMO channel estimation method that achieves nearly the same performance as the MMSE estimator without knowledge of the channel covariance matrix. However, this algorithm is generally restricted to ULA-based HMIMO systems.
An ML-based unsupervised near-field channel estimation method was developed in \cite{yu2023bayes} for HMIMO systems based on score matching and principal component analysis without prior knowledge of the true channel distribution.
%supervised training samples compared to the conventional Bayes-optimal estimators, which can be employed in practice with 
This method was shown to achieve a performance close to the MMSE bound with low computational complexity. 

In Table \ref{table:2}, we summarize the main design issues for new-form NGATs together with the most up-to-date solutions.

\subsubsection{Future Directions}
Several issues related to HMIMO are worthy of future investigation. First, the feasibility of HMIMO hardware architectures such as LWA proposed in existing works still needs to be verified in practice accounting for various hardware imperfections. For example, 
the EM fields radiated by HMIMO antennas generally have reverse effects on the reference waves, which may introduce substantial distortion for holographic imaging and degraded performance for holographic communications.   
In addition, although mutual coupling has been studied for  HMIMO systems (see, e.g., \cite{Williams_2022}), a comprehensive study for characterizing all important factors such as antenna spacing and array configurations and investigating the fundamental capacity limits in HMIMO systems is still lacking.     
Moreover, for large-aperture HMIMO, it is of paramount importance to design efficient beamforming schemes that strike a balance between complexity and performance.

	\begin{figure*} [!htbp]
		\begin{centering}
			\includegraphics[width=0.8\textwidth]{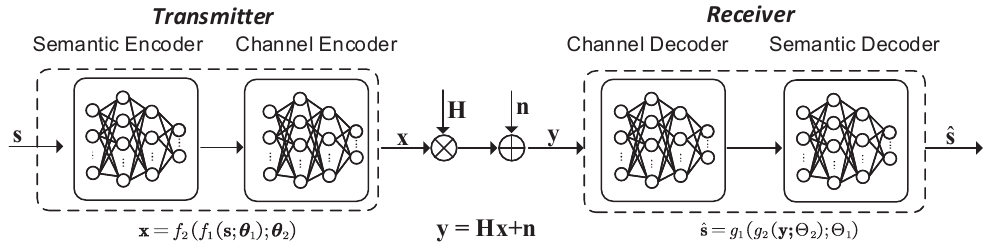}
			\par\end{centering}
		\caption{Detailed outline of a typical SC network.}
		\label{FrameArchitect2}
	\end{figure*}

\section{New Metric: Semantic-Aware Transceivers}\label{sec_nmetr}
	
	Over a number of decades, wireless communications systems designed on principles drawn from  Shannon's seminal work \cite{Shannon_BSTJ} have witnessed an evolution from analog first-generation (1G) systems to current 5G systems, and now research is progressing towards 6G. Looking to future applications and services, it is evident that 6G wireless networks will bridge the physical and cyber worlds, enabling human interactions with a multitude of devices, including those with integrated artificial intelligence, using diverse data modalities such as vision and text.  These resource demanding applications pose significant challenges for wireless networks with the need for ultra-high reliability, ultra-low latency, and extremely high data rates, as highlighted in \cite{NiuKai_ICM}. Accommodating and empowering these applications will necessitate addressing the exponential growth in both bandwidth and complexity, for example in, transmission of these massive datasets and large models. One promising approach to address this challenge is {\it semantic communications} (SC) \cite{semantic_index_assignment}, which focuses on the semantic information exchange between transmitters and receivers \cite{Guangming_ICM}. Essentially, SC aims to extract the semantics of the information source and perform transmission to convey the meaning and/or perform a certain task \cite{Deniz_JSAC,kalfa2021towards}. Towards this end, there are different approaches developed in the literature, including the design of communication transceivers in an end-to-end manner, similar to joint source and channel coding (JSCC), development of formal graph-based semantic language and goal filtering methods, and smoothing of extracted information \cite{kalfa2021towards,kalfa2022reliable}.
	
	 In this section, we first review existing DL-based approaches for SC systems and discuss three properties of SC, namely, performance, compatibility, and security. Then, we  elaborate on several key design issues in SC, including advanced coding methods, semantic-aware transceiver module design, and SC security. Furthermore, we  discuss potential solutions aimed at overcoming these challenges and meeting the evolving demands of next-generation communication systems.
	
	\subsection{SC Architecture and System Model}
	SC represents a paradigm shift from traditional bit-level transmissions to semantic-level transmissions, hence emerging as a more advanced type of design. The key elements and processes involved in SC are illustrated in Fig.~\ref{FrameArchitect1}, where the semantic encoder/decoder and channel encoder/decoder can be designed jointly or separately. In particular, the semantic information extracted from the source data undergoes transmission via physical channels with channel encoding and decoding. This source data can encompass text, vision, and audio data. Unlike traditional communication systems, SC systems introduce an extra semantic layer, encompassing a semantic encoder, a semantic decoder, and a shared/private knowledge base. This semantic layer facilitates the extraction of semantic features from the source information and the recovery of semantics from the received data, thereby enhancing transmission efficiency. By only transmitting the semantic information, information redundancy can be significantly reduced. The knowledge base can be viewed as a well-established context, consisting of both a public and a private knowledge base. The public knowledge base consists of information that is known and shared between both the transmitter and receiver, while the private knowledge base contains confidential information that cannot be shared by either party.
	
	We start by presenting a detailed framework for SC. As illustrated in Fig.~\ref{FrameArchitect2}, the transmitter utilizes a mapping to convert the source $\mathbf{s}$ into a symbol stream $\mathbf{x}$, which is subsequently transmitted through a physical channel affected by transceiver impairments. The received symbol $\mathbf{y}$ undergoes decoding at the receiver to generate an estimate of the source $\hat{\mathbf{s}}$. In its contemporary reincarnations, semantic communications and deep learning often go hand-in-hand together \cite{9450827}, although utilizing pre-trained models have recently started to materialize as demonstrated by \cite{emrecan_icc_ws}. In this tutorial, we shall consider the framework where the transmitter and receiver are implemented using DNNs. Specifically, the transmitter comprises a semantic encoder and a channel encoder, while the receiver incorporates a semantic decoder and a channel decoder. The semantic encoder is trained to transform transmitted data into channel symbols, while the semantic decoder is designed to reconstruct transmitted data from received symbols. %Additionally, the channel encoder and  decoder are dedicated to mitigating signal distortion introduced by  wireless channels. --this inherently suggests a separation approach.

	Consider an $N_{t}$-antenna transmitter communicating with a $N_{r}$-antenna receiver. The encoded symbol stream is given by
	\begin{equation} \label{SemCom01}
		\mathbf{x}=f_{2}\big( f_{1}(\mathbf{s};\bm{\theta}_{1} ); \bm{\theta}_{2}  \big),
	\end{equation}
	where $\mathbf{x}\in \mathbb{C}^{N_{t}\times 1}$, and  $\bm{\theta}_{1}$ and $\bm{\theta}_{2}$ represent the  parameters of the semantic encoder $f_{1}(\cdot)$ and the channel encoder $f_{2}(\cdot)$, respectively. Then, the received signal  is given by
	\begin{equation} \label{SemChannel}
		\mathbf{y}=\mathbf{H}\mathbf{x}+\mathbf{n},
	\end{equation}
	where $\mathbf{H}\in \mathbb{C}^{N_{r}\times N_{t}}$ represents the channel matrix and $\mathbf{n}\sim \mathcal{CN}(\bm{0}, \sigma^{2}\mathbf{I})$ denotes additive white Gaussian noise (AWGN). The decoded signal can be written as
	\begin{equation} \label{SemCom02}
		\hat{\mathbf{s}}=g_{1}\big( g_{2}(\mathbf{y};\bm{\Theta}_{2} ); \bm{\Theta}_{1}  \big),
	\end{equation}
	where $\bm{\Theta}_{1}$ and $\bm{\Theta}_{2}$ denote the trainable parameters of the semantic decoder, $g_{1}(\cdot)$, and the channel decoder, $g_{2}(\cdot)$, respectively.
	
	In general, the semantic encoder strives to extract useful meaning through semantic coding technology, while the semantic decoder aims to reconstruct the original information. Optimizing semantic coding technologies is pivotal for enhancing the performance of SC systems.  Consequently, efficient semantic transmission relies on the intricate design of semantic extraction, error protection, and network architecture. These factors play crucial roles in shaping the SC performance, which will be discussed in the following sections.

		\begin{figure*}[t]
		\begin{centering}
			\includegraphics[width=0.8\textwidth]{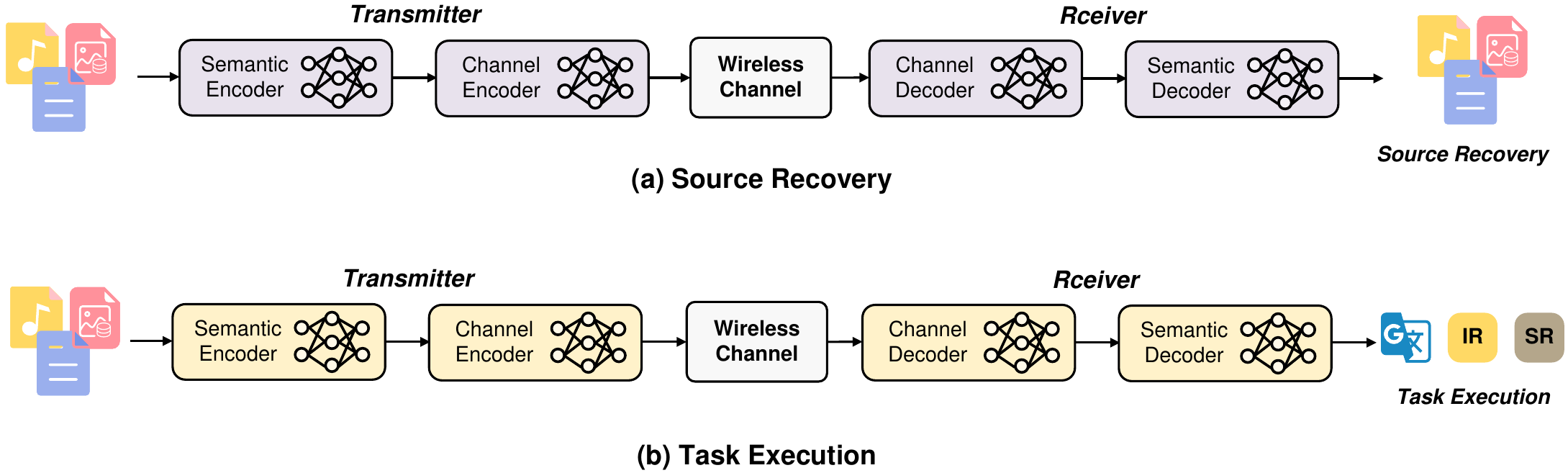}
			\par\end{centering}
		% \captionsetup{font=footnotesize}
		\caption{Illustration of two kinds of SC: (a) source recovery, (b) task execution.}
		\label{FrameKind}
	\end{figure*}
	
	\subsection{Performance: Advancements in Coding Mechanisms}
	Recently, various advanced coding mechanisms have received intensive interest in the research community, primarily focusing on improving the transmission performance of SC systems measured, e.g., in terms of reconstruction distortion or classification accuracy. These improvements are evaluated using various metrics, such as the peak signal-to-noise ratio (PSNR) and bilingual evaluation understudy (BLEU) metrics. Existing studies in the field of SC can  be categorized into three main groups, namely,  source recovery  \cite{ Jialong_TCVT, Tze-Yang_JSAC, Shuai_TWC, Kurka_JSIT,Sixian_JSAC,Bourtsoulatze_TCCN, Huiqiang_TSP, Lei_WCL_Text,ShengshiTCCN, zhenzi_JSAC,Tianxiao2023JSAC,Zixuan_Speech_2023, Shengshi_WCNC_Variational, Akhtar2019, liu2023semantic, Jiawei2021GNN,huang2022iscom}, task execution \cite{Jankowski_JSAC, ChiaHan_Access,  Qiyu_TWC, Huiqiang_JSAC,YangshuoTCOM2023,QiyuRobust}, and joint reconstruction and execution \cite{Jianhao_2024, Guangyi_Arxiv, wang2023privacypreserving, Zhonghao2024Semantic,Chamain2022End,Yufei_Arxiv}.
	
	\subsubsection{Source Recovery}
	For source recovery, SC systems strive to extract comprehensive semantic information from source data, which is akin to DL-based JSCC systems.  These methods primarily aim to compress the input source data directly into channel symbols using autoencoder architectures while incorporating redundancy to resist channel noise, as shown in Fig.~\ref{FrameKind}(a). In other words, a complete SC system should be able to simultaneously serve a combination of compression and error correction codes \cite{Deniz_JSAC}.

	A number of advancements have been made to improve the transmission performance for various data modalities including, e.g.,  vision \cite{Jialong_TCVT,Tze-Yang_JSAC,Shuai_TWC, Kurka_JSIT,  Bourtsoulatze_TCCN,Sixian_JSAC, Guangyi_CL}, text\cite{Huiqiang_TSP, Lei_WCL_Text,ShengshiTCCN}, audio \cite{zhenzi_JSAC,Tianxiao2023JSAC,Zixuan_Speech_2023, Shengshi_WCNC_Variational }, and point cloud data \cite{Akhtar2019, liu2023semantic, Jiawei2021GNN,huang2022iscom}.
	In particular, for vision data, pioneering work  \cite{Bourtsoulatze_TCCN, Jialong_TCVT, Kurka_JSIT,Sixian_JSAC} has employed DNNs to implement JSCC, demonstrating significantly higher performance compared to traditional separable approaches such as the combination of low-density parity-check codes (LDPC) and  joint photographic experts group (JPEG) encoding. Subsequently, inspired by entropy model-based compression methods \cite{balle2018variational, Balle_JSTSP,Cheng_2020_CVPR}, the authors of \cite{SixianNTSCCjj, Jincheng_JSAC_Nonlinear}  proposed a nonlinear transform source-channel coding (NTSCC) method for image transmission. \color{black}These methods have shown promising performance gains compared to traditional approaches. In Table \ref{BDResults}, we compare the image transmission performance of BD-CBR and BD-PSNR, computed from the PSNR-CBR curves, with ``BPG + LDPC" chosen as the baseline scheme. These metrics characterize the overhead savings and PSNR gains compared to the traditional separation-based baseline ``BPG + LDPC" approach. Specifically, a negative BD-CBR value indicates bandwidth savings, while a positive value signifies increased bandwidth costs. Meanwhile, BD-PSNR characterizes the PSNR gain, where a positive value indicates an improvement in image transmission quality, and vice versa. The results indicate that deep learning-based SC systems exhibit superior performance compared to traditional solutions.  \color{black}
 	
 		\begin{table*}[!htbp]\footnotesize \color{black}
 		\centering
 		\caption{\fontsize{10pt}{17bp}BD performance comparison.}
 		\begin{tabular}{| c | c  c  c | c  c  c|}
 			\hline
 			\multirow{2}{*}{Transmission scheme}     & \multicolumn{3}{c|}{BD-CBR $\downarrow$}  & \multicolumn{3}{c|}{BD-PSNR $\uparrow$}  \\    \cline{2-7}
 			&CIFAR10  & Kodak   & CLIC2021  &CIFAR10 & Kodak  & CLIC2021  \\   \hline
 			JPEG + LDPC    &--   & $167.27$\%  & --  &--   & $-4.40$dB  & --  \\
 			JPEG2000 + LDPC  &--  & $58.28$\%  & $184.71$\%  &--   & $-1.91$dB  & $-4.12$dB  \\
 			BPG + LDPC  &$\textbf{0}$\%  & $\textbf{0}$\%  &$\textbf{0}$\% &$\textbf{0dB}$    & $\textbf{0dB}$  & $\textbf{0dB}$  \\  \hline
 			\cite{balle2018variational} + LDPC    &--   & $7.19$\%  & $-6.76$\%  &--  & $-0.31$dB  & $0.31$dB  \\
 			\cite{Minnen2013nips} + LDPC    &--   & $-10.39$\%  & $-28.11$\% &--   & $0.48$dB  & $1.39$dB  \\   \hline    DeepJSCC\cite{Bourtsoulatze_TCCN}    &$-22.71$\%   & $36.94$\%  & $45.91$\%  &$1.71$dB   & $-1.41$dB  & $-1.74$dB  \\
 			%		\cite{Jialong_TCSVT}     & $30.22$ \%  & $37.33$\%  & $-1.01$dB  & $-1.27$dB  \\
    NTSCC\cite{Jincheng_JSAC_Nonlinear}  &$-28.91$\%   & $-17.96$\%   & $-28.09$\%  &$2.64$dB  & $0.81$dB  & $1.31$dB   \\
 			\hline
 		\end{tabular}
 		\label{BDResults}
 	\end{table*}

    Since different images contain different amounts of information and require different numbers of bits for their representation, NTSCC combines JSCC with an entropy model to determine the code length for each image, thus achieving variable-rate transmission. This approach was further extended to speech transmission \cite{Zixuan_Speech_2023, Shengshi_WCNC_Variational}. In addition, generative models such as the denoising diffusion probabilistic models, generative adversarial networks (GANs), variational adversarial autoencoders (VAEs), and normalizing flow have demonstrated remarkable success in various practical generation tasks  \cite{Ecenaz_JSAC_2023, xu2023latent,grassucci2023generative}. Intuitively, by viewing data transmission as a generation task, a novel communication paradigm capable of preserving semantic information can be developed by harnessing the capabilities of advanced deep generative models. The core idea revolves around finding low-dimensional representations for raw signals, such as segmentation maps of high-resolution images and motion keypoint representations of video frames \cite{Peiwen_JSAC_2023,sagduyu2023joint}, and then using generative models to reconstruct these raw signals.  In the context of other data modalities such as text, the so-called DL-enabled SC (DeepSC) framework in \cite{Huiqiang_TSP} encodes text into various lengths by employing sentence information.  Concurrently, the authors of \cite{Tianxiao2023JSAC}  proposed to  divide speech transmission into two distinct paths: text-to-speech (translation) and speech-to-speech, which achieves notable performance enhancements.
	
			\begin{table*}[!htp] \small
		\caption{Summary of existing semantic communication systems for three different categories.}
		\centering
		\renewcommand{\arraystretch}{1.2}
		\setlength{\arrayrulewidth}{0.75pt}
		\begin{tabular}{|c|c|c|}\hline
			\textbf{Category} 														&  			\textbf{Modality}   &  			\textbf{Reference}       	\\   \hline
			\multirow{4}{*}{Source Recovery }     &           Vision   		&		\cite{Jialong_TCVT,Tze-Yang_JSAC,Shuai_TWC, Kurka_JSIT,  Bourtsoulatze_TCCN,Sixian_JSAC}	\\   \cline{2-3}
			&           Text  		 &	 \cite{Huiqiang_TSP, Lei_WCL_Text,ShengshiTCCN}	  \\      \cline{2-3}
			&           Audio  		 &	\cite{zhenzi_JSAC,Tianxiao2023JSAC,Zixuan_Speech_2023, Shengshi_WCNC_Variational }			  \\    \cline{2-3}
			&			 Point cloud  &	 \cite{Akhtar2019, liu2023semantic, Jiawei2021GNN,huang2022iscom}										 \\       \hline
			
			\multirow{2}{*}{Intelligent Task}            &           Vision   	&	\cite{Qiyu_TWC, Huiqiang_JSAC,Shao_JSAC,XieJSACRobust,wang2023privacypreserving,Shao2023Task} \\    \cline{2-3}
			&           Text  		&	\cite{Huiguo2023TCCN,kutay2024classification,Huiqiang_JSAC}									  \\     \hline

			\multirow{2}{*}{Multitask}     					 &           Vision (Recovery and Classification)   		&		\cite{Jianhao_2024,  wang2023privacypreserving, Zhonghao2024Semantic,Chamain2022End,Yufei_Arxiv }								  						\\   \cline{2-3}
			
			&           Vision \& Text \& Audio  			   &			\cite{UDeepSC}							  \\   \cline{2-3}
			\hline      				
		\end{tabular}
		\label{table_s1}
	\end{table*}
	
	\subsubsection{Intelligent Task Execution}
	While JSCC-like SC has made significant strides in source recovery, it is crucial to acknowledge that its primary objective remains the recovery of the transmitted signals themselves. This inherent limitation arises from the need for a substantial transmission bandwidth to accomplish the reconstruction task. To further upgrade the effectiveness of SC, intelligent task-oriented SC has recently received significant attention. This paradigm seeks to address more specific tasks such as classification and object detection \cite{kutay2024classification}. In this context, only semantic information relevant to the intelligent task, such as class or object location, is considered, as shown in Fig.~\ref{FrameKind}(b).  Numerous studies have emerged in various domains to implement this approach for significantly reducing transmission overhead \cite{Jankowski_JSAC, ChiaHan_Access, Qiyu_TWC, Huiqiang_JSAC, QiyuRobust,sagduyu2023age}.
	
	Among others, some studies are dedicated to crafting effective models for diverse tasks across different modalities. In \cite{Qiyu_TWC}, an innovative SC system was devised for image retrieval and classification, leveraging the powerful Transformer architecture \cite{Han2023TPAMI}.  This approach employed vector quantization-variational autoencoder (VQ-VAE) techniques to realize digital semantic transmission. In another work \cite{Huiqiang_JSAC}, the authors introduced a multi-user SC system tailored for both text-only and text-and-image tasks. They designed three models to address translation, image retrieval, and visual question-answering problems. \color{black}It significantly reduces the transmission overhead compared to a traditional separation-based scheme, as presented in Table \ref{OverheadCompare}. \color{black}
		 \begin{table*} \color{black}
		\centering
		\caption{The number of transmitted symbols for one image in different tasks \cite{Huiqiang_JSAC}.}
		\label{OverheadCompare}
		\begin{tabular}{|c|c|c|c|c|}
			\hline
			Tasks & Datasets & Transmission Methods  & Overhead per image/word    \\ \hline
			Image retrieval  & Cars196 & DeepSC-IR / JPEG-LDPC-8QAM & $128/499, 920$   \\ \hline
			\multirow{2}{*}{VQA}      & CLEVR: Text & DeepSC-VQA / UTF-8-Turbo with BPSK  & $77/152$  \\ \cline{2-4}
			& CLEVR: Image & DeepSC-VQA / JPEG-LDPC with 8-QAM  & $25, 216/55, 624$  \\ \hline
		\end{tabular}
	\end{table*}
	
	Despite the performance improvement and overhead reduction, a fundamental question arises: \textit{How precisely can semantic information be characterized?} This is a critical consideration for several reasons, particularly in establishing foundational theories to analyze the system and determine an appropriate transmission rate. Several endeavors have aimed at meticulously determining transmission rates for different intelligent tasks. For example, in  \cite{Shao_JSAC,XieJSACRobust,wang2023privacypreserving}, the authors employed the information bottleneck (IB) principle to construct a classification-oriented system.  Specifically, IB can be used to design a semantic model for extracting the most relevant information by maximizing the mutual information between the received symbols and the expected task-specific information (e.g., the label for a classification task), while minimizing the transmission rate. This enables the model to achieve a tradeoff between task performance and transmission rate. This kind of principle was further extended to distributed scenarios in \cite{Shao2023Task}. While the aforementioned works primarily focused on rate optimization for tasks involving a single data modality, \cite{YangshuoTCOM2023} studied rate control in a distributed multimodal setting, and derived an importance metric based on robustness verification and unequally assigned coding rates for different modalities based on their semantic importance.
	
	\begin{figure*}[t]
		\begin{centering}
			\includegraphics[width=0.95\textwidth]{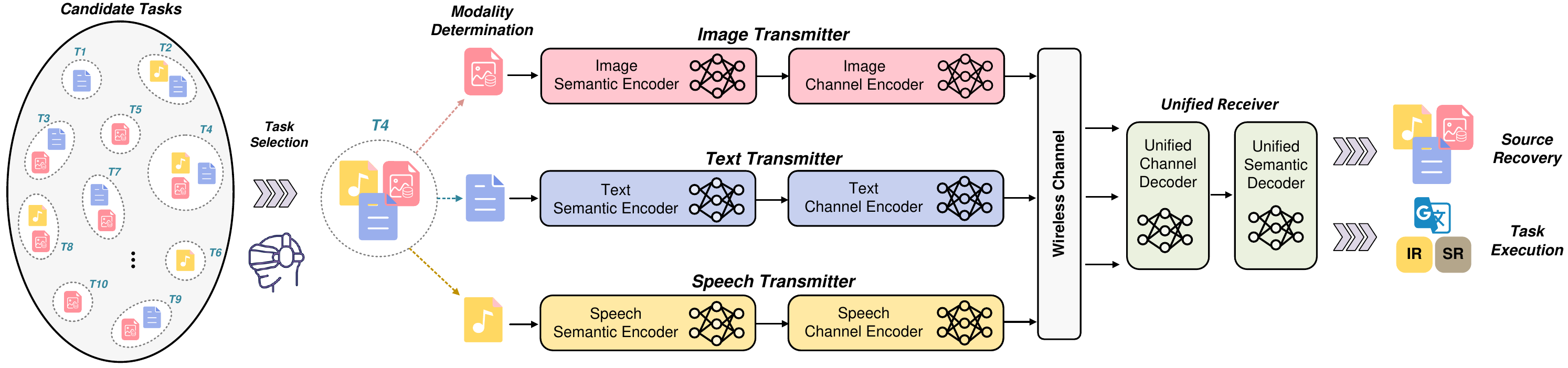}
			\par\end{centering}
		% \captionsetup{font=footnotesize}
		\caption{Illustration of multimodal data transmission.}
		\label{FrameMultimodal}
	\end{figure*}
	\subsubsection{Joint Recovery and Task Execution for Multitask Design}
	Existing SC systems have shown satisfactory performance in specific scenarios, but they typically specialize in handling specific tasks with a single data modality.  However, the demand for multitask execution is on the rise in practical applications. In a complete transceiver, both source recovery and other intelligent tasks may need to be performed simultaneously. Many devices now require the execution of multiple tasks within a single system, hence necessitating efficient handling of diverse data types. This poses great challenges for existing SC models, as they often require retraining when tasks change, resulting in significant gradient transfers and computational costs. Alternatively, the storage of multiple models may be required, which inevitably  increases  storage demands \cite{Guangyi_Arxiv}. Therefore, the development of multitask SC systems capable of unified processing for multimodal data transfer holds practical significance. As shown in  Fig.~\ref{FrameMultimodal}, multitask SC systems aim to automatically activate transmitters based on the selected task and efficiently handle data across different modalities.

	Notably, several studies \cite{Jianhao_2024, UDeepSC, wang2023privacypreserving, Zhonghao2024Semantic,Chamain2022End,Yufei_Arxiv,sagduyu2023multi} have addressed the implementation of multitask objectives. For example,  a unified SC system was proposed in \cite{Guangyi_Arxiv} to unify multimodal data and multi-task applications using a single DNN model, where the source recovery and intelligent task execution are simultaneously supported. The authors also determined the appropriate number of transmit symbols for different tasks by designing an adaptive transmission scheme. Then, \cite{Zhonghao2024Semantic} proposed a new end-to-end deep JSCC framework that is able to simultaneously perform image source recovery and classification by unifying the coding rate reduction maximization and the distortion. In \cite{Chamain2022End, Yufei_Arxiv, wang2023privacypreserving}, task-specific semantic-preserving compression was considered. The primary goal of these studies is to minimize the cost of source recovery while maximizing the performance of intelligent tasks. Moreover, \cite{sagduyu2023joint2} proposed a joint sensing and SC frameworks based on multi-task deep learing.
	
	\subsection{Security: Overcoming Threats in SC}
	Despite the promising performance of the systems proposed above, SC is still in its infancy and confronts a range of security challenges. Indeed, the security issues in semantic-based systems resemble many of the attack methods in conventional communication systems include eavesdropping, data fabrication, and jamming attacks \cite{yang2023secure}. However, security issues in SC have two unique features compared to traditional systems:
	\begin{itemize}
		\item {\bf Semantic information security:}
		An attacker in a semantic system considers not only the amount of stolen data but also the semantic meaning of the stolen data. A typical example in \cite{Tze-Yang_Arxiv} considers how an eavesdropper can theoretically recover the input sample in an unchallenged way simply by intercepting the semantic messages.
		In addition to semantic signals, knowledge bases are also susceptible to attacks. The shared knowledge base, employed by both the transmitter and receiver, facilitates the extraction of semantic information by the transmitter and enables the receiver to recover the input sample. However, this sharing also renders the data in the knowledge base vulnerable to leakage.
		
		\item {\bf Semantic model security:}
		Many ML techniques are perceived as important enablers for constructing basic components of SC systems. As such, an attacker in SC can target not only the transmission of semantic information but also the ML models employed for extracting that semantic information.  In particular, semantic models can be attacked by adding semantic noise to the raw signals \cite{Qiyu_TWC}, similar to  a jamming attack. Moreover, semantic models also face data poisoning risks. For example, some of the original data samples can be tampered with by poisoning during the model training process \cite{Goldblum_TAML}, which results in significant performance deterioration.
		\end{itemize}
	
	    As outlined in \cite{yang2023secure}, these issues pose several practical threats to semantic transceivers. The potential threats can be categorized based on their respective phases in preparation and deployment. In this section, we describe different threats faced by semantic transceivers during these phases and offer an overview of existing defense methods.
	
	\subsubsection{Threats and Countermeasures Before Deployment}
	 Training of SC systems encompasses various processes, including code implementation, data preparation, pre-training, and sometimes collaborative (distributed) learning. However, owing to the inherent fragility of DNN models, SC systems are susceptible to attacks, and potential threats may manifest themselves during the phases mentioned above.
	\begin{itemize}
		\item 	
		First, developers commonly implement SC systems using open-source frameworks such as TensorFlow and PyTorch, which often rely on third-party packages. In this scenario, an adversary could exploit vulnerabilities in these frameworks to launch various SC attacks.
		
		\item
		Second, model optimization involves a substantial number of data samples. If an adversary tampers with some original data samples by methods such as poisoning and pollution, the resulting gradient can be significantly disturbed, leading to a degradation in model performance. This scenario is particularly realistic when the transceiver collects data from multiple sources. For instance, popular datasets used in recent semantic works often rely on contributions from volunteers and the Internet.
		
		\item
		Third, pretraining is one of the most important techniques to upgrade the performance of advanced approaches such as large language models. These models may be released publicly by the opponent, which is likely to corrupt the semantic model being controlled.
				
		\item
		Finally, in the context of collaborative learning, the training processes of SC models are susceptible to attacks. The model undergoing training can be easily attacked if a number of participants are compromised or controlled by a malicious actor. Both data poisoning/pollution and gradient poisoning techniques are potential threats in such scenarios.
	\end{itemize}

	To address these security challenges, an increasing number of studies are dedicated to precise modeling and defending against potential threats. For example, \cite{Sagduyu2023Vul} investigated backdoor semantic attacks, which involve covertly embedding hidden associations or triggers within DL models, with the aim of manipulating model inferences such as classifications. These hidden elements are strategically designed to compel the system to act maliciously in alignment with the attacker's chosen objective while maintaining normal behavior in the absence of the trigger \cite{gao2020backdoor}. The work of \cite{Sagduyu2023Vul} studied a backdoor attack that is performed in the training phase by inserting triggers into training samples, demonstrating its effectiveness in altering semantic information for poisoned input samples.
	The authors of \cite{yang2023secure} provided a comprehensive list of  countermeasures against various threats, including trigger inversion and semantic watermarking for backdoor attacks, as well as authentication for code attacks. However, detailed modeling and countermeasures are still in their early stages, and deserve future investigation.
	
\begin{figure*}[t]
	\begin{centering}
		\includegraphics[width=0.55\textwidth]{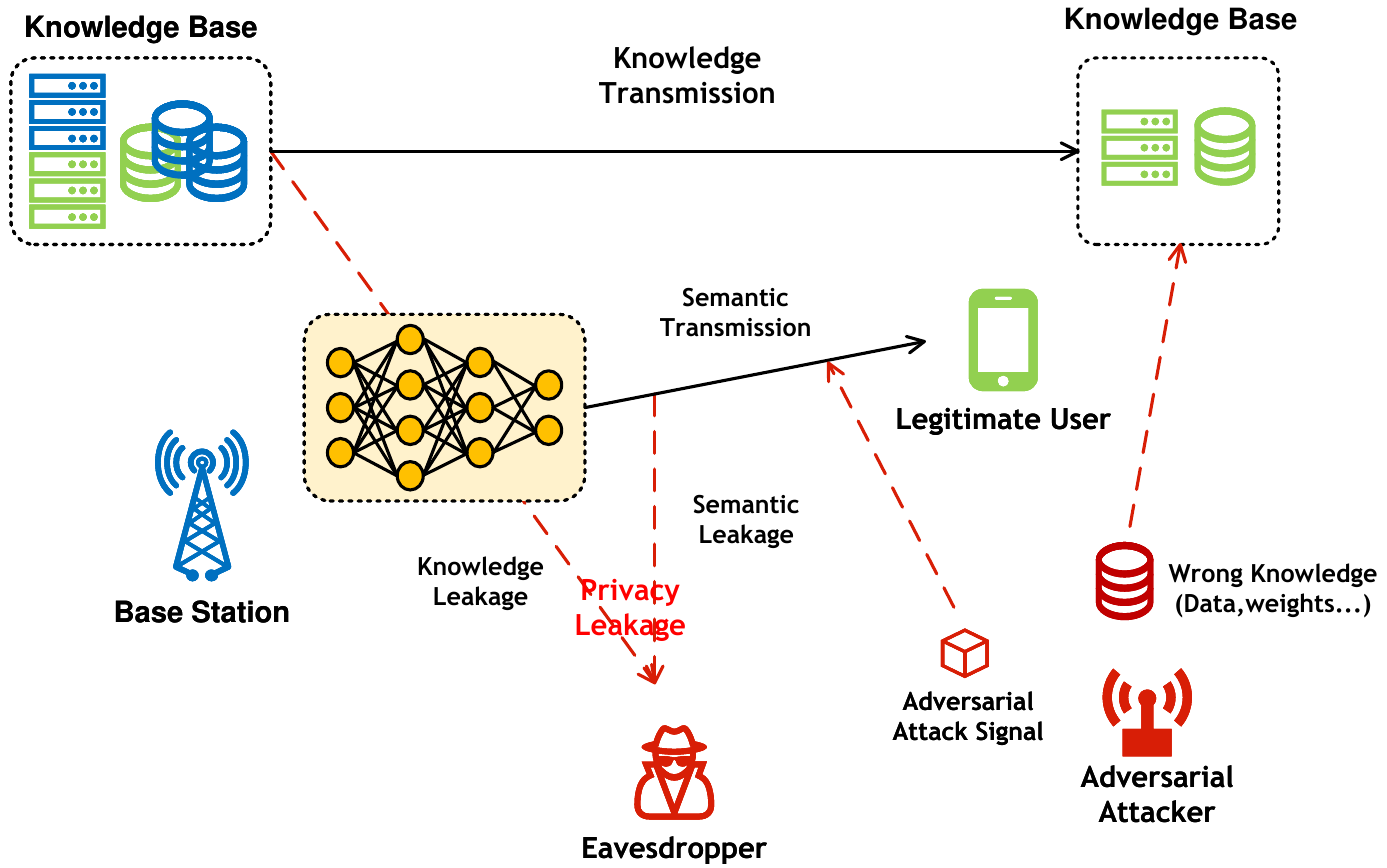}
		\par\end{centering}
	% \captionsetup{font=footnotesize}
	\caption{Overview of adversarial attacks and privacy leakage threats.}
	\label{Frameattack}
\end{figure*}
	
	\subsubsection{Threats and Countermeasures in Deployment Phase}
	Another prevalent threat that arises during the deployment phase focuses on attacking a trained model or pilfering semantic information. In this context, we identify two extensively studied threats: adversarial attacks, and privacy leakage, as illustrated in Fig.~\ref{Frameattack}.
	
	\paragraph{Adversarial Attacks} SC has demonstrated vulnerability to adversarial samples, where imperceptible or semantically consistent manipulations of inputs mislead the models to produce incorrect results \cite{Meng2023Secure,sagduyu2023tasknextG}. In \cite{QiyuRobust}, adversarial samples for SC are initially defined as the signals that are deliberately injected as semantic noise by a malicious attacker.  Adversarial attacks can exist at both the transmitter and receiver. The first type of adversarial attack is generated in the encoding stage. Semantic noise has a serious impact on the encoding process and will lead the DL models to generate incorrect results for different tasks. The second type occurs during the transmission process in the physical channel, causing decoding failures and misinterpretations at the receivers \cite{Qiyu_TWC}. This attack is analogous to a jamming attack in traditional communication systems, primarily focusing on inducing semantic errors rather than causing a reception failure for a legitimate user. To generate their attack, \cite{Qiyu_TWC} proposed a sample-independent method based on a universal attack, requiring no knowledge of the exact transmitted signal. Additionally, \cite{Guoshun2023Physical} developed an attack module referred to as SemAdv that is capable of generating perturbations for the second type of attack. In \cite{Tang2023GAN}, the attack is treated as a generation task, which can be implemented using a GAN model. Moreover, the authors of \cite{sagduyu2023semantic} introduced test-time (targeted and non-targeted) adversarial attacks on the DNNs by manipulating their inputs at different SC stages based on adversarial ML.
	Adversarial training is one of the most effective approaches for defending against such attacks and improving the robustness of SC systems\cite{Qiyu_TWC,Guoshun2023Physical,Tang2023GAN}. The idea of adversarial training is straightforward: Augment the training samples with adversarial examples in each training loop, enabling the trained models to behave more normally when facing adversarial examples compared to models trained without such data. Adversarial training design is normally formulated as a min-max problem, seeking the best solution to the worst-case optimum.
	
	\paragraph{Privacy Leakage} While SC systems are acknowledged for their efficiency compared to traditional implementations, this efficiency comes with an elevated risk of privacy leakage.
	In particular, DNN models extract semantic features by directly mapping the source data to the modulated channel input. Consequently, the channel input symbols maintain a direct correlation with the input source. In the realm of physical layer security, individuals within the physical range of a transmitter can receive the wireless signal and potentially decode the symbols. While the SC system prioritizes transmitting the intended meaning from the source rather than focusing on accurate transmission of raw bits, this emphasis renders it vulnerable to malicious eavesdropping attempts \cite{XinghanSem}. Furthermore, existing SC systems exhibit remarkable performance even in low-SNR regimes. While advantageous for SC, this characteristic implies that eavesdroppers can decipher semantic information even through highly noisy channels. For instance, a DL-based eavesdropper has been developed to exploit this vulnerability, demonstrating the risk of privacy leakage in SC \cite{MaojunIWCL}. The study designed a privacy-aware loss function to achieve a tradeoff between performance and secrecy. To address this issue, the authors of \cite{Xinlai2023Encrypted,XinghanSem} used semantic encryption and decryption modules that share the same DNN architecture, and an IB scheme was developed in \cite{wang2023privacypreserving} to defend against model inversion attacks.

	Another significant threat is the inference attack. Since an SC system learns from a vast number of data examples, including potentially sensitive information from the personal knowledge base (such as the user's health status and service access history), there is a risk of unintentionally revealing sensitive information in the semantic representations sent over wireless channels \cite{XinghanSem}. A representative type of attack in this class is the property inference attack, which aims to infer whether a given instance in the training data or model possesses specific properties. This type of attack can deduce sensitive information, including demographic details, medical conditions, economic status, and personal preferences.
	
	\subsection{Compatibility: Incorporation with Practical Transceivers}
	As discussed in the previous sections, SC significantly enhances transmission efficiency by prioritizing the transmission of semantic information, which can ensure minimal recovery and semantic distortion with extremely low transmission overhead. This approach results in a substantial reduction in transmission resource consumption \cite{PeiwenIWC2023}. However, previous work has predominantly concentrated on enhancing SC system performance by designing advanced encoding schemes. While effective, this approach tends to overlook the impact of the physical channels and does not consider existing system modules. Furthermore SC designs based on DL differ significantly from existing systems, hence raising concerns about compatibility and practical implementation. In addition to exploring various encoding and decoding methods, researchers have also focused on implementing SC by revising or redesigning the modules of existing communication systems, as discussed in  \cite{PeiwenIWC2023}.  In this section, we summarize various efforts to integrate SC into practical systems, including rate adaptation techniques, the design of joint multi-antenna transmission modules, digital modulation, hybrid automatic repeat request (HARQ) techniques, and resource allocation.

\subsubsection{Rate Adaptation}	
		In wireless communications, rate-adaptive techniques play a crucial role in enhancing transmission reliability and improving spectral efficiency. The concept of adaptive transmission involves estimating the CSI at the receiver, feeding back the estimated CSI to the transmitter, and then adjusting the transmission strategies based on the feedback. The adjustable parameters may include transmission power, data rate, and channel coding rate. Prior work on this topic for SC includes \cite{Jialong_TCVT}, which proposed a channel-adaptive SC system that supports a range of SNR values with the assistance of attention mechanisms. In \cite{Wenyu_TWC}, the authors predicted the image reconstruction quality at the transmitter for different channel bandwidths and selected an appropriate transmission rate based on the predicted values.  In \cite{YangWITT}, the authors introduced a novel attention module for adjusting SC systems based on the Swin Transformer. These approaches share similarities with adaptive modulation and coding approaches in traditional communication systems. Going a step further, an approach based on NTSCC++ was proposed in \cite{SixianNTSCCjj} that adjusts the transmission rate accounting for both semantic information and channel conditions, and achieves significant performance gains for image transmission.

\begin{table*}[!htp] \small
	\caption{Summary of semantic communication with physical layer techniques.}
	\centering
	\renewcommand{\arraystretch}{1.2}
	\setlength{\arrayrulewidth}{0.75pt}
    \resizebox{\textwidth}{!}{
	\begin{tabular}{|c|c|c|}\hline
		\textbf{Techniques} 											& \textbf{Characteristics}			        &  		\textbf{Reference}       	\\   \hline
		\multirow{1}{*}{Rate Adaptation}    				&	Adjust the transmission rates at the transmitter based on the channel state.				 &		\cite{ Jialong_TCVT, Wenyu_TWC, YangWITT, SixianNTSCCjj}							      \\      \hline
		
		\multirow{1}{*}{Multi-Antenna Transmission}    	& Learning to exploit the spatial diversity and the spatial multiplexing gain. 	         &		\cite{ Guangyi_Arxiv, Haotian_Arxiv, zhou2024feature,Shengshi2023PIMRC,AdaptMIMO}							      \\      \hline
		
		\color{black} \multirow{1}{*}{RIS-aided Transmission}    	&  \color{black}Enhance channel quality and significantly improve spectral and energy efficiency for SC. 	         &		\color{black}\cite{ Hongyangjsac2023, xujieWCL2024,shuyi2024ris}		\color{black}					      \\      \hline
		
		\multirow{1}{*}{Multi-carrier Techniques}    	& Superiority in high-speed data transmission and resistance to multi-path interference.		          &		\cite{Lan2023OFDM,9714510,Yulin_WCL,OFDM_Chuanhong2024}							      \\      \hline
		
		\multirow{1}{*}{Digital Modulation}    	&Higher error-correction capabilities and stronger resistance to interference. 			  &		\cite{ 	Yufei_Arxiv,TungDeepJSCCQ,Mengyang2022Q,Qiyu_TWC,Kristy2019,guo2024digitalsc}							      \\      \hline
		
		\multirow{1}{*}{Retransmission}    		&  Flexibility in adjusting the codeword length, and high transmission reliability.  							    &		\cite{wang2023spiking, PeiwenTCOM, Huiguo2023TCCN}							      \\      \hline

		\multirow{1}{*}{Resource Allocation}    		&  Allocate wireless resources to enhance power and spectral efficiency.					    &		\cite{mu2022semi_NOMA, mu2022mul_sem,liu2023semrelay,hu2023semrelay,yang2023energy,yan2022resource}							      \\      \hline
		      				
	\end{tabular}}
	\label{table_s2}
\end{table*}

\subsubsection{Multi-Antenna Transmission}
	While prior studies have demonstrated encouraging performance, most of them focused solely on SISO scenarios in additive white Gaussian noise (AWGN) channels. This simplified model fails to accurately capture the complexity of real-world communication environments. To address this limitation and better adapt to the complexity of actual channel conditions, researchers have explored various design methods for SC assuming multi-antenna transceivers. For example, \cite{Guangyi_Arxiv} introduced a MIMO SC system (DeepSC-MIMO) designed with channel-spatial attention technology. This model takes the CSI and noise variance as inputs, and adaptively generates an attention mask to assign different weights to different semantic features, thereby learning an appropriate power allocation strategy.
	In \cite{Haotian_Arxiv}, an adaptive JSCC scheme for wireless image transmission in MIMO systems was proposed based on the Vision Transformer (ViT) architecture, and referred to as ViT-MIMO. This model incorporates channel conditions into the encoding phase and significantly improves transmission quality. In \cite{zhou2024feature}, the authors explored feature-based importance and assigned high-importance features to more  sub-channels in both the time and antenna domains. In \cite{Shengshi2023PIMRC}, a spatial reuse mechanism was developed that achieves adaptive encoding rate allocation by jointly considering the entropy distribution of source semantic features and the wireless MIMO channel. In addition, \cite{AdaptMIMO} exploited spatial diversity and spatial multiplexing gain to upgrade the performance of semantic image transmission.
	These methods provide significant performance gains by jointly designing the multiple antenna transmission and semantic encoding modules.
	
	\color{black}
	\subsubsection{RIS-Aided Transmission} RIS techniques are highly valued for their ability to enhance channel quality and significantly improve spectral and energy efficiency. By leveraging RIS, a SC system can dynamically control and select the most favorable channels for transmitting semantic information. This dynamic control allows for the optimization of information entropy during communication, enabling more effective transmission of meaningful data while preserving the integrity and continuity of semantic information. Building upon the success of RIS in classical communications, several studies have investigated methodologies for RIS-aided SC. For example, the study in \cite{Jiajia2023ICSPCC} demonstrated that an RIS-aided SC system significantly outperforms a corresponding point-to-point SC system in terms of bilingual evaluation understudy (BLEU) scores in Rayleigh channels. Additionally, the RIS-aided system exhibits superior robustness against channel estimation errors compared to its point-to-point counterpart. The approach of establishing additional semantic transmission and reflection pathways by deploying RIS in different cells was explored in \cite{xujieWCL2024}. In this study, an optimization strategy for maximizing mutual information was proposed to enhance overall performance. The authors of \cite{shuyi2024ris} proposed RIS-aided on-the-air SC, which effectively addresses the incompatibility of SC systems with current digital hardware. This goal was achieved utilizing on-the-air or all-wave-based neural networks, specifically diffractional DNNs. In this framework, computations are inherently performed as wireless signals propagate through the RIS, offering significant benefits including rapid computation, low power consumption, and the capability to concurrently handle multiple tasks. In addition,  \cite{Hongyangjsac2023} introduced a new paradigm known as inverse semantic communications, leveraging innovative RIS hardware designs to dramatically reduces the sensing data volume, resulting in enhanced efficiency and resource utilization.
	\color{black}

	\subsubsection{Multi-Carrier  Transmission}
	Orthogonal frequency division multiplexing (OFDM) is a modulation technique widely used in wireless communication systems, excelling particularly for high-speed data transmission in multi-path channels \cite{Lan2023OFDM}. Combined with OFDM, an end-to-end SC system can be applied to multi-path channels, and several studies have explored incorporating multi-carrier techniques into SC. In \cite{9714510}, OFDM was combined with an autoencoder for wireless image transmission over multi-path fading channels. The multi-path channel and OFDM modulation are represented by differentiable layers, enabling the system to be trained in an end-to-end manner. The work in \cite{Yulin_WCL} highlighted that conventional DeepJSCC for image transmission \cite{Bourtsoulatze_TCCN} suffers from a high peak-to-average power ratio (PAPR). To address this issue, they developed a passband transceiver for  OFDM systems to minimize PAPR, which is considered as part of the loss function. Moreover, an OFDM-based digital SC framework was proposed in \cite{OFDM_Chuanhong2024} that uses deep reinforcement learning for sub-carrier and bit allocation.
	
	\subsubsection{Digital Modulation}
	Current research in SC primarily focuses on analog communication schemes, where semantic features are directly mapped to analog symbols for transmission without undergoing any bit-level conversion \cite{Yufei_Arxiv}. Although learning and transmitting semantic features via analog communication are feasible, this approach poses new challenges for hardware, protocols, and encryption. In contrast, digital communication has several advantages, including higher error-correction capabilities and stronger resistance to interference. However, the modulation process in digital communication involves mapping continuous variables to a finite number of discrete messages, presenting a challenge for neural networks due to the non-differentiability of this operation. This makes common optimization methods such as stochastic gradient descent (SGD) ineffective for training.
	
	In response to this challenge, some existing digital SC systems based on DL employ efficient strategies to make the modulation process trainable.
	Currently, the design of digital SC can be divided into two main categories.
	\begin{itemize}
		\item The first category involves directly mapping the source signals to discrete constellation points. For instance, \cite{TungDeepJSCCQ} introduced an end-to-end optimized JSCC scheme for digital wireless image transmission. This scheme employed a soft-to-hard quantization strategy, leveraging differentiable soft quantization during training and switching to non-differentiable soft quantization during testing. Additionally, a similar soft approximation method was proposed for backpropagation in \cite{Mengyang2022Q} that used a sigmoid function to approximate the quantization process. Furthermore, \cite{Yufei_Arxiv} proposed a joint coding and modulation learning scheme based on VAE that learns the transition probabilities from source data to discrete constellation symbols and employs the Gumbel-Softmax method to generate differentiable constellation symbols.
		
		\item The second category aims to map the source into a bit sequence and uses standard modulation approaches to transform the bit sequence into channel symbols. The main idea behind this category lies in its use of VAE. For instance, \cite{Qiyu_TWC} used VQVAE techniques to derive a free and learnable embedded space along with a discrete latent representation. By minimizing the distance between the encoder output and the embedded vector, the neural network was optimized to navigate the challenges associated with the ``quantization" process.
		In another work \cite{Kristy2019}, the authors modeled potential representations as sequences of bits following a Bernoulli distribution. They proposed a neural error correcting and source Trimming (NECST) algorithm for wireless image transmission. The NECST encoder outputs the probability that the bit is $0$ or $1$ and leverages a variational inference-based method to obtain a low-variance gradient, hence enabling the neural network optimization. In \cite{guo2024digitalsc}, a non-linear module was proposed to efficiently quantize semantic features with trainable quantization levels.
	\end{itemize}
	
	\subsubsection{Retransmission Design}
	Existing SC systems commonly employ end-to-end joint training by embedding channel environmental information into the neural networks. However, a limitation of these approaches is that the parameters of the trained network can only adapt to specific channel environments. When the environment changes significantly, retraining becomes necessary to accommodate the new channel conditions. Moreover, these networks are often unable to dynamically adjust the coding length based on the dynamic channels, resulting in limited flexibility and failure to ensure reliable transmission of semantic information.
	To address these challenges, researchers have explored the integration of SC with HARQ, which is widely used in modern mobile communication systems. One example is the incremental HARQ (IR-HARQ) architecture designed for text-based SC, referred to as SCHARQ \cite{PeiwenTCOM}. Using a Transformer-based JSCC, SCHARQ encodes the input sentence into various codewords to extract semantic correlation. The number of transmission codewords is determined based on the ACK feedback, resulting in an adaptive code rate. In \cite{wang2023spiking},  a spiking neural network (SNN)-based SC system was combined with HARQ to support the transmission of semantic features at varying bandwidths, with a policy model determining the appropriate bandwidth. Additionally, \cite{Huiguo2023TCCN} analyzed the robustness verification problem in SC systems and proposed a retransmission scheme based on the analysis.
	
	\subsubsection{Resource Allocation}
Efficient allocation of wireless resources such as spectrum and power is crucial and can significantly increase the spectral and energy efficiency.
However, performance evaluations of SC systems differ from those for conventional systems, with metrics tailored to specific tasks and constraints imposed by computing and storage requirements \cite{you2016energy,you2018asynchronous}. To address this issue, research has been recently been conducted on characterizing semantic performance metrics and designing resource allocation strategies tailored to SC.
For example, \cite{yan2022resource}, introduced the semantic rate (S-R) and semantic spectral efficiency (S-SE) to quantify the performance of SC systems, based on which the bandwidth allocation was optimized for achieving the maximum S-SE.
To further quantify the relationship between semantic information and wireless channels, \cite{mu2022semi_NOMA} derived an approximate closed-form expression for DeepSC semantic similarity, and developed optimal resource allocation strategies for  a heterogeneous semantic and bit transmission network. However, these studies assumed a DL-enabled semantic decoder at the receivers, which may not be realistic for resource-constrained mobile devices. To tackle this challenge, \cite{liu2023semrelay} proposed an innovative semantic relay (SemRelay)-assisted communication system that first decodes the semantic information and then forwards it to the user by adopting a conventional transmission scheme, effectively improving the text transmission efficiency.
By optimizing the placement of the SemRelay and system bandwidth allocation, the total effective rate of the resource-constrained text transmission system was maximized. This work was further extended in \cite{hu2023semrelay} for resource allocation in multi-user SemRelay-assisted text transmission systems. In addition, \cite{arda2024semantic} proposed two types of semantic forwarding schemes, called  semantic lossy forwarding and semantic predict-and-forward (SPF) to achieve cooperative semantic text communications.
In light of computing constraints, an efficient resource allocation algorithm  was proposed in \cite{yang2023energy} to minimize the total system energy consumption for semantic wireless networks with rate splitting.
	
	\subsection{Future Directions}
	\subsubsection{Large AI Model-Empowered SC Design}
		Motivated by the success of DL, large AI models have achieved great success in multimodal data processing, such as Segment Anything (SAM), DALL-E, and ChatGPT. Large AI models offer a multitude of advantages, prominently encompassing advanced semantic extraction and interpretation abilities, and rich background knowledge. These models excel in precisely extracting semantic information from diverse datasets, leveraging their extensive training to capture intricate patterns. Their vast reservoir of background knowledge enables contextual understanding. Furthermore, the robustness of large AI models allows them to interpret complex semantics in various contexts, effectively handling ambiguity and nuanced meanings \cite{LAI2023}. Overall, these capabilities make large AI models highly effective for tasks requiring advanced semantic analysis and interpretation. SC aims to achieve efficient and accurate information transmission between a transmitter and receiver by conveying semantic meaning in their messages. Large AI models possess robust semantic representation and understanding capabilities, allowing them to effectively extract the underlying meaning from data. Combining large AI models with SC can significantly enhance the performance of the wireless transceiver. For example, to improve the transmission efficiency, it is possible to generate low-dimensional representations for raw signals with large AI models, such as segmentation maps for high-resolution images and motion keypoint representations for video frames \cite{Peiwen_JSAC_2023}, presenting exciting opportunities for future exploration. These representations can also be effectively combined with large generative AI models such as the large latent diffusion model and DALL-E to reconstruct the original signals, thereby significantly improving the multimodal data transmission efficiency. Moreover, large AI models can also be used to determine SC transmission strategies, such as increasing the transmission power for more important words \cite{Shuaishuai2023SI}, image patches, or video frames. Additionally, due to the huge cost required for training and execution, research on improving the training efficiency and accelerating the inference speed of large AI models in SC systems is of great importance.
	
	\subsubsection{Advanced Coding Techniques}
	Recent studies in source coding have showcased promising performance using DL-based methods, primarily by leveraging the autoencoder architecture. However, conventional autoencoder designs have proven to be inefficient in various source coding tasks. Researchers often prefer alternatives such as VAE or learned entropy modeling methods \cite{balle2018variational, he2021checkerboard}. The shared characteristic of these successful approaches is their focus on entropy modeling and optimization methods. By adopting more advanced coding strategies, e.g., introducing hyper-priors such as intra-context and global context information from the source or channel statistics, the efficiency of source coding can be further improved. Additionally, recent efforts have concentrated solely on designing source and channel coding in a joint and ``black-box manner,'' potentially falling short of optimality due to the absence of specifically designed error protection through channel coding. Therefore, the integration of advanced source and channel coding strategies holds the potential to further improve the source coding performance.
	
	\subsubsection{Application Scenarios and Implementations}
	As intelligent services continue to grow and intelligent devices become increasingly prevalent, the landscape of future wireless networks is set to evolve from merely connecting things to fostering connected intelligence. In this context, the intelligent devices within future wireless networks must do more than simply enable various smart applications; they should also serve as a means to achieve more efficient and resilient communication. It is foreseen that SC holds great potential in different fields such as holographic communication, the industrial Internet-of-Things, smart transportation, digital twins, and more. Exploring specific potential scenarios for the future of SC is also a crucial issue.
	
	Notably, a successful implementation of SC relies not only on algorithm design but also  hardware realization. Several practical SC implementations  using dedicated hardware platforms such as the Universal Software Radio Peripheral, have demonstrated substantial performance enhancements compared to conventional systems \cite{Hanju2023Access, Dong2022WCSP}. These successes underscore the immense potential of incorporating semantic transceivers into everyday devices. Moreover, research on practical hardware implementations also faces numerous critical challenges, as listed below.
	\begin{itemize}
		\item \textbf{Performance optimization:} Implementation optimization can enhance system performance by accelerating specific tasks like semantic encoding and data processing, improving the efficiency and response time of communication systems.
		
		\item \textbf{Energy efficiency:}
		Specially designed hardware can optimize power consumption, increasing the energy efficiency of the system. This is crucial for resource-constrained mobile devices and embedded systems.
		
		\item \textbf{Real-time capability:} Hardware implementations should provide better real-time performance, ensuring that the system can meet specific time requirements when processing semantic information. This is particularly important for applications requiring rapid responses such as autonomous driving and medical devices.
		
		\item \textbf{Security enhancement:}
		Practical implementations should consider enhancing security, e.g., by providing an additional layer of protection against various attacks.
		
		\item \textbf{Customization and specialization:} SC implementations should be able to allow customization based on the specific requirements of an application, focusing the system on SC tasks. This helps improve overall efficiency by reducing redundancy and unnecessary computations.
	\end{itemize}
	
	\subsubsection{Semantic Information Theory}
	Semantic information theory involves the mathematical representation of semantics and serves as a guide for the design of SC systems. However, a comprehensive and effective framework for semantic information theory is still in its infancy. The primary challenges in establishing a fundamental theory lie in addressing the impact of semantic ambiguity, where definitions significantly vary across tasks. Overcoming these challenges is expected to drive further advancements in the field of SC. Additionally, the design and optimization of SC networks directly impact task completion and performance, which could benefit from such a theory.  This could possibly draw inspiration from research related to interpretability in machine learning. Exploring and providing further theoretical guidance for an SC information theory framework based on DL, including appropriate performance metrics and semantic similarity measures, will have a profound impact on advancing the effective transmission of semantics.
	
	\color{black}
	\subsubsection{Protocol Implementation}
	The problem of adapting current communication protocols to support SC transceivers is a key challenge. For instance, 5G New Radio (5GNR) protocols are the cutting-edge standards driving the evolution of mobile networks. 5GNR protocols encompass various layers of the protocol stacks, each serving a distinct function in facilitating seamless communication. SC will inevitably require the joint optimization of different layers, which deviates from conventional approach of layer separation. Since semantic information always exists in the application layer, it will be necessary to establish new connections between the application layer and lower layers of the protocol stack \cite{Utkovski_IOTJ}.  Despite the required complexity, this could lead to significant gains.
	
	In addition to the joint optimization of different layers, an important question is information security, as it may be unsafe for the routers or switches to have access to data that has been encrypted by other layers. This issue necessitates the development of innovative encryption algorithms and protocols capable of safeguarding semantic information while preserving the efficiency and performance of the communication system. Additionally, addressing the security concerns associated with SC protocols entails not only encryption but also authentication mechanisms to verify the integrity and authenticity of the transmitted data.
	\color{black}
	
	\color{black}
	\subsubsection{Key Performance Indicators}
     The KPIs for SC will likely revolve around measuring the effectiveness of conveying meaning and understanding between the transmitter and receiver. Some potential KPIs can be surmised, as follows.
	\begin{itemize}
		\item  \textbf{SC adaptability:} Most of the previous studies on SC are implemented based on deep learning models, which generally suffer from poor generalizability. However, the transmission conditions are typically time-varying. As a result, the ability to adapt communication strategies to changing environments is essential. KPIs could include the speed of adapting communication strategies to feedback or changes in the operating environment.
		
		\item \textbf{SC efficiency:} Efficiency measures how effectively meaning is conveyed with a given set of resources (time, power, bandwidth, etc.). KPIs could include transmission latency in communication, and time taken to comprehend messages (model execution time).
		
		\item \textbf{Semantic accuracy:} This KPI would measure the percentage of communication instances where the intended meaning is accurately understood by the recipient(s). 
		
		\item \textbf{Semantic scope:} Another important indicator of a SC system is its scope of understanding. This includes the system's ability to process information from various domains, topics, and contexts. A broader understanding of the scope implies that the system can handle more diverse and complex scenarios.
	\end{itemize}
	\color{black}

\section{Other Technologies}\label{sec_extens}
In this section, we discuss other technologies related to advanced transceiver designs in future wireless systems.

\subsection{NOMA}
The significant increase in the number of smartphones and devices imposes stringent connectivity requirements for future wireless networks. Among others, NOMA is a promising candidate for providing flexible resource allocation among different terminals and supporting massive connectivity~\cite{2022liunoma,10024901}. The key idea of NOMA is to multiplex users over the same radio resources and mitigate the resulting co-channel interference with the aid of successive interference cancellation (SIC) \cite{2022liunoma}. However, spectrum sharing via NOMA inevitably leads to complicated user scheduling and resource management to realize its full potential. To overcome these challenges, NGAT technologies provide new opportunities for advanced NOMA designs, as discussed below.
\subsubsection{Near-field NOMA}
The unique location-based beamfocusing capability achieved in the near-field region provides two new NOMA modalities, namely ``far-to-near SIC decoding'' and ``distance-based user clustering''. The key idea behind far-to-near SIC decoding is to design the beamformers to focus the signal primiarily on the users who are far from the BS rather than those near the BS. By doing so, far users can have a higher effective channel gain than the near users, and thus SIC can be carried out at the far users to eliminate co-channel interference from near users. This operation enables the far users to receive their intended signals with little interference, which is useful when the far users have stricter communication requirements than the near users~\cite{Zuo}. Note that far-to-near SIC decoding cannot be achieved with beamsteering in the far-field, since the effective channel gain decreases with increasing distance to the BS. On the other hand, distance-based user clustering exploits beamfocusing to group users at the same angle into different clusters in the distance domain. In doing so, the co-channel interference is further mitigated and the number of SIC stages required at each user can be reduced, which leads to a low-complexity solution.

The use of NOMA can also benefit near-field communications. For example, \cite{10129111} investigated the coexistence of near- and far-field users with the aid of NOMA, demonstrating that the beamformer designed for the near users can serve additional far-field users. Therefore, the connectivity and spectral efficiency of the near-field system can be enhanced with NOMA. The authors of \cite{10315058} illustrated that the resolution of near-field beamfocusing is not always perfect, which provides possibilities for using NOMA to serve additional users via near-field beamfocusing. This opens up new research directions for near-field NOMA that require further research efforts.

\subsubsection{New Forms of Antennas with NOMA}
New types of antennas and arrays also facilitate flexible NOMA. Smart radio environments enabled by RISs can beneficially adjust the users' channel conditions to improve NOMA performance \cite{9801736}. Motivated by this benefit, growing research efforts have been devoted to the investigation of RIS-aided NOMA designs. For example, joint active and passive beamforming for an RIS-aided MISO-NOMA system was studied in \cite{9139273}, which confirmed the benefits of using RISs to optimize the channel-based SIC decoding order for NOMA. The authors of \cite{9240028} explored the use of RISs to create wireless channels for ensuring that NOMA can achieve the same capacity region as optimal dirty paper coding with high probability. The full-space coverage of STAR-RISs facilitates advanced `transmission-reflection NOMA' operations in which users on different sides of the STAR-RIS are paired together for resource sharing. The channel disparity between the two users can be enlarged via the amplitude control of the STAR-RIS, which can significantly increase the performance gain of NOMA over OMA~\cite{9740451}. Compared to extensive research contributions on RIS/STAR-RIS-aided NOMA, studies using FA or HMIMO for NOMA are still in an early stage. The authors of \cite{Zheng} have explored FA-aided short-packet NOMA systems, deriving a corresponding block error rate expression and showing the effectiveness of FA for NOMA. The investigation of FA/HMIMO-aided NOMA constitutes an interesting avenue of research for the future.

\subsubsection{SC for NOMA}
Since NOMA users compete for the available radio resources, some users only achieve limited performance due to severe co-channel interference. To address this issue, the new information provided by SC can be exploited to enhance the effectiveness of NOMA. Given that SC is in general more robust and requires less radio resources compared to conventional systems, an opportunistic SC and bit communication strategy was proposed for uplink NOMA in \cite{10158994}. The principle is that a secondary NOMA user can select from SC or conventional bit communication to control the introduced co-channel interference when reusing the spectrum of the primary NOMA user. The results obtained in \cite{10158994} showed that the proposed opportunistic SC and bit communication strategy can guarantee the performance of the secondary user compared with the conventional communication scheme, and the performance gain is more pronounced when the primary user has a stringent communication requirement. In this case, SC enables the secondary user to still achieve an acceptable performance with the limited permitted transmit power budget.
{\color{black} Moreover, the authors in \cite{li2023non} proposed a NOMA-enhanced multi-user SC system that enables semantic transmission for multiple users, accommodating diverse modalities of source information. An asymmetric quantizer was used to discretize the semantic features, thereby decreasing the hardware cost and complexity. For intelligent multi-user detection, neural networks were further employed to map the quantized semantic features into self-learned symbols.}

{\color{black}{\subsubsection{Discussion and Outlook} IMT-2030 has identified NOMA as an important multiple access technique for future wireless networks~\cite{ITU_NOMA}, underscoring its great potential. However, to unlock the aforementioned benefits of NOMA, several practical issues have to be addressed:
\begin{itemize}
\item \textbf{{Low-complexity user clustering}:} User clustering is an important aspect that determines the maximum performance that can be achieved by NOMA and the operational complexity for carrying out SIC. The user clustering will become much more complicated given the enhanced DoFs introduced by near-field channels and new forms of antennas. To address this issue, conventional mathematical methods, such as convex optimization and match theory, may become inefficient, which calls for other advanced ML tools. This constitutes an interesting future research direction for NOMA.
\item \textbf{{CSI estimation issues}:} As discussed in previous sections, CSI estimation becomes more challenging for near-field communications and higher frequency bands. The resulting high overhead and complexity will greatly limit the performance of NOMA. To address this issue, efficient tailored CSI estimation schemes must be developed, and NOMA communication design under imperfect or limited CSI should be further studied. 
\item \textbf{{Error propagation issues}:} Due to the use of SIC, the mitigation of error propagation has been a long-term research topic for practical NOMA implementations. Fortunately, the advancement of SC could come to the rescue. Recall the fact that SC is more robust to interference than conventional bit-level communication \cite{mu2022semi_NOMA,10158994}. Allowing some NOMA users with higher SIC decoding order to use SC would be helpful for combating SIC error propagation. This is another interesting research topic for NOMA. 
\end{itemize}
}}

\subsection{Localization and Sensing}
Localization and sensing (L\&S) are increasingly important in the evolution of next-generation wireless networks, driven by the emergence of various location and environment-aware applications for smart cities, smart homes, XR, and vehicle-to-everything (V2X) communication and sensing \cite{liu2022integrated}. NGAT plays a key role in achieving robust L\&S performance and seamlessly integrating L\&S functionalities within next-generation wireless networks, as explored in the following.

\subsubsection{Near-field L\&S}
In conventional far-field L\&S systems, the propagation of planar waves means that the transceiver antenna arrays are limited to resolving the direction of incoming signals. Consequently, pinpointing targets in far-field L\&S systems necessitates either a large bandwidth or the deployment of multiple anchors to accurately estimate the target range information \cite{lu2024integrated}. These demands can lead to considerable system overhead, particularly in terms of signal synchronization. The properties of EM wave propagation have recently been leveraged for near-field localization and attitude sensing \cite{chen2023cramer,chen2023near}. Since the near-field channel enables estimation of target range, the need for increased bandwidth and multiple anchors is reduced \cite{8736783, wang2023nearisac, wang2024cramer}. 

The near-field effect also introduces transformative advancements in L\&S by enabling estimation of the complete motion status of mobile targets, which is determined by both the radial and transverse velocities. In conventional far-field systems, estimates of the target's velocity are based on Doppler frequency shifts, and thus are limited to describing the radial motion of the target \cite{lu2024integrated}. However, in near-field L\&S systems, the Doppler frequencies involve information about both the radial and transverse velocities, offering a more detailed motion profile of the targets \cite{wang2023near_v, wang2023rethinking}. With knowledge of both the radial and transverse velocities of the targets, it is easier to predict and track the targets' trajectories, enhancing the L\&S performance in dynamic environments. 

\subsubsection{RIS/STAR-RIS for L\&S} L\&S typically rely on LoS links between transceivers and targets. Obstructions in the LoS link pose considerable challenges for effective L\&S. However, leveraging RISs presents a promising solution by creating virtual LoS paths, thereby enhancing L\&S performance \cite{shao2022target,shao2024intelligent,hua2023intelligent,hua2023secure}. Additionally, even in scenarios where the LoS is unobstructed, an RIS can provide additional observations, further improving L\&S accuracy \cite{10243495}. However, a conventional RIS only reflects signals and requires the transceivers and targets to be positioned on the same side of the surface, which limits the coverage area and deployment flexibility. To address this issue, STAR-RIS can provide full 360$^\circ$ coverage and overcome LoS blockages in more challenging scenarios, such as indoors and for indoor-to-outdoor L\&S scenarios \cite{wang2023stars}.

However, for either RIS or STAR-RIS, the sensing performance may be limited by the cumulative loss of signals that follow the echo path from the passive target through the virtual LoS link. A promising strategy to address this issue is to install dedicated active sensors on the RIS and STAR-RIS \cite{shao2022target, wang2023stars, zhang2023stars} so that the signals can be received with higher SNR due to the reduced path loss. Another effective strategy to counteract the cumulative path loss is leveraging active RIS or STAR-RIS, which incorporate active amplifiers within each element to enhance the amplitude of the reflected or refracted echo signals \cite{rihan2023passive}.

\subsubsection{ISAC}
ISAC has emerged as a promising technology for providing both communication and sensing functions on the same shared platform \cite{liu2022integrated}.
A key design issue for ISAC is how to develop efficient multi-function transceivers. Specifically, for joint communication and sensing, it is necessary to balance the trade-off between improving the communication and sensing performance since they usually compete for the same resources. In addition, new types of waveforms must be designed since communication systems usually employ random signals for carrying information, while sensing systems such as radar use deterministic signals to recover the information to be sensed \cite{zhang2022integrated}. This trade-off was studied in \cite{xiong2023fundamental}, which designed the transmitted waveforms to share a deterministic sample covariance for delivering communication information, while assuming the realizations of the random waveform to be perfectly known at the sensing receivers. One application for sensing-assisted communication systems is to exploit sensing information such as echo signals to estimate the communication channels, thus avoiding the need for dedicated pilots and hence achieving enhanced spectral efficiency. For communication-assisted sensing, one could use OFDM waveforms common to communication systems for target detection and tracking. Efficient signal processing techniques must be developed for such ISAC systems and deserve future investigation.

\subsection{Circuit Theory Based Transceiver Design}
{\color{black}
To achieve demanding performance requirements in future wireless systems, new cross-disciplinary techniques are needed that better account for the effects of real hardware on communication systems. For example, circuit and EM theory can be used to establish the fundamental relationship between the digital signals processed by the computing hardware of a communication system, and the corresponding EM fields produced by the system's antennas \cite{abrardo2023design,bjornson2024towards}. Compared with traditional simplified models, an end-to-end channel model based on circuit theory that takes into consideration fundamental physical laws such as conservation of energy and the superposition principle will provide more accurate predictions of performance.

For example, in circuit theory, an antenna can be modeled as a wave transformer that converts guided waves at its terminal ports into EM fields. As such, a multi-antenna communication system can be modeled as a multi-port network:
\begin{equation}
	\left[\begin{array}{l}
	\boldsymbol{V}_{\rm{t}} \\
	\boldsymbol{V}_{\rm{r}}
	\end{array}\right]=\left[\begin{array}{ll}
	\boldsymbol{Z}_{\rm {t t}} & \boldsymbol{Z}_{\rm {t r}} \\
	\boldsymbol{Z}_{\rm{r t}} & \boldsymbol{Z}_{\rm{r r}}
	\end{array}\right]\left[\begin{array}{l}
	\boldsymbol{I}_{\rm{t}} \\
	\boldsymbol{I}_{\rm{r}}
	\end{array}\right],  
\end{equation}
where $ \boldsymbol{V}_{\rm{t}} $ and $ \boldsymbol{I}_{\rm{t}} $ ($ \boldsymbol{V}_{\rm{r}}$ and $ \boldsymbol{I}_{\rm{r}} $) respectively denote the circuit voltages and currents for the transmitter (receiver), $ \boldsymbol{Z}_{\rm {t t}} $ and $ \boldsymbol{Z}_{\rm {r r}} $ denote the mutual impedance matrix of the transmit and the receive array, respectively, $ \boldsymbol{Z}_{\rm{r t}} $ denotes the mutual impedance between the transmit and receive array, and $ \boldsymbol{Z}_{\rm{t r}} $ represents the back-scattering impedance from the receive antennas back to the transmitter. This latter term can typically be neglected for far-field scenarios since the relatively weak currents received at the users cause negligible effects at the transmitter \cite{Gong_Vinieratou_Ji_Huang_Alexandropoulos_Wei_Zhang_Debbah_Poor_Yuen_2022,bjornson2024towards}.
Such a multi-port network model based on circuit theory can facilitate in-depth analysis of impedance matching, antenna mutual coupling, structural losses, and both intrinsic and extrinsic sources of noise \cite{ivrlavc2010toward}. For example, \cite{gradoni2021end} proposed a circuit-based EM-compliant communication model for RIS assisted wireless communication systems, which characterizes the end-to-end mutual coupling in RIS elements based on mutual impedance. In \cite{nerini2023universal}, a universally EM-compliant framework was proposed for RIS-assisted wireless systems based on impedance, admittance, and scattering parameter analysis, revealing the effects of impedance mismatches and mutual coupling on RIS communication performance. In addition, the authors of \cite{abrardo2023design} showed that the typical scattering models utilized for RIS-aided channels largely ignored the structural scattering from the RIS, hence resulting in an unwanted specular reflection. To address this issue, an iterative algorithm was developed in \cite{abrardo2023design} 
to  optimize the tunable loads of RISs, based on the S-parameter representation in the presence of electromagnetic mutual coupling.  Moreover, multi-port network models were used in \cite{Williams_2022,Marzetta_2019} to reveal the negative and positive aspects of mutual coupling on wireless communication systems. The authors of
\cite{akrout2022achievable} used multi-port models and accurate antenna characterizations to determine the effect of mutual coupling on the achievable rate of SISO near-field communication systems.
However, applying circuit theory to more complex wireless systems, possibly with multi-user interference, is an open problem that deserves more in-depth study in the future. }

\section{Conclusions}\label{sec_con}
In this paper, we have provided a comprehensive overview of the recent progress on NGAT with an emphasis on three new aspects. Specifically, we first introduced new channel models and transceiver design issues for near-field communication systems, such as beamforming, beam training and channel estimation. Then, three new-form NGAT technologies were presented, including RISs (from metals to metamaterials), FAs (from static to movable), and HMIMO (from discrete to continuous). Furthermore, we elaborated recent advances in semantic-aware NGAT technology and its open challenges. Finally,  other promising transceiver technologies were discussed to inspire future research. It is hoped that this paper will provide valuable insights as well as up-to-date guidance for future research on NGAT to unlock its full potential in next generation wireless networks.

\bibliographystyle{IEEEtran}
\bibliography{IEEEabrv,mybib}

\begin{thebibliography}{100}
\providecommand{\url}[1]{#1}
\csname url@samestyle\endcsname
\providecommand{\newblock}{\relax}
\providecommand{\bibinfo}[2]{#2}
\providecommand{\BIBentrySTDinterwordspacing}{\spaceskip=0pt\relax}
\providecommand{\BIBentryALTinterwordstretchfactor}{4}
\providecommand{\BIBentryALTinterwordspacing}{\spaceskip=\fontdimen2\font plus
\BIBentryALTinterwordstretchfactor\fontdimen3\font minus
  \fontdimen4\font\relax}
\providecommand{\BIBforeignlanguage}[2]{{%
\expandafter\ifx\csname l@#1\endcsname\relax
\typeout{** WARNING: IEEEtran.bst: No hyphenation pattern has been}%
\typeout{** loaded for the language `#1'. Using the pattern for}%
\typeout{** the default language instead.}%
\else
\language=\csname l@#1\endcsname
\fi
#2}}
\providecommand{\BIBdecl}{\relax}
\BIBdecl

\bibitem{6gusage}
{ITU-R WP5D}, ``Future technology trends of terrestrial international mobile
  telecommunications systems towards 2030 and beyond,'' [Online]. Available:
  https://www.itu.int/pub/R-REP-M.2516, Nov. 2023.

\bibitem{you2021towards}
X.~You, C.-X. Wang, J.~Huang, X.~Gao, Z.~Zhang, M.~Wang, Y.~Huang, C.~Zhang,
  Y.~Jiang, J.~Wang \emph{et~al.}, ``Towards $6g$ wireless communication
  networks: Vision, enabling technologies, and new paradigm shifts,''
  \emph{Science China Info. Sci.}, vol.~64, pp. 1--74, 2021.

\bibitem{9349624}
W.~Jiang, B.~Han, M.~A. Habibi, and H.~D. Schotten, ``The road towards {6G}: A
  comprehensive survey,'' \emph{IEEE Open J. Commun. Soc.}, vol.~2, pp.
  334--366, Feb. 2021.

\bibitem{VIMT}
``Views towards {IMT} for 2030 and beyond,'' Available:
  \url{https://hexa-x.eu/wp-content/uploads/2023/01/IMT-20306GPromotionGroup-QingyangWang-Workshop-Jan-2023.pdf}.

\bibitem{8808168}
K.~B. Letaief, W.~Chen, Y.~Shi, J.~Zhang, and Y.-J.~A. Zhang, ``The roadmap to
  {6G}: {AI} empowered wireless networks,'' \emph{IEEE Commun. Mag.}, vol.~57,
  no.~8, pp. 84--90, Aug. 2019.

\bibitem{8869705}
W.~Saad, M.~Bennis, and M.~Chen, ``A vision of {6G} wireless systems:
  Applications, trends, technologies, and open research problems,'' \emph{IEEE
  Netw.}, vol.~34, no.~3, pp. 134--142, Mar. 2020.

\bibitem{shafi20175g}
M.~Shafi, A.~F. Molisch, P.~J. Smith, T.~Haustein, P.~Zhu, P.~De~Silva,
  F.~Tufvesson, A.~Benjebbour, and G.~Wunder, ``{5G}: A tutorial overview of
  standards, trials, challenges, deployment, and practice,'' \emph{IEEE J. Sel.
  Areas Commun.}, vol.~35, no.~6, pp. 1201--1221, Jun. 2017.

\bibitem{boccardi2014five}
F.~Boccardi, R.~W. Heath, A.~Lozano, T.~L. Marzetta, and P.~Popovski, ``Five
  disruptive technology directions for {5G},'' \emph{IEEE Commu. Mag.},
  vol.~52, no.~2, pp. 74--80, Feb. 2014.

\bibitem{cui2022near}
M.~Cui, Z.~Wu, Y.~Lu, X.~Wei, and L.~Dai, ``Near-field {MIMO} communications
  for {6G}: Fundamentals, challenges, potentials, and future directions,''
  \emph{IEEE Commu. Mag.}, vol.~61, no.~1, pp. 40--46, Jan. 2022.

\bibitem{liu2023near}
Y.~Liu, Z.~Wang, J.~Xu, C.~Ouyang, X.~Mu, and R.~Schober, ``Near-field
  communications: A tutorial review,'' \emph{IEEE Open J. Commun. Soc.},
  vol.~4, pp. 1999--2049, Aug. 2023.

\bibitem{lu2023tutorial}
H.~Lu, Y.~Zeng, C.~You, Y.~Han, J.~Zhang, Z.~Wang, Z.~Dong, S.~Jin, C.-X. Wang,
  T.~Jiang \emph{et~al.}, ``A tutorial on near-field {XL-MIMO} communications
  towards {6G},'' \emph{IEEE Commun. Surveys Tuts.}, Early Access, 2024.

\bibitem{you2023near}
C.~You, Y.~Zhang, C.~Wu, Y.~Zeng, B.~Zheng, L.~Chen, L.~Dai, and A.~L.
  Swindlehurst, ``Near-field beam management for extremely large-scale array
  communications,'' \emph{arXiv preprint arXiv:2306.16206}, 2023.

\bibitem{liu2024near}
Y.~Liu, C.~Ouyang, Z.~Wang, J.~Xu, X.~Mu, and A.~L. Swindlehurst, ``Near-field
  communications: A comprehensive survey,'' \emph{arXiv preprint
  arXiv:2401.05900}, 2024.

\bibitem{wu2019towards}
Q.~Wu and R.~Zhang, ``Towards smart and reconfigurable environment: Intelligent
  reflecting surface aided wireless network,'' \emph{IEEE Commun. Mag.},
  vol.~58, no.~1, pp. 106--112, Jan. 2019.

\bibitem{wu2021intelligent}
Q.~Wu, S.~Zhang, B.~Zheng, C.~You, and R.~Zhang, ``Intelligent reflecting
  surface-aided wireless communications: A tutorial,'' \emph{IEEE Trans.
  Commun.}, vol.~69, no.~5, pp. 3313--3351, May 2021.

\bibitem{di2020smart}
M.~Di~Renzo, A.~Zappone, M.~Debbah, M.-S. Alouini, C.~Yuen, J.~De~Rosny, and
  S.~Tretyakov, ``Smart radio environments empowered by reconfigurable
  intelligent surfaces: How it works, state of research, and the road ahead,''
  \emph{IEEE J. Sel. Areas Commun.}, vol.~38, no.~11, pp. 2450--2525, Nov.
  2020.

\bibitem{8796365}
E.~Basar, M.~Di~Renzo, J.~De~Rosny, M.~Debbah, M.-S. Alouini, and R.~Zhang,
  ``Wireless communications through reconfigurable intelligent surfaces,''
  \emph{IEEE Access}, vol.~7, pp. 116\,753--116\,773, Aug. 2019.

\bibitem{liu2021reconfigurable}
Y.~Liu, X.~Liu, X.~Mu, T.~Hou, J.~Xu, M.~Di~Renzo, and N.~Al-Dhahir,
  ``Reconfigurable intelligent surfaces: Principles and opportunities,''
  \emph{IEEE Commun. Surveys Tuts.}, vol.~23, no.~3, pp. 1546--1577, May 2021.

\bibitem{swindlehurst2022channel}
A.~L. Swindlehurst, G.~Zhou, R.~Liu, C.~Pan, and M.~Li, ``Channel estimation
  with reconfigurable intelligent surfaces---{A} general framework,''
  \emph{Proc. IEEE}, vol. 110, no.~9, pp. 1312--1338, Sept. 2022.

\bibitem{wong2022bruce}
K.-K. Wong, K.-F. Tong, Y.~Shen, Y.~Chen, and Y.~Zhang, ``Bruce lee-inspired
  fluid antenna system: Six research topics and the potentials for {6G},''
  \emph{Front. Comms. Net.}, vol.~3, p. 853416, Mar. 2022.

\bibitem{zhu2023movable}
L.~Zhu, W.~Ma, and R.~Zhang, ``Movable antennas for wireless communication:
  Opportunities and challenges,'' \emph{arXiv preprint arXiv:2306.02331}, 2023.

\bibitem{huang2020holographic}
C.~Huang, S.~Hu, G.~C. Alexandropoulos, A.~Zappone, C.~Yuen, R.~Zhang,
  M.~Di~Renzo, and M.~Debbah, ``Holographic {MIMO} surfaces for {6G} wireless
  networks: Opportunities, challenges, and trends,'' \emph{IEEE Wireless
  Commun.}, vol.~27, no.~5, pp. 118--125, May 2020.

\bibitem{an2023tutorial}
J.~An, C.~Yuen, C.~Huang, M.~Debbah, H.~V. Poor, and L.~Hanzo, ``A tutorial on
  holographic {MIMO} communications---{Part I}: Channel modeling and channel
  estimation,'' \emph{IEEE Commun. Lett.}, Jul. 2023.

\bibitem{9509294}
D.~C. Nguyen, M.~Ding, P.~N. Pathirana, A.~Seneviratne, J.~Li, D.~Niyato,
  O.~Dobre, and H.~V. Poor, ``{6G Internet of Things}: A comprehensive
  survey,'' \emph{IEEE Internet Things J.}, vol.~9, no.~1, pp. 359--383, Jan.
  2022.

\bibitem{8732419}
T.~S. Rappaport, Y.~Xing, O.~Kanhere, S.~Ju, A.~Madanayake, S.~Mandal,
  A.~Alkhateeb, and G.~C. Trichopoulos, ``Wireless communications and
  applications above 100 {GHz}: Opportunities and challenges for {6G} and
  beyond,'' \emph{IEEE Access}, vol.~7, pp. 78\,729--78\,757, Jun. 2019.

\bibitem{2022liunoma}
Y.~Liu, S.~Zhang, X.~Mu, Z.~Ding, R.~Schober, N.~Al-Dhahir, E.~Hossain, and
  X.~Shen, ``Evolution of {NOMA} toward next generation multiple access
  ({NGMA}) for {6G},'' \emph{IEEE J. Sel. Areas Commun.}, vol.~40, no.~4, pp.
  1037--1071, Jan. 2022.

\bibitem{Shannon_BSTJ}
C.~E. Shannon, ``A mathematical theory of communication,'' \emph{Bell Syst.
  Tech. J.}, vol.~27, pp. 379--423, Jul. 1948.

\bibitem{guler_13_globalsip}
B.~Guler, A.~Yener, and P.~Basu, ``A study of semantic data compression,'' in
  \emph{Proc. IEEE Global Conf. Signal Info. Process. (GlobalSIP)}, 2013, pp.
  887--890.

\bibitem{semantic_index_assignment}
B.~Guler and A.~Yener, ``Semantic index assignment,'' in \emph{Proc. IEEE Int.
  Conf. Pervasive Comput. Commun. Workshops (PERCOM WORKSHOPS)}, 2014, pp.
  431--436.

\bibitem{guler_18_semantic_game}
B.~Guler, A.~Yener, and A.~Swami, ``The semantic communication game,''
  \emph{IEEE Trans. Cogn. Commun. Netw.}, vol.~4, no.~4, pp. 787--802, Apr.
  2018.

\bibitem{sagduyu2024will}
Y.~E. Sagduyu, T.~Erpek, A.~Yener, and S.~Ulukus, ``Will {6G} be semantic
  communications? opportunities and challenges from task oriented and secure
  communications to integrated sensing,'' \emph{arXiv preprint
  arXiv:2401.01531}, 2024.

\bibitem{9450827}
Z.~Weng and Z.~Qin, ``Semantic communication systems for speech transmission,''
  \emph{IEEE J. Sel. Areas Commun.}, vol.~39, no.~8, pp. 2434--2444, Aug. 2021.

\bibitem{emrecan_semantic_overview}
E.~Kutay and A.~Yener, ``Semantic communications: A paradigm whose time has
  come,'' in \emph{Proc. IEEE Int. Conf. Collab. Inter. Comput. (CIC)}, 2022,
  pp. 68--71.

\bibitem{NiuKai_ICM}
K.~Niu, J.~Dai, S.~Yao, S.~Wang, Z.~Si, X.~Qin, and P.~Zhang, ``A paradigm
  shift toward semantic communications,'' \emph{IEEE Commun. Mag.}, vol.~60,
  no.~11, pp. 113--119, Nov. 2022.

\bibitem{lan2021semantic}
Q.~Lan, D.~Wen, Z.~Zhang, Q.~Zeng, X.~Chen, P.~Popovski, and K.~Huang, ``What
  is semantic communication? {A} view on conveying meaning in the era of
  machine intelligence,'' \emph{J. Commun. Info. Netw.}, vol.~6, no.~4, pp.
  336--371, Apr. 2021.

\bibitem{selvan2017fraunhofer}
K.~T. Selvan and R.~Janaswamy, ``Fraunhofer and {Fresnel} distances: Unified
  derivation for aperture antennas,'' \emph{IEEE Antennas Propag. Mag.},
  vol.~59, no.~4, pp. 12--15, Aug. 2017.

\bibitem{lu2021communicating}
H.~Lu and Y.~Zeng, ``Communicating with extremely large-scale array/surface:
  Unified modeling and performance analysis,'' \emph{IEEE Trans. Wireless
  Commun.}, vol.~21, no.~6, pp. 4039--4053, Jun. 2021.

\bibitem{lu2021does}
------, ``How does performance scale with antenna number for extremely
  large-scale {MIMO}?'' in \emph{Proc. IEEE Int. Conf. Commun. (ICC)}, Jun.
  2021.

\bibitem{cui2021near}
M.~Cui, L.~Dai, R.~Schober, and L.~Hanzo, ``Near-field wideband beamforming for
  extremely large antenna arrays,'' \emph{arXiv preprint arXiv:2109.10054},
  2021.

\bibitem{bjornson2021primer}
E.~Bjornson, {\"O}.~T. Demir, and L.~Sanguinetti, ``A primer on near-field
  beamforming for arrays and reconfigurable intelligent surfaces,'' in
  \emph{Proc. Asilomar Conf. Signals Syst. Comp.}, Oct. 2021, pp. 105--112.

\bibitem{zhang20236g}
H.~Zhang, N.~Shlezinger, F.~Guidi, D.~Dardari, and Y.~C. Eldar, ``6{G} wireless
  communications: From far-field beam steering to near-field beam focusing,''
  \emph{IEEE Commun. Mag.}, vol.~61, no.~4, pp. 72--77, Apr. 2023.

\bibitem{deutschmann2022location}
B.~J. Deutschmann, T.~Wilding, E.~G. Larsson, and K.~Witrisal, ``Location-based
  initial access for wireless power transfer with physically large arrays,'' in
  \emph{Proc. IEEE Int. Conf. Commun. Workshops (ICC Wkshps)}, May. 2022, pp.
  127--132.

\bibitem{guerra2021near}
A.~Guerra, F.~Guidi, D.~Dardari, and P.~M. Djuri{\'c}, ``Near-field tracking
  with large antenna arrays: Fundamental limits and practical algorithms,''
  \emph{IEEE Trans. Signal Process.}, vol.~69, pp. 5723--5738, Jan. 2021.

\bibitem{cong2023near}
J.~Cong, C.~You, J.~Li, L.~Chen, B.~Zheng, Y.~Liu, W.~Wu, Y.~Gong, S.~Jin, and
  R.~Zhang, ``Near-field integrated sensing and communication: Opportunities
  and challenges,'' \emph{arXiv preprint arXiv:2310.01342}, 2023.

\bibitem{zhang2023physical}
Z.~Zhang, Y.~Liu, Z.~Wang, X.~Mu, and J.~Chen, ``Physical layer security in
  near-field communications,'' \emph{IEEE Trans. Veh. Technol.}, Early Access,
  2024. doi:10.1109/TVT.2024.3366115.

\bibitem{cui2022channel}
M.~Cui and L.~Dai, ``Channel estimation for extremely large-scale {MIMO}:
  Far-field or near-field?'' \emph{IEEE Trans. Commun.}, vol.~70, no.~4, pp.
  2663--2677, Apr. 2022.

\bibitem{wei2021channel}
X.~Wei and L.~Dai, ``Channel estimation for extremely large-scale massive
  {MIMO}: Far-field, near-field, or hybrid-field?'' \emph{IEEE Commun. Lett.},
  vol.~26, no.~1, pp. 177--181, Jan. 2021.

\bibitem{lu2023near}
Y.~Lu and L.~Dai, ``Near-field channel estimation in mixed {L}o{S}/{NL}o{S}
  environments for extremely large-scale {MIMO} systems,'' \emph{IEEE Trans.
  Commun.}, vol.~71, no.~6, pp. 3694--3707, Jun. 2023.

\bibitem{dong2022near}
Z.~Dong and Y.~Zeng, ``Near-field spatial correlation for extremely large-scale
  array communications,'' \emph{IEEE Commun. Lett.}, vol.~26, no.~7, pp.
  1534--1538, Jul. 2022.

\bibitem{bjornson2019massive}
E.~Bjornson, L.~Sanguinetti, H.~Wymeersch, J.~Hoydis, and T.~L. Marzetta,
  ``Massive {MIMO} is a reality---what is next?: Five promising research
  directions for antenna arrays,'' \emph{Digit. Signal Process.}, vol.~94, pp.
  3--20, Nov. 2019.

\bibitem{ali2019linear}
A.~Ali, E.~De~Carvalho, and R.~W. Heath, ``Linear receivers in non-stationary
  massive {MIMO} channels with visibility regions,'' \emph{IEEE Wireless
  Commun. Lett.}, vol.~8, no.~3, pp. 885--888, Jun. 2019.

\bibitem{li2015capacity}
X.~Li, S.~Zhou, E.~Bj{\"o}rnson, and J.~Wang, ``Capacity analysis for spatially
  non-wide sense stationary uplink massive {MIMO} systems,'' \emph{IEEE Trans.
  Wireless Commun.}, vol.~14, no.~12, pp. 7044--7056, Dec. 2015.

\bibitem{feng2022mutual}
R.~Feng, C.-X. Wang, J.~Huang, Y.~Zheng, F.~Lai, and W.~Zhou, ``Mutual coupling
  analysis of {6G} ultra-massive {MIMO} channel measurements and models,'' in
  \emph{Proc. IEEE Int. Conf. Commun. (ICC)}, May. 2022, pp. 956--961.

\bibitem{may2020stochastic}
L.~May~Taniguchi and T.~Abr{\~a}o, ``Stochastic channel models for massive and
  extreme large multiple-input multiple-output systems,'' \emph{Trans. Emerg.
  Telecommun. Technol}, vol.~31, no.~9, p. e4099, Aug. 2020.

\bibitem{han2023towards}
Y.~Han, S.~Jin, M.~Matthaiou, T.~Q. Quek, and C.-K. Wen, ``Towards extra
  large-scale {MIMO}: New channel properties and low-cost designs,'' \emph{IEEE
  Internet Things J.}, vol.~10, no.~16, pp. 14\,569--14\,594, Aug. 2023.

\bibitem{zhu2021bayesian}
Y.~Zhu, H.~Guo, and V.~K. Lau, ``Bayesian channel estimation in multi-user
  massive {MIMO} with extremely large antenna array,'' \emph{IEEE Trans. Signal
  Process.}, vol.~69, pp. 5463--5478, Oct. 2021.

\bibitem{han2020channel}
Y.~Han, S.~Jin, C.-K. Wen, and X.~Ma, ``Channel estimation for extremely
  large-scale massive {MIMO} systems,'' \emph{IEEE Wireless Commun. Lett.},
  vol.~9, no.~5, pp. 633--637, May. 2020.

\bibitem{han2020deep}
Y.~Han, M.~Li, S.~Jin, C.-K. Wen, and X.~Ma, ``Deep learning-based fdd
  non-stationary massive {MIMO} downlink channel reconstruction,'' \emph{IEEE
  J. Sel. Areas Commun.}, vol.~38, no.~9, pp. 1980--1993, Sept. 2020.

\bibitem{yang2020uplink}
X.~Yang, F.~Cao, M.~Matthaiou, and S.~Jin, ``On the uplink transmission of
  extra-large scale massive {MIMO} systems,'' \emph{IEEE Trans. Veh. Technol.},
  vol.~69, no.~12, pp. 15\,229--15\,243, Dec. 2020.

\bibitem{rodrigues2020low}
V.~C. Rodrigues, A.~Amiri, T.~Abrao, E.~De~Carvalho, and P.~Popovski,
  ``Low-complexity distributed {XL-MIMO} for multiuser detection,'' in
  \emph{Proc. IEEE Int. Conf. Commun. Wkshps. (ICC Wkshps)}, Jun. 2020.

\bibitem{croisfelt2021accelerated}
V.~Croisfelt, A.~Amiri, T.~Abr{\~a}o, E.~De~Carvalho, and P.~Popovski,
  ``Accelerated randomized methods for receiver design in extra-large scale
  {MIMO} arrays,'' \emph{IEEE Trans. Veh. Technol.}, vol.~70, no.~7, pp.
  6788--6799, Jul. 2021.

\bibitem{Z2023near}
Z.~Dong, X.~Li, Y.~Zeng, S.~Jin, and T.~Jiang, ``Near-field spatial correlation
  for multi-path {XL}-array communications with partial visibility,'' in
  \emph{Proc. IEEE Global Commun. Conf. (GLOBECOM)}, 2023.

\bibitem{amiri2021distributed}
A.~Amiri, S.~Rezaie, C.~N. Manch{\'o}n, and E.~De~Carvalho, ``Distributed
  receiver processing for extra-large {MIMO} arrays: A message passing
  approach,'' \emph{IEEE Trans. Wireless Commun.}, vol.~21, no.~4, pp.
  2654--2667, Apr. 2021.

\bibitem{guerra2022clustered}
D.~W.~M. Guerra and T.~Abr{\~a}o, ``Clustered double-scattering channel
  modeling for {XL-MIMO} with uniform arrays,'' \emph{IEEE Access}, vol.~10,
  pp. 20\,173--20\,186, Mar. 2022.

\bibitem{wu2023location}
Z.~Wu and L.~Dai, ``Location division multiple access for near-field
  communications,'' \emph{arXiv preprint arXiv:2301.09082}, 2023.

\bibitem{kosasih2023finite}
A.~Kosasih and E.~Bj{\"o}rnson, ``Finite beam depth analysis for large
  arrays,'' \emph{arXiv preprint arXiv:2306.12367}, 2023.

\bibitem{ding2023resolution}
Z.~Ding, ``Resolution of near-field beamforming and its impact on {NOMA},''
  \emph{IEEE Wireless Commun. Lett.}, vol.~13, no.~2, pp. 456--460, Feb. 2024.

\bibitem{li2022analytical}
L.~Li, H.~Li, Z.~Chen, W.~Chen, and S.~Li, ``An analytical range-angle
  dependent beam focusing model for terahertz linear antenna array,''
  \emph{IEEE Wireless Commun. Lett.}, vol.~11, no.~9, pp. 1870--1874, Sept.
  2022.

\bibitem{xie2023near}
Y.~Xie, B.~Ning, L.~Li, and Z.~Chen, ``Near-field beam training in {THz}
  communications: The merits of uniform circular array,'' \emph{IEEE Wireless
  Commun. Lett.}, vol.~12, no.~4, pp. 575--579, Apr. 2023.

\bibitem{zhang2023mixed}
Y.~Zhang, C.~You, L.~Chen, and B.~Zheng, ``Mixed near- and far-field
  communications for extremely large-scale array: An interference
  perspective,'' \emph{IEEE Commun. Lett.}, vol.~27, no.~9, pp. 2496--2500,
  Sept. 2023.

\bibitem{zhang2023swipt}
Y.~Zhang and C.~You, ``{SWIPT} in mixed near-and far-field channels: Joint beam
  scheduling and power allocation,'' \emph{IEEE J. Sel. Areas Commun.},
  vol.~42, no.~6, pp. 1583--1597, Jun. 2024.

\bibitem{li2022near}
X.~Li, H.~Lu, Y.~Zeng, S.~Jin, and R.~Zhang, ``Near-field modeling and
  performance analysis of modular extremely large-scale array communications,''
  \emph{IEEE Commun. Lett.}, vol.~26, no.~7, pp. 1529--1533, Jul. 2022.

\bibitem{zhi2023performance}
K.~Zhi, C.~Pan, H.~Ren, K.~K. Chai, C.-X. Wang, R.~Schober, and X.~You,
  ``Performance analysis and low-complexity design for {XL-MIMO} with
  near-field spatial non-stationarities,'' \emph{arXiv preprint
  arXiv:2304.00172}, 2023.

\bibitem{zhou2024sparse}
C.~Zhou, C.~You, H.~Zhang, L.~Chen, and S.~Shi, ``Sparse array enabled
  near-field communications: Beam pattern analysis and hybrid beamforming
  design,'' \emph{arXiv preprint arXiv:2401.05690}, 2024.

\bibitem{wang2023ttd}
Z.~Wang, X.~Mu, Y.~Liu, and R.~Schober, ``{TTD} configurations for near-field
  beamforming: Parallel, serial, or hybrid?'' \emph{IEEE Trans. Commun.}, Early
  Access, 2024.

\bibitem{wang2024tutorial}
Z.~Wang, J.~Zhang, H.~Du, D.~Niyato, S.~Cui, B.~Ai, M.~Debbah, K.~B. Letaief,
  and H.~V. Poor, ``A tutorial on extremely large-scale {MIMO} for {6G}:
  Fundamentals, signal processing, and applications,'' \emph{IEEE Commun.
  Surveys Tuts.}, Early Access, 2024.

\bibitem{xie2023performance}
Z.~Xie, Y.~Liu, J.~Xu, X.~Wu, and A.~Nallanathan, ``Performance analysis for
  near-field {MIMO}: Discrete and continuous aperture antennas,'' \emph{arXiv
  preprint arXiv:2304.06141}, 2023.

\bibitem{xu2023low}
B.~Xu, Z.~Wang, H.~Xiao, J.~Zhang, B.~Ai, and D.~W.~K. Ng, ``Low-complexity
  precoding for extremely large-scale {MIMO} over non-stationary channels,''
  \emph{arXiv preprint arXiv:2302.00847}, 2023.

\bibitem{sun2021low}
Z.~Sun, X.~Pu, S.~Shao, S.~Jin, and Q.~Chen, ``A low complexity expectation
  propagation detector for extra-large scale massive {MIMO},'' in \emph{Proc.
  IEEE/CIC Int. Conf. Commun. China (ICCC)}, Jul. 2021, pp. 746--751.

\bibitem{wang2020expectation}
H.~Wang, A.~Kosasih, C.-K. Wen, S.~Jin, and W.~Hardjawana, ``Expectation
  propagation detector for extra-large scale massive {MIMO},'' \emph{IEEE
  Trans. Wireless Commun.}, vol.~19, no.~3, pp. 2036--2051, Mar. 2020.

\bibitem{amiri2019message}
A.~Amiri, C.~N. Manch{\'o}n, and E.~De~Carvalho, ``A message passing based
  receiver for extra-large scale {MIMO},'' in \emph{Proc. IEEE Int. Workshop
  Comput. Adv. Multi-Sensor Adapt. Process.}, Dec. 2019, pp. 564--568.

\bibitem{chen2020hybrid}
Y.~Chen, Y.~Xiong, D.~Chen, T.~Jiang, S.~X. Ng, and L.~Hanzo, ``Hybrid
  precoding for wideband millimeter wave {MIMO} systems in the face of beam
  squint,'' \emph{IEEE Trans. Wireless Commun.}, vol.~20, no.~3, pp.
  1847--1860, Mar. 2020.

\bibitem{dai2022delay}
L.~Dai, J.~Tan, Z.~Chen, and H.~V. Poor, ``Delay-phase precoding for wideband
  {THz} massive {MIMO},'' \emph{IEEE Trans. Wireless Commun.}, vol.~21, no.~9,
  pp. 7271--7286, Sept. 2022.

\bibitem{zhai2020thzprism}
B.~Zhai, Y.~Zhu, A.~Tang, and X.~Wang, ``{THzPrism}: Frequency-based beam
  spreading for terahertz communication systems,'' \emph{IEEE Wireless Commun.
  Lett.}, vol.~9, no.~6, pp. 897--900, Jun. 2020.

\bibitem{rotman2016true}
R.~Rotman, M.~Tur, and L.~Yaron, ``True time delay in phased arrays,''
  \emph{Proc. IEEE}, vol. 104, no.~3, pp. 504--518, Mar. 2016.

\bibitem{wang2023beamfocusing}
Z.~Wang, X.~Mu, and Y.~Liu, ``Beamfocusing optimization for near-field wideband
  multi-user communications,'' \emph{arXiv preprint arXiv:2306.16861}, 2023.

\bibitem{wei2022codebook}
X.~Wei, L.~Dai, Y.~Zhao, G.~Yu, and X.~Duan, ``Codebook design and beam
  training for extremely large-scale {RIS}: Far-field or near-field?''
  \emph{China Commun.}, vol.~19, no.~6, pp. 193--204, Jun. 2022.

\bibitem{shi2023chirp}
X.~Shi, J.~Wang, Z.~Sun, and J.~Song, ``Chirp-based hierarchical beam training
  for extremely large-scale massive {MIMO},'' \emph{arXiv preprint
  arXiv:2301.11570}, 2023.

\bibitem{zhang2022fast}
Y.~Zhang, X.~Wu, and C.~You, ``Fast near-field beam training for extremely
  large-scale array,'' \emph{IEEE Wireless Commun. Lett.}, vol.~11, no.~12, pp.
  2625--2629, Dec. 2022.

\bibitem{wu2023near}
X.~Wu, C.~You, J.~Li, and Y.~Zhang, ``Near-field beam training: Joint angle and
  range estimation with {DFT} codebook,'' \emph{IEEE Trans. Wireless Commun.},
  Early Access, 2024.

\bibitem{hu2023design}
S.~Hu, H.~Wang, and M.~C. Ilter, ``Design of near-field beamforming for large
  intelligent surfaces,'' \emph{IEEE Trans. Wireless Commun.}, vol.~23, no.~1,
  pp. 762--774, Jan. 2024.

\bibitem{wang2023near}
T.~Wang, J.~Lv, H.~Tong, C.~You, and C.~Yin, ``Near-field beam training for
  extremely large-scale {IRS},'' \emph{arXiv preprint arXiv:2303.06962}, 2023.

\bibitem{lu2023hierarchical}
Y.~Lu, Z.~Zhang, and L.~Dai, ``Hierarchical beam training for extremely
  large-scale {MIMO}: From far-field to near-field,'' \emph{IEEE Trans.
  Commun.}, Early Access, 2023.

\bibitem{chen2023hierarchical}
J.~Chen, F.~Gao, M.~Jian, and W.~Yuan, ``Hierarchical codebook design for
  near-field mmwave {MIMO} communications systems,'' \emph{IEEE Wireless
  Commun. Lett.}, vol.~12, no.~11, pp. 1926--1930, Nov. 2023.

\bibitem{wu2023twonfh}
C.~Wu, C.~You, Y.~Liu, L.~Chen, and S.~Shi, ``Two-stage hierarchical beam
  training for near-field communications,'' \emph{IEEE Trans. Veh. Technol.},
  vol.~73, no.~2, pp. 2032--2044, Feb. 2024.

\bibitem{liu2022deep}
W.~Liu, H.~Ren, C.~Pan, and J.~Wang, ``Deep learning based beam training for
  extremely large-scale massive {MIMO} in near-field domain,'' \emph{IEEE
  Commun. Lett.}, vol.~27, no.~1, pp. 170--174, Jan. 2022.

\bibitem{jiang2023near}
G.~Jiang and C.~Qi, ``Near-field beam training based on deep learning for
  extremely large-scale {MIMO},'' \emph{IEEE Commun. Lett.}, vol.~27, no.~8,
  pp. 2063--2067, Aug. 2023.

\bibitem{he2015suboptimal}
T.~He and Z.~Xiao, ``Suboptimal beam search algorithm and codebook design for
  millimeter-wave communications,'' \emph{Mobile Netw. Appl.}, vol.~20, no.~1,
  pp. 86--97, Feb. 2015.

\bibitem{zhou2024near2}
C.~Zhou, C.~You, Z.~Huang, S.~Shi, Y.~Gong, C.-B. Chae, and K.~Huang,
  ``Multi-beam training for near-field communications in high-frequency
  bands,'' \emph{arXiv preprint arXiv:2406.14931}, 2024.

\bibitem{zhou2024near}
C.~Zhou, C.~Wu, C.~You, and S.~Shi, ``Near-field beam training with sparse
  {DFT} codebook,'' \emph{arXiv preprint arXiv:2406.04262}, 2024.

\bibitem{guo2023compressed}
X.~Guo, Y.~Chen, and Y.~Wang, ``Compressed channel estimation for near-field
  {XL-MIMO} using triple parametric decomposition,'' \emph{IEEE Trans. Veh.
  Technol.}, vol.~72, no.~11, pp. 15\,040--15\,045, Nov. 2023.

\bibitem{lei2023channel}
H.~Lei, J.~Zhang, H.~Xiao, X.~Zhang, B.~Ai, and D.~W.~K. Ng, ``Channel
  estimation for {XL-MIMO} systems with polar-domain multi-scale residual dense
  network,'' \emph{IEEE Trans. Veh. Technol.}, vol.~73, no.~1, pp. 1479--1484,
  Jan. 2024.

\bibitem{cui2023near}
M.~Cui and L.~Dai, ``Near-field wideband channel estimation for extremely
  large-scale {MIMO},'' \emph{Sci. China. Inf. Sci.}, vol.~66, no.~7, p.
  172303, Jul. 2023.

\bibitem{elbir2023near}
A.~M. Elbir, W.~Shi, A.~K. Papazafeiropoulos, P.~Kourtessis, and
  S.~Chatzinotas, ``Near-field terahertz communications: Model-based and
  model-free channel estimation,'' \emph{IEEE Access}, vol.~11, pp.
  36\,409--36\,420, May. 2023.

\bibitem{hu2022hybrid}
Z.~Hu, C.~Chen, Y.~Jin, L.~Zhou, and Q.~Wei, ``Hybrid-field channel estimation
  for extremely large-scale massive {MIMO} system,'' \emph{IEEE Commun. Lett.},
  vol.~27, no.~1, pp. 303--307, Jan. 2022.

\bibitem{yang2023practical}
W.~Yang, M.~Li, and Q.~Liu, ``A practical channel estimation strategy for
  {XL-MIMO} communication systems,'' \emph{IEEE Commun. Lett.}, vol.~27, no.~6,
  pp. 1580--1583, Jun. 2023.

\bibitem{tarboush2024cross}
S.~Tarboush, A.~Ali, and T.~Y. Al-Naffouri, ``Cross-field channel estimation
  for ultra massive-{MIMO} {THz} systems,'' \emph{arXiv preprint
  arXiv:2305.13757}, 2023.

\bibitem{chen2021hybrid}
Y.~Chen, L.~Yan, and C.~Han, ``Hybrid spherical- and planar-wave modeling and
  {DCNN}-powered estimation of terahertz ultra-massive {MIMO} channels,''
  \emph{IEEE Trans. Commun.}, vol.~69, no.~10, pp. 7063--7076, Oct. 2021.

\bibitem{yu2023adaptive}
W.~Yu, Y.~Shen, H.~He, X.~Yu, S.~Song, J.~Zhang, and K.~B. Letaief, ``An
  adaptive and robust deep learning framework for {THz} ultra-massive {MIMO}
  channel estimation,'' \emph{IEEE J. Sel. Topics Signal Process.}, vol.~17,
  no.~4, pp. 761--776, Jul. 2023.

\bibitem{tian2023low}
J.~Tian, Y.~Han, S.~Jin, and M.~Matthaiou, ``Low-overhead localization and {VR}
  identification for subarray-based {ELAA} systems,'' \emph{IEEE Wireless
  Commun. Lett.}, vol.~12, no.~5, pp. 784--788, May. 2023.

\bibitem{liu2023location}
D.~Liu, J.~Wang, Y.~Li, Y.~Han, R.~Ding, J.~Zhang, S.~Jin, and T.~Q.~S. Quek,
  ``Location-based visible region recognition in extra-large massive {MIMO}
  systems,'' \emph{IEEE Trans. Veh. Technol.}, vol.~72, no.~6, pp. 8186--8191,
  Jun. 2023.

\bibitem{iimori2022joint}
H.~Iimori, T.~Takahashi, K.~Ishibashi, G.~T.~F. de~Abreu, D.~Gonz{\'a}lez, and
  O.~Gonsa, ``Joint activity and channel estimation for extra-large {MIMO}
  systems,'' \emph{IEEE Trans. Wireless Commun.}, vol.~21, no.~9, pp.
  7253--7270, Sept. 2022.

\bibitem{iimori2022grant}
H.~Iimori, T.~Takahashi, H.~S. Rou, K.~Ishibashi, G.~T.~F. de~Abreu,
  D.~Gonz{\'a}lez, and O.~Gonsa, ``Grant-free access for extra-large {MIMO}
  systems subject to spatial non-stationarity,'' in \emph{Proc. IEEE Int. Conf.
  Commun. (ICC)}, Seoul, Korea, Republic of, May 2022, pp. 1758--1762.

\bibitem{chen2023beam}
K.~Chen, C.~Qi, C.-X. Wang, and G.~Y. Li, ``Beam training and tracking for
  extremely large-scale {MIMO} communications,'' \emph{{IEEE} Trans. Wireless
  Commun.}, Early Access, 2023.

\bibitem{wu2023intelligent}
Q.~Wu, B.~Zheng, C.~You, L.~Zhu, K.~Shen, X.~Shao, W.~Mei, B.~Di, H.~Zhang,
  E.~Basar \emph{et~al.}, ``Intelligent surfaces empowered wireless network:
  Recent advances and the road to {6G},'' \emph{arXiv preprint
  arXiv:2312.16918}, 2023.

\bibitem{10380596}
X.~Mu, J.~Xu, Y.~Liu, and L.~Hanzo, ``Reconfigurable intelligent surface-aided
  near-field communications for {6G}: Opportunities and challenges,''
  \emph{IEEE Veh. Technol. Mag.}, vol.~19, no.~1, pp. 65--74, 2024.

\bibitem{you2021enabling}
C.~You, Z.~Kang, Y.~Zeng, and R.~Zhang, ``Enabling smart reflection in
  integrated air-ground wireless network: {IRS} meets {UAV},'' \emph{IEEE
  Wireless Commun.}, vol.~28, no.~6, pp. 138--144, Jun. 2021.

\bibitem{kang2021irs}
Z.~Kang, C.~You, and R.~Zhang, ``{IRS}-aided wireless relaying: Deployment
  strategy and capacity scaling,'' \emph{IEEE Wireless Commun. Lett.}, vol.~11,
  no.~2, pp. 215--219, Feb. 2021.

\bibitem{zhou2020delay}
F.~Zhou, C.~You, and R.~Zhang, ``Delay-optimal scheduling for {IRS}-aided
  mobile edge computing,'' \emph{IEEE Wireless Commun. Lett.}, vol.~10, no.~4,
  pp. 740--744, Apr. 2020.

\bibitem{mao2023roar}
J.~Mao and A.~Yener, ``{ROAR-Fed}: {RIS}-assisted over-the-air adaptive
  resource allocation for federated learning,'' in \emph{Proc. IEEE Int. Conf.
  Commun.}, May 2023, pp. 4341--4346.

\bibitem{wu2019intelligent}
Q.~Wu and R.~Zhang, ``Intelligent reflecting surface enhanced wireless network
  via joint active and passive beamforming,'' \emph{IEEE Trans. Wireless
  Commun.}, vol.~18, no.~11, pp. 5394--5409, Nov. 2019.

\bibitem{huang2019reconfigurable}
C.~Huang, A.~Zappone, G.~C. Alexandropoulos, M.~Debbah, and C.~Yuen,
  ``Reconfigurable intelligent surfaces for energy efficiency in wireless
  communication,'' \emph{IEEE Trans. Wireless Commun.}, vol.~18, no.~8, pp.
  4157--4170, Aug. 2019.

\bibitem{nayeri2018reflectarray}
P.~Nayeri, F.~Yang, and A.~Z. Elsherbeni, \emph{Reflectarray Antennas: Theory,
  Designs, and Applications}.\hskip 1em plus 0.5em minus 0.4em\relax John Wiley
  \& Sons, 2018.

\bibitem{9020088}
L.~Dai, B.~Wang, M.~Wang, X.~Yang, J.~Tan, S.~Bi, S.~Xu, F.~Yang, Z.~Chen,
  M.~D. Renzo, C.-B. Chae, and L.~Hanzo, ``Reconfigurable intelligent
  surface-based wireless communications: Antenna design, prototyping, and
  experimental results,'' \emph{IEEE Access}, vol.~8, pp. 45\,913--45\,923,
  Mar. 2020.

\bibitem{9551980}
X.~Pei, H.~Yin, L.~Tan, L.~Cao, Z.~Li, K.~Wang, K.~Zhang, and E.~Björnson,
  ``Ris-aided wireless communications: Prototyping, adaptive beamforming, and
  indoor/outdoor field trials,'' \emph{IEEE Trans. Commun.}, vol.~69, no.~12,
  pp. 8627--8640, Dec. 2021.

\bibitem{9749219}
R.~Fara, P.~Ratajczak, D.-T. Phan-Huy, A.~Ourir, M.~Di~Renzo, and J.~de~Rosny,
  ``A prototype of reconfigurable intelligent surface with continuous control
  of the reflection phase,'' \emph{IEEE Wireless Commun.}, vol.~29, no.~1, pp.
  70--77, Feb. 2022.

\bibitem{10373827}
H.~Yang, S.~Kim, H.~Kim, S.~Bang, Y.~Kim, S.~Kim, K.~Park, D.~Kwon, and J.~Oh,
  ``Beyond limitations of {5G} with {RIS}: Field trial in a commercial network,
  recent advances, and future directions,'' \emph{IEEE Commun. Mag.}, pp. 1--7,
  Early Access, 2023.

\bibitem{9999288}
J.~Sang, Y.~Yuan, W.~Tang, Y.~Li, X.~Li, S.~Jin, Q.~Cheng, and T.~J. Cui,
  ``Coverage enhancement by deploying {RIS} in {5G} commercial mobile networks:
  Field trials,'' \emph{IEEE Wireless Commun.}, vol.~31, no.~1, pp. 172--180,
  Feb. 2024.

\bibitem{zheng2022survey}
B.~Zheng, C.~You, W.~Mei, and R.~Zhang, ``A survey on channel estimation and
  practical passive beamforming design for intelligent reflecting surface aided
  wireless communications,'' \emph{IEEE Commun. Surveys Tuts.}, vol.~24, no.~2,
  pp. 1035--1071, Feb. 2022.

\bibitem{yang2020intelligent}
Y.~Yang, B.~Zheng, S.~Zhang, and R.~Zhang, ``Intelligent reflecting surface
  meets {OFDM}: Protocol design and rate maximization,'' \emph{IEEE Trans.
  Commun.}, vol.~68, no.~7, pp. 4522--4535, Jul. 2020.

\bibitem{you2020channel}
C.~You, B.~Zheng, and R.~Zhang, ``Channel estimation and passive beamforming
  for intelligent reflecting surface: Discrete phase shift and progressive
  refinement,'' \emph{IEEE J. Sel. Areas Commun.}, vol.~38, no.~11, pp.
  2604--2620, Nov. 2020.

\bibitem{wang2020channel}
Z.~Wang, L.~Liu, and S.~Cui, ``Channel estimation for intelligent reflecting
  surface assisted multiuser communications: Framework, algorithms, and
  analysis,'' \emph{IEEE Trans. Wireless Commun.}, vol.~19, no.~10, pp.
  6607--6620, Oct. 2020.

\bibitem{zheng2020intelligent}
B.~Zheng, C.~You, and R.~Zhang, ``Intelligent reflecting surface assisted
  multi-user {OFDMA}: Channel estimation and training design,'' \emph{IEEE
  Trans. Wireless Commun.}, vol.~19, no.~12, pp. 8315--8329, Dec. 2020.

\bibitem{10028753}
Z.~Huang, B.~Zheng, and R.~Zhang, ``Roadside {IRS}-aided vehicular
  communication: Efficient channel estimation and low-complexity beamforming
  design,'' \emph{IEEE Trans. Wireless Commun.}, vol.~22, no.~9, pp.
  5976--5989, Sept. 2023.

\bibitem{zhou2022channel}
G.~Zhou, C.~Pan, H.~Ren, P.~Popovski, and A.~L. Swindlehurst, ``Channel
  estimation for {RIS}-aided multiuser millimeter-wave systems,'' \emph{IEEE
  Trans. Signal Process.}, vol.~70, pp. 1478--1492, Mar. 2022.

\bibitem{wang2023hierarchical}
J.~Wang, W.~Tang, S.~Jin, C.-K. Wen, X.~Li, and X.~Hou, ``Hierarchical
  codebook-based beam training for {RIS}-assisted {mmWave} communication
  systems,'' \emph{IEEE Trans. Commun.}, vol.~72, no.~6, pp. 3650--3662, Jun.
  2023.

\bibitem{alexandropoulos2022near}
G.~C. Alexandropoulos, V.~Jamali, R.~Schober, and H.~V. Poor, ``Near-field
  hierarchical beam management for {RIS}-enabled millimeter wave multi-antenna
  systems,'' in \emph{Proc. IEEE Sensor Array Multichannel Signal Process.
  Workshop (SAM)}, Jun. 2022, pp. 460--464.

\bibitem{you2020fast}
C.~You, B.~Zheng, and R.~Zhang, ``Fast beam training for {IRS}-assisted
  multiuser communications,'' \emph{IEEE Wireless Commun. Lett.}, vol.~9,
  no.~11, pp. 1845--1849, Sept. 2020.

\bibitem{10100676}
M.~Ouyang, F.~Gao, Y.~Wang, S.~Zhang, P.~Li, and J.~Ren, ``Computer
  vision-aided reconfigurable intelligent surface-based beam tracking:
  Prototyping and experimental results,'' \emph{IEEE Trans. Wireless Commun.},
  vol.~22, no.~12, pp. 8681--8693, Dec. 2023.

\bibitem{ren2022configuring}
S.~Ren, K.~Shen, Y.~Zhang, X.~Li, X.~Chen, and Z.-Q. Luo, ``Configuring
  intelligent reflecting surface with performance guarantees: Blind
  beamforming,'' \emph{IEEE Trans. Wireless Commun.}, vol.~22, no.~5, pp.
  3355--3370, May 2023.

\bibitem{yan2023channel}
G.~Yan, L.~Zhu, and R.~Zhang, ``Channel autocorrelation estimation for
  {IRS}-aided wireless communications based on power measurements,''
  \emph{arXiv preprint arXiv:2310.11038}, 2023.

\bibitem{sun2023user}
H.~Sun, W.~Mei, L.~Zhu, and R.~Zhang, ``User power measurement based {IRS}
  channel estimation via single-layer neural network,'' \emph{arXiv preprint
  arXiv:2309.08275}, 2023.

\bibitem{wu2019beamforming}
Q.~Wu and R.~Zhang, ``Beamforming optimization for wireless network aided by
  intelligent reflecting surface with discrete phase shifts,'' \emph{IEEE
  Trans. Commun.}, vol.~68, no.~3, pp. 1838--1851, Mar. 2019.

\bibitem{di2020hybrid}
B.~Di, H.~Zhang, L.~Song, Y.~Li, Z.~Han, and H.~V. Poor, ``Hybrid beamforming
  for reconfigurable intelligent surface based multi-user communications:
  Achievable rates with limited discrete phase shifts,'' \emph{IEEE J. Sel.
  Areas Commun.}, vol.~38, no.~8, pp. 1809--1822, Aug. 2020.

\bibitem{xiu2021secrecy}
Y.~Xiu, J.~Zhao, W.~Sun, and Z.~Zhang, ``Secrecy rate maximization for
  reconfigurable intelligent surface aided millimeter wave system with
  low-resolution {DACs},'' \emph{IEEE Commun. Lett.}, vol.~25, no.~7, pp.
  2166--2170, Jul. 2021.

\bibitem{zhao2021exploiting}
M.-M. Zhao, Q.~Wu, M.-J. Zhao, and R.~Zhang, ``Exploiting amplitude control in
  intelligent reflecting surface aided wireless communication with imperfect
  {CSI},'' \emph{IEEE Trans. Commun.}, vol.~69, no.~6, pp. 4216--4231, Jun.
  2021.

\bibitem{abeywickrama2020intelligent}
S.~Abeywickrama, R.~Zhang, Q.~Wu, and C.~Yuen, ``Intelligent reflecting
  surface: Practical phase shift model and beamforming optimization,''
  \emph{IEEE Trans. Commun.}, vol.~68, no.~9, pp. 5849--5863, Sept. 2020.

\bibitem{zeng2023resource}
M.~Zeng, W.~Hao, Z.~Peng, Z.~Chu, X.~Li, C.~You, and C.~Pan, ``Resource
  allocation for {RIS}-empowered wireless communications: Low-complexity and
  robust designs,'' \emph{arXiv preprint arXiv:2311.03282}, 2023.

\bibitem{li2021intelligent}
H.~Li, W.~Cai, Y.~Liu, M.~Li, Q.~Liu, and Q.~Wu, ``Intelligent reflecting
  surface enhanced wideband {MIMO-OFDM} communications: From practical model to
  reflection optimization,'' \emph{IEEE Trans. Commun.}, vol.~69, no.~7, pp.
  4807--4820, Jul. 2021.

\bibitem{khel2021effects}
A.~M.~T. Khel and K.~A. Hamdi, ``Effects of hardware impairments on
  {IRS}-enabled {MISO} wireless communication systems,'' \emph{IEEE Commun.
  Lett.}, vol.~26, no.~2, pp. 259--263, Feb. 2021.

\bibitem{badiu2019communication}
M.-A. Badiu and J.~P. Coon, ``Communication through a large reflecting surface
  with phase errors,'' \emph{IEEE Wireless Commun. Lett.}, vol.~9, no.~2, pp.
  184--188, Feb. 2019.

\bibitem{zhou2020spectral}
S.~Zhou, W.~Xu, K.~Wang, M.~Di~Renzo, and M.-S. Alouini, ``Spectral and energy
  efficiency of {IRS}-assisted {MISO} communication with hardware
  impairments,'' \emph{IEEE Wireless Commun. Lett.}, vol.~9, no.~9, pp.
  1366--1369, Sept. 2020.

\bibitem{9293148}
S.~Hong, C.~Pan, H.~Ren, K.~Wang, K.~K. Chai, and A.~Nallanathan, ``Robust
  transmission design for intelligent reflecting surface-aided secure
  communication systems with imperfect cascaded {CSI},'' \emph{IEEE Trans.
  Wireless Commun.}, vol.~20, no.~4, pp. 2487--2501, Apr. 2021.

\bibitem{9316283}
Y.~Omid, S.~M. Shahabi, C.~Pan, Y.~Deng, and A.~Nallanathan, ``Low-complexity
  robust beamforming design for {IRS}-aided {MISO} systems with imperfect
  channels,'' \emph{IEEE Commun. Lett.}, vol.~25, no.~5, pp. 1697--1701, May
  2021.

\bibitem{9973349}
K.~Zhi, C.~Pan, H.~Ren, K.~Wang, M.~Elkashlan, M.~D. Renzo, R.~Schober, H.~V.
  Poor, J.~Wang, and L.~Hanzo, ``Two-timescale design for reconfigurable
  intelligent surface-aided massive {MIMO} systems with imperfect {CSI},''
  \emph{IEEE Trans. Inf. Theory}, vol.~69, no.~5, pp. 3001--3033, May 2023.

\bibitem{10504275}
D.~Gürgünoğlu, E.~Björnson, and G.~Fodor, ``Combating inter-operator pilot
  contamination in reconfigurable intelligent surfaces assisted multi-operator
  networks,'' \emph{IEEE Trans. Commun.}, Early Access, 2024.

\bibitem{9206044}
W.~Tang, M.~Z. Chen, X.~Chen, J.~Y. Dai, Y.~Han, M.~Di~Renzo, Y.~Zeng, S.~Jin,
  Q.~Cheng, and T.~J. Cui, ``Wireless communications with reconfigurable
  intelligent surface: Path loss modeling and experimental measurement,''
  \emph{IEEE Trans. Wireless Commun.}, vol.~20, no.~1, pp. 421--439, 2021.

\bibitem{9433568}
F.~H. Danufane, M.~D. Renzo, J.~de~Rosny, and S.~Tretyakov, ``On the path-loss
  of reconfigurable intelligent surfaces: An approach based on {Green's}
  theorem applied to vector fields,'' \emph{IEEE Trans. Commun.}, vol.~69,
  no.~8, pp. 5573--5592, Aug. 2021.

\bibitem{9119122}
M.~Di~Renzo, K.~Ntontin, J.~Song, F.~H. Danufane, X.~Qian, F.~Lazarakis,
  J.~De~Rosny, D.-T. Phan-Huy, O.~Simeone, R.~Zhang, M.~Debbah, G.~Lerosey,
  M.~Fink, S.~Tretyakov, and S.~Shamai, ``Reconfigurable intelligent surfaces
  vs. relaying: Differences, similarities, and performance comparison,''
  \emph{IEEE Open J. Commun. Soc.}, vol.~1, pp. 798--807, Jun. 2020.

\bibitem{long2021active}
R.~Long, Y.-C. Liang, Y.~Pei, and E.~G. Larsson, ``Active reconfigurable
  intelligent surface-aided wireless communications,'' \emph{IEEE Trans.
  Wireless Commun.}, vol.~20, no.~8, pp. 4962--4975, Aug. 2021.

\bibitem{you2021wireless}
C.~You and R.~Zhang, ``Wireless communication aided by intelligent reflecting
  surface: Active or passive?'' \emph{IEEE Wireless Commun. Lett.}, vol.~10,
  no.~12, pp. 2659--2663, Dec. 2021.

\bibitem{zhang2022active}
Z.~Zhang, L.~Dai, X.~Chen, C.~Liu, F.~Yang, R.~Schober, and H.~V. Poor,
  ``Active {RIS} vs. passive {RIS}: Which will prevail in {6G}?'' \emph{IEEE
  Trans. Commun.}, vol.~71, no.~3, pp. 1707--1725, Mar. 2022.

\bibitem{kang2023activemag}
Z.~Kang, C.~You, and R.~Zhang, ``Active-{IRS}-aided wireless communication:
  Fundamentals, designs and open issues,'' \emph{arXiv preprint
  arXiv:2301.04311}, 2023.

\bibitem{kishor2011amplifying}
K.~K. Kishor and S.~V. Hum, ``An amplifying reconfigurable reflectarray
  antenna,'' \emph{IEEE Trans. Antennas Propag.}, vol.~60, no.~1, pp. 197--205,
  Jan. 2011.

\bibitem{amato2018tunneling}
F.~Amato, C.~W. Peterson, B.~P. Degnan, and G.~D. Durgin, ``Tunneling {RFID}
  tags for long-range and low-power microwave applications,'' \emph{IEEE J.
  Radio Freq. Identif.}, vol.~2, no.~2, pp. 93--103, Feb. 2018.

\bibitem{landsberg2017low}
N.~Landsberg and E.~Socher, ``A low-power 28-nm {CMOS} {FD-SOI} reflection
  amplifier for an active {F}-band reflectarray,'' \emph{IEEE Trans. Microw.
  Theory Tech.}, vol.~65, no.~10, pp. 3910--3921, Oct. 2017.

\bibitem{10001687}
Z.~Zhang, L.~Dai, X.~Chen, C.~Liu, F.~Yang, R.~Schober, and H.~V. Poor,
  ``Active {RISs}: Signal modeling, asymptotic analysis, and beamforming
  design,'' in \emph{Proc. IEEE Global Commun. Conf. (GLOBECOM)}, 2022, pp.
  1618--1624.

\bibitem{li2022active}
Y.~Li, C.~You, and Y.~J. Chun, ``Active-{IRS} aided wireless network: System
  modeling and performance analysis,'' \emph{IEEE Commun. Lett.}, vol.~27,
  no.~2, pp. 487--491, Feb. 2022.

\bibitem{kang2023double}
Z.~Kang, C.~You, and R.~Zhang, ``Double-active-{IRS} aided wireless
  communication: Deployment optimization and capacity scaling,'' \emph{IEEE
  Wireless Commun. Lett.}, vol.~12, no.~11, pp. 1821--1825, Nov. 2023.

\bibitem{li2023double}
X.~Li, C.~You, Z.~Kang, Y.~Zhang, and B.~Zheng, ``Double-active-{IRS} aided
  wireless communication with total amplification power constraint,''
  \emph{IEEE Communi. Lett.}, vol.~27, no.~10, pp. 2817--2821, Oct. 2023.

\bibitem{nguyen2022hybrid}
N.~T. Nguyen, Q.-D. Vu, K.~Lee, and M.~Juntti, ``Hybrid relay-reflecting
  intelligent surface-assisted wireless communications,'' \emph{IEEE Trans.
  Veh. Technol.}, vol.~71, no.~6, pp. 6228--6244, Jun. 2022.

\bibitem{liu2021active}
K.~Liu, Z.~Zhang, L.~Dai, S.~Xu, and F.~Yang, ``Active reconfigurable
  intelligent surface: Fully-connected or sub-connected?'' \emph{IEEE Commun.
  Lett.}, vol.~26, no.~1, pp. 167--171, Jan. 2021.

\bibitem{kang2023active}
Z.~Kang, C.~You, and R.~Zhang, ``Active-passive {IRS} aidedwireless
  communication: New hybrid architecture and elements allocation
  optimization,'' \emph{IEEE Trans. Wireless Commun.}, Early Access, 2023.
  doi:10.1109/TWC.2023.3308373.

\bibitem{lioliou2010self}
P.~Lioliou, M.~Viberg, M.~Coldrey, and F.~Athley, ``Self-interference
  suppression in full-duplex {MIMO} relays,'' in \emph{Proc. Asilomar Conf.
  Signals Syst. Comp.}, Nov. 2010, pp. 658--662.

\bibitem{xu2021star}
J.~Xu, Y.~Liu, X.~Mu, and O.~A. Dobre, ``{STAR-RISs}: Simultaneous transmitting
  and reflecting reconfigurable intelligent surfaces,'' \emph{IEEE Commun.
  Lett.}, vol.~25, no.~9, pp. 3134--3138, Sept. 2021.

\bibitem{10133841}
M.~Ahmed, A.~Wahid, S.~S. Laique, W.~U. Khan, A.~Ihsan, F.~Xu, S.~Chatzinotas,
  and Z.~Han, ``A survey on {STAR-RIS}: Use cases, recent advances, and future
  research challenges,'' \emph{IEEE Internet Things J.}, vol.~10, no.~16, pp.
  14\,689--14\,711, Aug. 2023.

\bibitem{zhang2020beyond}
S.~Zhang, H.~Zhang, B.~Di, Y.~Tan, Z.~Han, and L.~Song, ``Beyond intelligent
  reflecting surfaces: Reflective-transmissive metasurface aided communications
  for full-dimensional coverage extension,'' \emph{IEEE Trans. Veh. Technol.},
  vol.~69, no.~11, pp. 13\,905--13\,909, Nov. 2020.

\bibitem{mu2021simultaneously}
X.~Mu, Y.~Liu, L.~Guo, J.~Lin, and R.~Schober, ``Simultaneously transmitting
  and reflecting ({STAR}) {RIS} aided wireless communications,'' \emph{IEEE
  Trans. Wireless Commun.}, vol.~21, no.~5, pp. 3083--3098, May 2021.

\bibitem{9895224}
S.~Zeng, H.~Zhang, B.~Di, Y.~Liu, M.~D. Renzo, Z.~Han, H.~V. Poor, and L.~Song,
  ``Intelligent omni-surfaces: Reflection-refraction circuit model,
  full-dimensional beamforming, and system implementation,'' \emph{IEEE Trans.
  Commun.}, vol.~70, no.~11, pp. 7711--7727, Nov. 2022.

\bibitem{wu2021coverage}
C.~Wu, Y.~Liu, X.~Mu, X.~Gu, and O.~A. Dobre, ``Coverage characterization of
  {STAR-RIS} networks: {NOMA} and {OMA},'' \emph{IEEE Commun. Lett.}, vol.~25,
  no.~9, pp. 3036--3040, Sept. 2021.

\bibitem{wu2021channel}
C.~Wu, C.~You, Y.~Liu, X.~Gu, and Y.~Cai, ``Channel estimation for
  {STAR-RIS-aided} wireless communication,'' \emph{IEEE Commun. Lett.},
  vol.~26, no.~3, pp. 652--656, Mar. 2021.

\bibitem{wu2023two}
C.~Wu, C.~You, Y.~Liu, S.~Han, and M.~D. Renzo, ``Two-timescale design for
  {STAR-RIS-Aided} {NOMA} systems,'' \emph{IEEE Trans. Commun.}, vol.~72,
  no.~1, pp. 585--600, Nov. 2024.

\bibitem{liu2022simultaneously}
Y.~Liu, X.~Mu, R.~Schober, and H.~V. Poor, ``Simultaneously transmitting and
  reflecting {(STAR)-RISs}: A coupled phase-shift model,'' in \emph{Proc. IEEE
  Int. Conf. Commun. (ICC)}, May 2022, pp. 2840--2845.

\bibitem{10133914}
Z.~Wang, X.~Mu, J.~Xu, and Y.~Liu, ``Simultaneously transmitting and reflecting
  surface ({STARS}) for terahertz communications,'' \emph{IEEE J. Sel. Topics
  Signal Process.}, vol.~17, no.~4, pp. 861--877, Jul. 2023.

\bibitem{9774942}
J.~Xu, Y.~Liu, X.~Mu, R.~Schober, and H.~V. Poor, ``{STAR-RISs}: A correlated
  {T\&R} phase-shift model and practical phase-shift configuration
  strategies,'' \emph{IEEE J. Sel. Topics Signal Process.}, vol.~16, no.~5, pp.
  1097--1111, May 2022.

\bibitem{9751144}
H.~Niu and X.~Liang, ``Weighted sum-rate maximization for {STAR-RISs}-aided
  networks with coupled phase-shifters,'' \emph{IEEE Syst. J.}, vol.~17, no.~1,
  pp. 1083--1086, Jan. 2023.

\bibitem{9935266}
Z.~Wang, X.~Mu, Y.~Liu, and R.~Schober, ``Coupled phase-shift {STAR-RISs}: A
  general optimization framework,'' \emph{IEEE Wireless Commun. Lett.},
  vol.~12, no.~2, pp. 207--211, Feb. 2023.

\bibitem{9786807}
M.~Aldababsa, A.~Khaleel, and E.~Basar, ``{STAR-RIS-NOMA} networks: An error
  performance perspective,'' \emph{IEEE Commun. Lett.}, vol.~26, no.~8, pp.
  1784--1788, Aug. 2022.

\bibitem{9740451}
C.~Wu, X.~Mu, Y.~Liu, X.~Gu, and X.~Wang, ``Resource allocation in
  {STAR-RIS-aided} networks: {OMA} and {NOMA},'' \emph{IEEE Trans. Wireless
  Commun.}, vol.~21, no.~9, pp. 7653--7667, Sept. 2022.

\bibitem{9863732}
J.~Zuo, Y.~Liu, Z.~Ding, L.~Song, and H.~V. Poor, ``Joint design for
  simultaneously transmitting and reflecting ({STAR}) {RIS} assisted {NOMA}
  systems,'' \emph{IEEE Trans. Wireless Commun.}, vol.~22, no.~1, pp. 611--626,
  Jan. 2023.

\bibitem{mei2022intelligent}
W.~Mei, B.~Zheng, C.~You, and R.~Zhang, ``Intelligent reflecting surface-aided
  wireless networks: From single-reflection to multireflection design and
  optimization,'' \emph{Proc. IEEE}, vol. 110, no.~9, pp. 1380--1400, May 2022.

\bibitem{you2020wireless}
C.~You, B.~Zheng, and R.~Zhang, ``Wireless communication via double {IRS}:
  Channel estimation and passive beamforming designs,'' \emph{IEEE Wireless
  Commun. Lett.}, vol.~10, no.~2, pp. 431--435, Feb. 2020.

\bibitem{zheng2021double}
B.~Zheng, C.~You, and R.~Zhang, ``Double-{IRS} assisted multi-user {MIMO}:
  Cooperative passive beamforming design,'' \emph{IEEE Trans. Wireless
  Commun.}, vol.~20, no.~7, pp. 4513--4526, Jul. 2021.

\bibitem{zhang2023multi}
Y.~Zhang, C.~You, and B.~Zheng, ``Multi-active multi-passive ({MAMP})-{IRS}
  aided wireless communication: A multi-hop beam routing design,'' \emph{IEEE
  J. Selet. Areas Commun.}, vol.~41, no.~8, pp. 2497--2513, Aug. 2023.

\bibitem{9661068}
Z.~Huang, B.~Zheng, and R.~Zhang, ``Transforming fading channel from fast to
  slow: Intelligent refracting surface aided high-mobility communication,''
  \emph{IEEE Trans. Wireless Commun.}, vol.~21, no.~7, pp. 4989--5003, Jul.
  2022.

\bibitem{you2022deploy}
C.~You, B.~Zheng, W.~Mei, and R.~Zhang, ``How to deploy intelligent reflecting
  surfaces in wireless network: {BS}-side, user-side, or both sides?'' \emph{J.
  Commun. Info. Netw.}, vol.~7, no.~1, pp. 1--10, Jan. 2022.

\bibitem{10316541}
Z.~Huang, L.~Zhu, and R.~Zhang, ``Intelligent surfaces aided high-mobility
  communications: Opportunities and design issues,'' \emph{IEEE Commun. Mag.},
  Early Access, 2023. doi: 10.1109/MCOM.003.2300213.

\bibitem{9913356}
H.~Li, S.~Shen, and B.~Clerckx, ``Beyond diagonal reconfigurable intelligent
  surfaces: From transmitting and reflecting modes to single-, group-, and
  fully-connected architectures,'' \emph{IEEE Trans. Wireless Commun.},
  vol.~22, no.~4, pp. 2311--2324, Apr. 2023.

\bibitem{10316535}
H.~Li, S.~Shen, M.~Nerini, and B.~Clerckx, ``Reconfigurable intelligent
  surfaces 2.0: Beyond diagonal phase shift matrices,'' \emph{IEEE Commun.
  Mag.}, vol.~62, no.~3, pp. 102--108, Mar. 2024.

\bibitem{wong2020fluid}
K.-K. Wong, A.~Shojaeifard, K.-F. Tong, and Y.~Zhang, ``Fluid antenna
  systems,'' \emph{IEEE Trans. Wireless Commun.}, vol.~20, no.~3, pp.
  1950--1962, Mar. 2020.

\bibitem{borda2019low}
C.~Borda-Fortuny, L.~Cai, K.~F. Tong, and K.-K. Wong, ``Low-cost {3D}-printed
  coupling-fed frequency agile fluidic monopole antenna system,'' \emph{IEEE
  Access}, vol.~7, pp. 95\,058--95\,064, Jul. 2019.

\bibitem{singh2019multistate}
A.~Singh, I.~Goode, and C.~E. Saavedra, ``A multistate frequency reconfigurable
  monopole antenna using fluidic channels,'' \emph{IEEE Antennas Wireless
  Propag. Lett.}, vol.~18, no.~5, pp. 856--860, May 2019.

\bibitem{shen2020beam}
Y.~Shen, K.-F. Tong, and K.-K. Wong, ``Beam-steering surface wave fluid
  antennas for {MIMO} applications,'' in \emph{Proc. IEEE Asia-Pacific Micro.
  Conf. (APMC)}, Dec. 2020, pp. 634--636.

\bibitem{shen2021reconfigurable}
------, ``Reconfigurable surface wave fluid antenna for spatial {MIMO}
  applications,'' in \emph{Proc. IEEE-APS Topical Conf. Antennas Propag.
  Wireless Commun. (APWC)}, Aug. 2021, pp. 150--152.

\bibitem{castanheira2010distributed}
D.~Castanheira and A.~Gameiro, ``Distributed antenna system capacity scaling,''
  \emph{IEEE Wireless Commun.}, vol.~17, no.~3, pp. 68--75, Mar. 2010.

\bibitem{122111}
Y.~Shen, B.~Tang, S.~Gao, K.-F. Tong, H.~Wong, K.-K. Wong, and Y.~Zhang,
  ``Design and implementation of mmwave surface wave enabled fluid antennas and
  experimental results for fluid antenna multiple access,'' \emph{arXiv
  preprint arXiv:2405.09663}, 2024.

\bibitem{zhu2023movableNull}
L.~Zhu, W.~Ma, and R.~Zhang, ``Movable-antenna array enhanced beamforming:
  Achieving full array gain with null steering,'' \emph{IEEE Commun. Lett.},
  vol.~27, no.~12, pp. 3340--3344, Dec. 2023.

\bibitem{zhu2023modeling}
------, ``Modeling and performance analysis for movable antenna enabled
  wireless communications,'' \emph{IEEE Trans. Wireless Commun.}, Early Access,
  2023. doi:10.1109/TWC.2023.3330887.

\bibitem{ma2023mimo}
W.~Ma, L.~Zhu, and R.~Zhang, ``{MIMO} capacity characterization for movable
  antenna systems,'' \emph{IEEE Trans. Wireless Commun.}, Early Access, 2023.
  doi:10.1109/TWC.2023.3307696.

\bibitem{new2023information}
W.~K. New, K.-K. Wong, X.~Hao, K.-F. Tong, and C.-B. Chae, ``An
  information-theoretic characterization of {MIMO-FAS}: Optimization,
  diversity-multiplexing tradeoff and $ q $-outage capacity,'' \emph{arXiv
  preprint arXiv:2303.02269}, 2023.

\bibitem{skouroumounis2022fluid}
C.~Skouroumounis and I.~Krikidis, ``Fluid antenna with linear {MMSE} channel
  estimation for large-scale cellular networks,'' \emph{IEEE Trans. Commun.},
  vol.~71, no.~2, pp. 1112--1125, Feb. 2022.

\bibitem{wang2023estimation}
R.~Wang, Y.~Chen, Y.~Hou, K.-K. Wong, and X.~Tao, ``Estimation of channel
  parameters for port selection in millimeter-wave fluid antenna systems,'' in
  \emph{Proc. IEEE/CIC Int. Conf. Commun. China Workshops (ICCC Workshops)},
  Aug. 2023.

\bibitem{ma2023compressed}
W.~Ma, L.~Zhu, and R.~Zhang, ``Compressed sensing based channel estimation for
  movable antenna communications,'' \emph{arXiv preprint arXiv:2306.04333},
  2023.

\bibitem{zhang2023successive}
Z.~Zhang, J.~Zhu, L.~Dai, and R.~W. Heath~Jr, ``Successive bayesian
  reconstructor for channel estimation in flexible antenna systems,''
  \emph{arXiv preprint arXiv:2312.06551}, 2023.

\bibitem{10497534}
Z.~Xiao, S.~Cao, L.~Zhu, Y.~Liu, B.~Ning, X.-G. Xia, and R.~Zhang, ``Channel
  estimation for movable antenna communication systems: A framework based on
  compressed sensing,'' \emph{IEEE Trans. Wireless Commun.}, Early Access,
  2024.

\bibitem{zhu2024performance}
L.~Zhu, W.~Ma, Z.~Xiao, and R.~Zhang, ``Performance analysis and optimization
  for movable antenna aided wideband communications,'' \emph{arXiv preprint
  arXiv:2401.08974}, 2024.

\bibitem{10354003}
L.~Zhu, W.~Ma, B.~Ning, and R.~Zhang, ``Movable-antenna enhanced multiuser
  communication via antenna position optimization,'' \emph{IEEE Trans. Wireless
  Commun.}, Early Access, 2023.

\bibitem{chai2022port}
Z.~Chai, K.-K. Wong, K.-F. Tong, Y.~Chen, and Y.~Zhang, ``Port selection for
  fluid antenna systems,'' \emph{IEEE Commun. Lett.}, vol.~26, no.~5, pp.
  1180--1184, May 2022.

\bibitem{waqar2023deep}
N.~Waqar, K.-K. Wong, K.-F. Tong, A.~Sharples, and Y.~Zhang, ``Deep learning
  enabled slow fluid antenna multiple access,'' \emph{IEEE Commun. Lett.},
  vol.~27, no.~3, pp. 861--865, Mar. 2023.

\bibitem{hu2023movable}
G.~Hu, Q.~Wu, K.~Xu, J.~Ouyang, J.~Si, Y.~Cai, and N.~Al-Dhahir,
  ``Movable-antenna array enabled multiuser uplink: {A} low-complexity gradient
  descent for total transmit power minimization,'' \emph{arXiv preprint
  arXiv:2312.05763}, 2023.

\bibitem{9977471}
H.~Wang, Y.~Shen, K.-F. Tong, and K.-K. Wong, ``Continuous electrowetting
  surface-wave fluid antenna for mobile communications,'' in \emph{Proc. IEEE
  Region 10 Conference (TENCON)}, 2022, pp. 1--3.

\bibitem{checcacci1970holographic}
P.~Checcacci, V.~Russo, and A.~Scheggi, ``Holographic antennas,'' \emph{IEEE
  Trans. on Antennas and Propag.}, vol.~18, no.~6, pp. 811--813, Nov. 1970.

\bibitem{pereda2018experimental}
A.~T. Pereda, F.~Caminita, E.~Martini, I.~Ederra, J.~Teniente, J.~C. Iriarte,
  R.~Gonzalo, and S.~Maci, ``Experimental validation of a ku-band
  dual-circularly polarized metasurface antenna,'' \emph{IEEE Trans. on
  Antennas and Propag.}, vol.~66, no.~3, pp. 1153--1159, Mar. 2018.

\bibitem{araghi2021holographic}
A.~Araghi, M.~Khalily, P.~Xiao, and R.~Tafazolli, ``Holographic-based
  leaky-wave structures: Transformation of guided waves to leaky waves,''
  \emph{IEEE Microw. Mag.}, vol.~22, no.~6, pp. 49--63, Jun. 2021.

\bibitem{zong20196g}
B.~Zong, C.~Fan, X.~Wang, X.~Duan, B.~Wang, and J.~Wang, ``{6G} technologies:
  Key drivers, core requirements, system architectures, and enabling
  technologies,'' \emph{IEEE Veh. Technol. Mag.}, vol.~14, no.~3, pp. 18--27,
  Jul. 2019.

\bibitem{Gong_Vinieratou_Ji_Huang_Alexandropoulos_Wei_Zhang_Debbah_Poor_Yuen_2022}
T.~Gong, P.~Gavriilidis, R.~Ji, C.~Huang, G.~C. Alexandropoulos, L.~Wei,
  Z.~Zhang, M.~Debbah, H.~V. Poor, and C.~Yuen, ``Holographic {MIMO}
  communications: Theoretical foundations, enabling technologies, and future
  directions,'' \emph{IEEE Commun. Surveys Tuts.}, vol.~26, no.~1, pp.
  196--257, Aug. 2023.

\bibitem{Balanis_1982}
C.~A. Balanis, \emph{Antenna theory: analysis and design}.\hskip 1em plus 0.5em
  minus 0.4em\relax John wiley \& sons, 2016.

\bibitem{demir2022channel}
{\"O}.~T. Demir, E.~Bj{\"o}rnson, and L.~Sanguinetti, ``Channel modeling and
  channel estimation for holographic massive {MIMO} with planar arrays,''
  \emph{IEEE Wireless Commun. Lett.}, vol.~11, no.~5, pp. 997--1001, May 2022.

\bibitem{sun2022characteristics}
S.~Sun and M.~Tao, ``Characteristics of channel eigenvalues and mutual coupling
  effects for holographic reconfigurable intelligent surfaces,''
  \emph{Sensors}, vol.~22, no.~14, p. 5297, Jul. 2022.

\bibitem{Williams_2022}
R.~J. Williams, P.~Ram{\'\i}rez-Espinosa, J.~Yuan, and E.~De~Carvalho,
  ``Electromagnetic based communication model for dynamic metasurface
  antennas,'' \emph{IEEE Trans. Wireless Commun.}, vol.~21, no.~10, pp.
  8616--8630, Oct. 2022.

\bibitem{wei2023tri}
L.~Wei, C.~Huang, G.~C. Alexandropoulos, Z.~Yang, J.~Yang, E.~Wei, Z.~Zhang,
  M.~Debbah, and C.~Yuen, ``Tri-polarized holographic {MIMO} surfaces for
  near-field communications: Channel modeling and precoding design,''
  \emph{IEEE Trans. Wireless Commun.}, Dec. 2023.

\bibitem{gong2023holographic}
T.~Gong, L.~Wei, C.~Huang, Z.~Yang, J.~He, M.~Debbah, and C.~Yuen,
  ``Holographic {MIMO} communications with arbitrary surface placements:
  Near-field {LoS} channel model and capacity limit,'' \emph{arXiv preprint
  arXiv:2304.05259}, 2023.

\bibitem{pizzo2022fourier}
A.~Pizzo, L.~Sanguinetti, and T.~L. Marzetta, ``Fourier plane-wave series
  expansion for holographic {MIMO} communications,'' \emph{IEEE Trans. Wireless
  Commun.}, vol.~21, no.~9, pp. 6890--6905, Sept. 2022.

\bibitem{pizzo2020spatially}
A.~Pizzo, T.~L. Marzetta, and L.~Sanguinetti, ``Spatially-stationary model for
  holographic {MIMO} small-scale fading,'' \emph{IEEE J. Sel. Areas Commun.},
  vol.~38, no.~9, pp. 1964--1979, Jun. 2020.

\bibitem{wang2022electromagnetic}
T.~Wang, W.~Han, Z.~Zhong, J.~Pang, G.~Zhou, S.~Wang, and Q.~Li,
  ``Electromagnetic-compliant channel modeling and performance evaluation for
  holographic {MIMO},'' in \emph{Proc. IEEE Global Commun. Conf. (GLOBECOM)},
  Dec. 2022, pp. 747--752.

\bibitem{bjornson2024towards}
E.~Bj{\"o}rnson, C.-B. Chae, R.~W. Heath~Jr, T.~L. Marzetta, A.~Mezghani,
  L.~Sanguinetti, F.~Rusek, M.~R. Castellanos, D.~Jun, and {\"O}.~T. Demir,
  ``Towards {6G} {MIMO}: Massive spatial multiplexing, dense arrays, and
  interplay between electromagnetics and processing,'' \emph{arXiv preprint
  arXiv:2401.02844}, 2024.

\bibitem{Marzetta_2019}
T.~L. Marzetta, ``Super-directive antenna arrays: Fundamentals and new
  perspectives,'' in \emph{Proc. Asilomar Conf. Signals, Syst., Comp.}, Nov.
  2019.

\bibitem{An_Yuen_Huang_Debbah_Poor_Hanzo_2023}
J.~An, C.~Yuen, C.~Huang, M.~Debbah, H.~V. Poor, and L.~Hanzo, ``A tutorial on
  holographic {MIMO} communications---{Part III}: Open opportunities and
  challenges,'' \emph{IEEE Commun. Lett.}, May 2023.

\bibitem{Han_Yin_Marzetta_2022}
L.~Han, H.~Yin, and T.~L. Marzetta, ``Coupling matrix-based beamforming for
  superdirective antenna arrays,'' in \emph{Proc. IEEE Int. Conf. Commun.
  (ICC)}, May 2022, pp. 5159--5164.

\bibitem{sanguinetti2022wavenumber}
L.~Sanguinetti, A.~A. D'Amico, and M.~Debbah, ``Wavenumber-division
  multiplexing in line-of-sight holographic {MIMO} communications,'' \emph{IEEE
  Trans. Wireless Commun.}, Sept. 2022.

\bibitem{zhang2023pattern}
Z.~Zhang and L.~Dai, ``Pattern-division multiplexing for multi-user
  continuous-aperture {MIMO},'' \emph{IEEE J. Sel. Areas Commun.}, Jun. 2023.

\bibitem{pizzo2022nyquist}
A.~Pizzo, A.~de~Jesus~Torres, L.~Sanguinetti, and T.~L. Marzetta, ``Nyquist
  sampling and degrees of freedom of electromagnetic fields,'' \emph{IEEE
  Trans. Signal Process.}, vol.~70, pp. 3935--3947, Jun. 2022.

\bibitem{wan2021terahertz}
Z.~Wan, Z.~Gao, F.~Gao, M.~Di~Renzo, and M.-S. Alouini, ``Terahertz massive
  {MIMO} with holographic reconfigurable intelligent surfaces,'' \emph{IEEE
  Trans. Commun.}, vol.~69, no.~7, pp. 4732--4750, Mar. 2021.

\bibitem{ghermezcheshmeh2023parametric}
M.~Ghermezcheshmeh and N.~Zlatanov, ``Parametric channel estimation for {LoS}
  dominated holographic massive {MIMO} systems,'' \emph{IEEE Access}, May 2023.

\bibitem{d2023dft}
A.~A. D'Amico, G.~Bacci, and L.~Sanguinetti, ``{DFT}-based channel estimation
  for holographic {MIMO},'' \emph{arXiv preprint arXiv:2306.05156}, 2023.

\bibitem{yu2023bayes}
W.~Yu, H.~He, X.~Yu, S.~Song, J.~Zhang, R.~D. Murch, and K.~B. Letaief,
  ``Bayes-optimal unsupervised learning for channel estimation in near-field
  holographic {MIMO},'' \emph{arXiv preprint arXiv:2312.10438}, 2023.

\bibitem{Kildal_Vosoogh_Maci_2016}
P.-S. Kildal, A.~Vosoogh, and S.~Maci, ``Fundamental directivity limitations of
  dense array antennas: A numerical study using {Hannan's} embedded element
  efficiency,'' \emph{IEEE Antennas Wireless Propag. Lett.}, vol.~15, pp.
  766--769, Aug. 2015.

\bibitem{9779586}
L.~Wei, C.~Huang, G.~C. Alexandropoulos, W.~E.~I. Sha, Z.~Zhang, M.~Debbah, and
  C.~Yuen, ``Multi-user holographic {MIMO} surfaces: Channel modeling and
  spectral efficiency analysis,'' \emph{IEEE J. Sel. Topics Signal Process.},
  vol.~16, no.~5, pp. 1112--1124, May 2022.

\bibitem{di2024electromagnetic}
M.~Di~Renzo and M.~D. Migliore, ``Electromagnetic signal and information
  theory,'' \emph{IEEE BITS Info. Theory Mag.}, Early Access, 2024.

\bibitem{Guangming_ICM}
G.~Shi, Y.~Xiao, Y.~Li, and X.~Xie, ``From semantic communication to
  semantic-aware networking: Model, architecture, and open problems,''
  \emph{IEEE Commun. Mag.}, vol.~59, no.~8, pp. 44--50, Aug. 2021.

\bibitem{Deniz_JSAC}
D.~G{\"u}nd{\"u}z, Z.~Qin, I.~E. Aguerri, H.~S. Dhillon, Z.~Yang, A.~Yener,
  K.~K. Wong, and C.-B. Chae, ``Beyond transmitting bits: Context, semantics,
  and task-oriented communications,'' \emph{{IEEE} J. Sel. Areas Commun.},
  vol.~41, no.~1, pp. 5--41, Jan. 2023.

\bibitem{kalfa2021towards}
M.~Kalfa, M.~Gok, A.~Atalik, B.~Tegin, T.~M. Duman, and O.~Arikan, ``Towards
  goal-oriented semantic signal processing: {Applications} and future
  challenges,'' \emph{Digit. Signal Process.}, vol. 119, p. 103134, Dec. 2021.

\bibitem{kalfa2022reliable}
M.~Kalfa, S.~Y. Yetim, A.~Atalik, M.~G{\"o}k, Y.~Ge, R.~Li, W.~Tong, T.~M.
  Duman, and O.~Ar{\i}kan, ``Reliable extraction of semantic information and
  rate of innovation estimation for graph signals,'' \emph{IEEE J. Selet. Areas
  Commun.}, vol.~41, no.~1, pp. 119--140, Jan. 2022.

\bibitem{emrecan_icc_ws}
E.~Kutay and A.~Yener, ``Semantic text compression for classification,'' in
  \emph{Proc. IEEE Int. Conf. Commun. Workshops (ICC Workshops)}, 2023, pp.
  1368--1373.

\bibitem{Jialong_TCVT}
J.~Xu, B.~Ai, W.~Chen, A.~Yang, P.~Sun, and M.~Rodrigues, ``Wireless image
  transmission using deep source channel coding with attention modules,''
  \emph{IEEE Trans. Circuits Syst. Video Technol.}, vol.~32, no.~4, pp.
  2315--2328, Apr. 2022.

\bibitem{Tze-Yang_JSAC}
T.-Y. Tung and D.~G{\"u}nd{\"u}z, ``Deep{W}i{V}e: Deep-learning-aided wireless
  video transmission,'' \emph{IEEE J. Select. Areas Commun.}, vol.~40, no.~9,
  pp. 2570--2583, Sept. 2022.

\bibitem{Shuai_TWC}
S.~Ma, W.~Qiao, Y.~Wu, H.~Li, G.~Shi, D.~Gao, Y.~Shi, S.~Li, and N.~Al-Dhahir,
  ``Task-oriented explainable semantic communications,'' \emph{IEEE Trans.
  Wireless Commun.}, vol.~22, no.~12, pp. 9248--9262, Apr. 2023.

\bibitem{Kurka_JSIT}
D.~B. Kurka and D.~G{\"u}nd{\"u}z, ``{DeepJSCC}-f: Deep joint source-channel
  coding of images with feedback,'' \emph{IEEE J. Sel. Areas Inf. Theory},
  vol.~1, no.~1, pp. 178--193, May 2020.

\bibitem{Sixian_JSAC}
S.~Wang, J.~Dai, Z.~Liang, K.~Niu, Z.~Si, C.~Dong, X.~Qin, and P.~Zhang,
  ``Wireless deep video semantic transmission,'' \emph{IEEE J. Sel. Areas
  Commun.}, vol.~41, no.~1, pp. 214--229, Jan. 2023.

\bibitem{Bourtsoulatze_TCCN}
E.~Bourtsoulatze, D.~Burth~Kurka, and D.~G{\"u}nd{\"u}z, ``Deep joint
  source-channel coding for wireless image transmission,'' \emph{IEEE Trans.
  Cogn. Commun. Netw.}, vol.~5, no.~3, pp. 567--579, Sept. 2019.

\bibitem{Huiqiang_TSP}
H.~Xie, Z.~Qin, G.~Y. Li, and B.~Juang, ``Deep learning enabled semantic
  communication systems,'' \emph{IEEE Trans. Signal Process.}, vol.~69, pp.
  2663--2675, Apr. 2021.

\bibitem{Lei_WCL_Text}
L.~Yan, Z.~Qin, R.~Zhang, Y.~Li, and G.~Y. Li, ``Resource allocation for text
  semantic communications,'' \emph{IEEE Wireless Commun. Lett.}, vol.~11,
  no.~7, pp. 1394--1398, Jul. 2022.

\bibitem{ShengshiTCCN}
S.~Yao, K.~Niu, S.~Wang, and J.~Dai, ``Semantic coding for text transmission:
  An iterative design,'' \emph{IEEE Trans. Cogn. Commun. Netw.}, vol.~8, no.~4,
  pp. 1594--1603, Dec. 2022.

\bibitem{zhenzi_JSAC}
Z.~Weng and Z.~Qin, ``Semantic communication systems for speech transmission,''
  \emph{IEEE J. Sel. Areas Commun.}, vol.~39, no.~8, pp. 2434--2444, Aug. 2021.

\bibitem{Tianxiao2023JSAC}
T.~Han, Q.~Yang, Z.~Shi, S.~He, and Z.~Zhang, ``Semantic-preserved
  communication system for highly efficient speech transmission,'' \emph{IEEE
  J. Sel. Areas Commun.}, vol.~41, no.~1, pp. 245--259, Jan. 2023.

\bibitem{Zixuan_Speech_2023}
Z.~Xiao, S.~Yao, J.~Dai, S.~Wang, K.~Niu, and P.~Zhang, ``Wireless deep speech
  semantic transmission,'' in \emph{Proc. IEEE Int. Conf. Acoust. Speech Signal
  Process. (ICASSP)}, Jun. 2023.

\bibitem{Shengshi_WCNC_Variational}
S.~Yao, Z.~Xiao, S.~Wang, J.~Dai, K.~Niu, and P.~Zhang, ``Variational speech
  waveform compression to catalyze semantic communications,'' in \emph{Proc.
  IEEE Wireless Commun. Netw. Conf. (WCNC)}, Mar. 2023.

\bibitem{Akhtar2019}
A.~Akhtar, B.~Kathariya, and Z.~Li, ``Low latency scalable point cloud
  communication,'' in \emph{Proc. IEEE Int. Conf. Image Process. (ICIP)}, Sept.
  2019, pp. 2369--2373.

\bibitem{liu2023semantic}
X.~Liu, H.~Liang, Z.~Bao, C.~Dong, and X.~Xu, ``Semantic communications system
  with model division multiple access and controllable coding rate for point
  cloud,'' \emph{arXiv preprint arXiv:2307.06027}, 2023.

\bibitem{Jiawei2021GNN}
J.~Shao, H.~Zhang, Y.~Mao, and J.~Zhang, ``Branchy-{GNN}: A device-edge
  co-inference framework for efficient point cloud processing,'' in \emph{Proc.
  IEEE Int. Conf. Acoust. Speech Signal Process. (ICASSP)}, Jun. 2021, pp.
  8488--8492.

\bibitem{huang2022iscom}
Y.~Huang, B.~Bai, Y.~Zhu, X.~Qiao, X.~Su, and P.~Zhang, ``{ISCom}:
  Interest-aware semantic communication scheme for point cloud video
  streaming,'' \emph{arXiv preprint arXiv:2210.06808}, 2022.

\bibitem{Jankowski_JSAC}
M.~Jankowski, D.~G{\"u}nd{\"u}z, and K.~Mikolajczyk, ``Wireless image retrieval
  at the edge,'' \emph{IEEE J. Sel. Areas Commun.}, vol.~39, no.~1, pp.
  89--100, Jan. 2021.

\bibitem{ChiaHan_Access}
C.~H. Lee, J.~W. Lin, P.~H. Chen, and Y.~C. Chang, ``Deep learning-constructed
  joint transmission-recognition for {Internet of Things},'' \emph{IEEE
  Access}, vol.~7, pp. 76\,547--76\,561, Jun. 2019.

\bibitem{Qiyu_TWC}
Q.~Hu, G.~Zhang, Z.~Qin, Y.~Cai, G.~Yu, and G.~Y. Li, ``Robust semantic
  communications with masked {VQ-VAE} enabled codebook,'' \emph{IEEE Trans.
  Wireless Commun.}, Apr. 2023.

\bibitem{Huiqiang_JSAC}
H.~Xie, Z.~Qin, X.~Tao, and K.~B. Letaief, ``Task-oriented multi-user semantic
  communications,'' \emph{IEEE J. Sel. Areas Commun.}, vol.~40, no.~9, pp.
  2584--2597, Sept. 2022.

\bibitem{YangshuoTCOM2023}
Y.~He, G.~Yu, and Y.~Cai, ``Rate-adaptive coding mechanism for semantic
  communications with multi-modal data,'' \emph{IEEE Trans. Commun.}, Early
  Access, 2023.

\bibitem{QiyuRobust}
Q.~Hu, G.~Zhang, Z.~Qin, Y.~Cai, G.~Yu, and G.~Y. Li, ``Robust semantic
  communications against semantic noise,'' in \emph{Proc. Veh. Technol. Conf.
  (VTC-Fall)}, Sept. 2022.

\bibitem{Jianhao_2024}
J.~Huang, D.~Li, C.~Huang, X.~Qin, and W.~Zhang, ``Joint task and data-oriented
  semantic communications: A deep separate source-channel coding scheme,''
  \emph{IEEE Internet Things J.}, vol.~11, no.~2, pp. 2255--2272, Jan. 2024.

\bibitem{Guangyi_Arxiv}
G.~Zhang, Q.~Hu, Y.~Cai, and G.~Yu, ``{SCAN}: Semantic communication with
  adaptive channel feedback,'' \emph{arXiv preprint arXiv:2306.15534}, 2023.

\bibitem{wang2023privacypreserving}
Y.~Wang, S.~Guo, Y.~Deng, H.~Zhang, and Y.~Fang, ``Privacy-preserving
  task-oriented semantic communications against model inversion attacks,''
  \emph{arXiv preprint arXiv:2312.03252}, 2023.

\bibitem{Zhonghao2024Semantic}
Z.~Lyu, G.~Zhu, J.~Xu, B.~Ai, and S.~Cui, ``Semantic communications for image
  recovery and classification via deep joint source and channel coding,''
  \emph{IEEE Trans. Wireless Commun.}, Early Access, 2024.

\bibitem{Chamain2022End}
L.~D. Chamain, S.~Qi, and Z.~Ding, ``End-to-end image classification and
  compression with variational autoencoders,'' \emph{IEEE Internet Things J.},
  vol.~9, no.~21, pp. 21\,916--21\,931, Nov. 2022.

\bibitem{Yufei_Arxiv}
Y.~Bo, Y.~Duan, S.~Shao, and M.~Tao, ``Joint coding-modulation for digital
  semantic communications via variational autoencoder,'' \emph{arXiv preprint
  arXiv:2310.06690}, 2023.

\bibitem{Guangyi_CL}
G.~Zhang, Q.~Hu, Y.~Cai, and G.~Yu, ``Alleviating distortion accumulation in
  multi-hop semantic communication,'' \emph{IEEE Comm. Lett.}, Early Access,
  2024. doi:10.1109/LCOMM.2023.3339776.

\bibitem{balle2018variational}
J.~Ball{\'e}, D.~Minnen, S.~Singh, S.~J. Hwang, and N.~Johnston, ``Variational
  image compression with a scale hyperprior,'' \emph{arXiv preprint
  arXiv:1802.01436}, 2018.

\bibitem{Balle_JSTSP}
J.~Ball{\'e}, P.~A. Chou, D.~Minnen, S.~Singh, N.~Johnston, E.~Agustsson, S.~J.
  Hwang, and G.~Toderici, ``Nonlinear transform coding,'' \emph{IEEE J. Sel.
  Topics Signal Process.}, vol.~15, no.~2, pp. 339--353, Oct. 2020.

\bibitem{Cheng_2020_CVPR}
Z.~Cheng, H.~Sun, M.~Takeuchi, and J.~Katto, ``Learned image compression with
  discretized gaussian mixture likelihoods and attention modules,'' in
  \emph{Proc. IEEE Conf. Comput. Vis. Pattern Recognit. (CVPR)}, Jun. 2020.

\bibitem{SixianNTSCCjj}
S.~Wang, J.~Dai, X.~Qin, Z.~Si, K.~Niu, and P.~Zhang, ``Improved nonlinear
  transform source-channel coding to catalyze semantic communications,''
  \emph{IEEE J. Sel. Topics Signal Process.}, vol.~17, no.~5, pp. 1022--1037,
  Sept. 2023.

\bibitem{Jincheng_JSAC_Nonlinear}
J.~Dai, S.~Wang, K.~Tan, Z.~Si, X.~Qin, K.~Niu, and P.~Zhang, ``Nonlinear
  transform source-channel coding for semantic communications,'' \emph{IEEE J.
  Sel. Areas Commun.}, vol.~40, no.~8, pp. 2300--2316, Aug. 2022.

\bibitem{Minnen2013nips}
D.~Minnen, J.~Ball\'{e}, and G.~D. Toderici, ``Joint autoregressive and
  hierarchical priors for learned image compression,'' in \emph{Advances in
  Neural Information Processing Systems}, S.~Bengio, H.~Wallach, H.~Larochelle,
  K.~Grauman, N.~Cesa-Bianchi, and R.~Garnett, Eds., vol.~31.\hskip 1em plus
  0.5em minus 0.4em\relax Curran Associates, Inc., 2018.

\bibitem{Ecenaz_JSAC_2023}
E.~Erdemir, T.-Y. Tung, P.~L. Dragotti, and D.~G{\"u}nd{\"u}z, ``Generative
  joint source-channel coding for semantic image transmission,'' \emph{IEEE J.
  Sel. Areas Commun.}, vol.~41, no.~8, pp. 2645--2657, Aug. 2023.

\bibitem{xu2023latent}
B.~Xu, R.~Meng, Y.~Chen, X.~Xu, C.~Dong, and H.~Sun, ``Latent semantic
  diffusion-based channel adaptive de-noising {SemCom} for future {6G}
  systems,'' \emph{arXiv preprint arXiv:2304.09420}, 2023.

\bibitem{grassucci2023generative}
E.~Grassucci, S.~Barbarossa, and D.~Comminiello, ``Generative semantic
  communication: Diffusion models beyond bit recovery,'' \emph{arXiv preprint
  arXiv:2306.04321}, 2023.

\bibitem{Peiwen_JSAC_2023}
P.~Jiang, C.-K. Wen, S.~Jin, and G.~Y. Li, ``Wireless semantic communications
  for video conferencing,'' \emph{IEEE J. Sel. Areas Commun.}, vol.~41, no.~1,
  pp. 230--244, Jan. 2023.

\bibitem{sagduyu2023joint}
Y.~E. Sagduyu, T.~Erpek, A.~Yener, and S.~Ulukus, ``Joint sensing and
  task-oriented communications with image and wireless data modalities for
  dynamic spectrum access,'' \emph{arXiv preprint arXiv:2312.13931}, 2023.

\bibitem{Shao_JSAC}
J.~Shao, Y.~Mao, and J.~Zhang, ``Learning task-oriented communication for edge
  inference: An information bottleneck approach,'' \emph{{IEEE} J. Sel. Areas
  Commun.}, vol.~40, no.~1, pp. 197--211, Jan. 2022.

\bibitem{XieJSACRobust}
S.~Xie, S.~Ma, M.~Ding, Y.~Shi, M.~Tang, and Y.~Wu, ``Robust information
  bottleneck for task-oriented communication with digital modulation,''
  \emph{{IEEE} J. Sel. Areas Commun.}, vol.~41, no.~8, pp. 2577--2591, Aug.
  2023.

\bibitem{Shao2023Task}
J.~Shao, Y.~Mao, and J.~Zhang, ``Task-oriented communication for multidevice
  cooperative edge inference,'' \emph{IEEE Trans. Wireless Commun.}, vol.~22,
  no.~1, pp. 73--87, Jan. 2023.

\bibitem{Huiguo2023TCCN}
H.~Gao, G.~Yu, and Y.~Cai, ``Adaptive modulation and retransmission scheme for
  semantic communication systems,'' \emph{IEEE Trans. Cogn. Commun. Netw.},
  vol.~10, no.~1, pp. 150--163, 2024.

\bibitem{kutay2024classification}
E.~Kutay and A.~Yener, ``Classification-oriented semantic wireless
  communications,'' in \emph{Proc. IEEE Int. Conf. Acoust., Speech Signal
  Process. (ICASSP)}, Apr. 2024.

\bibitem{UDeepSC}
G.~Zhang, Q.~Hu, Z.~Qin, Y.~Cai, G.~Yu, X.~Tao, and G.~Y. Li, ``A unified
  multi-task semantic communication system for multimodal data,'' \emph{arXiv
  preprint arXiv:2209.07689}, 2022.

\bibitem{sagduyu2023age}
Y.~E. Sagduyu, S.~Ulukus, and A.~Yener, ``Age of information in deep
  learning-driven task-oriented communications,'' in \emph{IEEE Conf. Comput.
  Commun. Workshops (INFOCOM WKSHPS)}, 2023, pp. 1--6.

\bibitem{Han2023TPAMI}
K.~Han, Y.~Wang, H.~Chen, X.~Chen, J.~Guo, Z.~Liu, Y.~Tang, A.~Xiao, C.~Xu,
  Y.~Xu, Z.~Yang, Y.~Zhang, and D.~Tao, ``A survey on vision transformer,''
  \emph{IEEE Trans. Pattern Anal. Mach. Intell.}, vol.~45, no.~1, pp. 87--110,
  Jan. 2023.

\bibitem{sagduyu2023multi}
Y.~E. Sagduyu, T.~Erpek, A.~Yener, and S.~Ulukus, ``Multi-receiver
  task-oriented communications via multi-task deep learning,'' \emph{EEE Future
  Netw. World Forum, Baltimore, MD}, Nov. 2023.

\bibitem{sagduyu2023joint2}
------, ``Joint sensing and semantic communications with multi-task deep
  learning,'' \emph{arXiv preprint arXiv:2311.05017}, 2023.

\bibitem{yang2023secure}
Z.~Yang, M.~Chen, G.~Li, Y.~Yang, and Z.~Zhang, ``Secure semantic
  communications: Fundamentals and challenges,'' \emph{arXiv preprint
  arXiv:2301.01421}, 2023.

\bibitem{Tze-Yang_Arxiv}
T.-Y. Tung and D.~G{\"u}nd{\"u}z, ``Deep joint source-channel and encryption
  coding: Secure semantic communications,'' \emph{arXiv preprint
  arXiv:2208.09245}, 2022.

\bibitem{Goldblum_TAML}
M.~Goldblum, D.~Tsipras, C.~Xie, X.~Chen, A.~Schwarzschild, D.~Song, A.~Madry,
  B.~Li, and T.~Goldstein, ``Dataset security for machine learning: Data
  poisoning, backdoor attacks, and defenses,'' \emph{IEEE Trans. Pattern Anal.
  Mach. Intell.}, vol.~45, no.~2, pp. 1563--1580, Mar. 2023.

\bibitem{Sagduyu2023Vul}
Y.~E. Sagduyu, T.~Erpek, S.~Ulukus, and A.~Yener, ``Vulnerabilities of deep
  learning-driven semantic communications to backdoor (trojan) attacks,'' in
  \emph{Proc. Annu. Conf. Inf. Sci. Syst. (CISS)}, Mar. 2023.

\bibitem{gao2020backdoor}
Y.~Gao, B.~G. Doan, Z.~Zhang, S.~Ma, J.~Zhang, A.~Fu, S.~Nepal, and H.~Kim,
  ``Backdoor attacks and countermeasures on deep learning: A comprehensive
  review,'' \emph{arXiv preprint arXiv:2007.10760}, 2020.

\bibitem{Meng2023Secure}
M.~Shen, J.~Wang, H.~Du, D.~Niyato, X.~Tang, J.~Kang, Y.~Ding, and L.~Zhu,
  ``Secure semantic communications: Challenges, approaches, and
  opportunities,'' \emph{IEEE Netw.}, Early Access, 2023.

\bibitem{sagduyu2023tasknextG}
Y.~E. Sagduyu, S.~Ulukus, and A.~Yener, ``Task-oriented communications for
  nextg: {End}-to-end deep learning and ai security aspects,'' \emph{IEEE
  Wireless Commun.}, vol.~30, no.~3, pp. 52--60, Mar. 2023.

\bibitem{Guoshun2023Physical}
G.~Nan, Z.~Li, J.~Zhai, Q.~Cui, G.~Chen, X.~Du, X.~Zhang, X.~Tao, Z.~Han, and
  T.~Q.~S. Quek, ``Physical-layer adversarial robustness for deep
  learning-based semantic communications,'' \emph{IEEE J. Sel. Areas Commun.},
  vol.~41, no.~8, pp. 2592--2608, Jun. 2023.

\bibitem{Tang2023GAN}
R.~Tang, D.~Gao, M.~Yang, T.~Guo, H.~Wu, and G.~Shi, ``Gan-inspired intelligent
  jamming and anti-jamming strategy for semantic communication systems,'' in
  \emph{Proc. IEEE Int. Conf. Commun. Workshops (ICC Workshops)}, Jun. 2023,
  pp. 1623--1628.

\bibitem{sagduyu2023semantic}
Y.~E. Sagduyu, T.~Erpek, S.~Ulukus, and A.~Yener, ``Is semantic communication
  secure? {A} tale of multi-domain adversarial attacks,'' \emph{IEEE Commun.
  Mag.}, vol.~61, no.~11, pp. 50--55, Nov. 2023.

\bibitem{XinghanSem}
X.~Liu, G.~Nan, Q.~Cui, Z.~Li, P.~Liu, Z.~Xing, H.~Mu, X.~Tao, and T.~Q.
  S.~Quek, ``Semprotector: A unified framework for semantic protection in deep
  learning-based semantic communication systems,'' \emph{IEEE Commun. Mag.},
  vol.~61, no.~11, pp. 56--62, Nov. 2023.

\bibitem{MaojunIWCL}
M.~Zhang, Y.~Li, Z.~Zhang, G.~Zhu, and C.~Zhong, ``Wireless image transmission
  with semantic and security awareness,'' \emph{IEEE Wireless Commun. Lett.},
  vol.~12, no.~8, pp. 1389--1393, Aug. 2023.

\bibitem{Xinlai2023Encrypted}
X.~Luo, Z.~Chen, M.~Tao, and F.~Yang, ``Encrypted semantic communication using
  adversarial training for privacy preserving,'' \emph{IEEE Commun. Lett.},
  vol.~27, no.~6, pp. 1486--1490, Jun. 2023.

\bibitem{PeiwenIWC2023}
P.~Jiang, C.-K. Wen, S.~Jin, and G.~Y. Li, ``Wireless semantic transmission via
  revising modules in conventional communications,'' \emph{IEEE Wireless
  Commun.}, vol.~30, no.~3, pp. 28--34, Jun. 2023.

\bibitem{Wenyu_TWC}
W.~Zhang, H.~Zhang, H.~Ma, H.~Shao, N.~Wang, and V.~C.~M. Leung, ``Predictive
  and adaptive deep coding for wireless image transmission in semantic
  communication,'' \emph{IEEE Trans. Wireless Commun.}, vol.~22, no.~8, pp.
  5486--5501, Jan. 2023.

\bibitem{YangWITT}
K.~Yang, S.~Wang, J.~Dai, K.~Tan, K.~Niu, and P.~Zhang, ``{WITT}: A wireless
  image transmission transformer for semantic communications,'' in \emph{Proc.
  IEEE Int. Conf. Acoust. Speech Signal Process. (ICASSP)}, Jun. 2023.

\bibitem{Haotian_Arxiv}
H.~Wu, Y.~Shao, C.~Bian, K.~Mikolajczyk, and G{\"u}nd{\"u}z, ``Vision
  transformer for adaptive image transmission over {MIMO} channels,''
  \emph{arXiv preprint arXiv:2210.15347}, 2022.

\bibitem{zhou2024feature}
K.~Zhou, G.~Zhang, Y.~Cai, Q.~Hu, G.~Yu, and A.~L. Swindlehurst, ``Feature
  allocation for semantic communication with space-time importance awareness,''
  \emph{arXiv preprint arXiv:2401.14614}, 2024.

\bibitem{Shengshi2023PIMRC}
S.~Yao, S.~Wang, J.~Dai, and K.~Niu, ``Learned image transmission over {MIMO}
  fading channels,'' in \emph{Proc. IEEE Annual Int. Symposium Personal, Indoor
  and Mobile Radio Commun. (PIMRC)}, 2023.

\bibitem{AdaptMIMO}
C.~Bian, Y.~Shao, H.~Wu, and D.~G{\"u}nd{\"u}z, ``Space-time design for deep
  joint source channel coding of images over {MIMO} channels,'' in \emph{Proc.
  IEEE Int. Workshop Signal Process. Adv. Wireless Commun. (SPAWC)}, Sept.
  2023, pp. 616--620.

\bibitem{Hongyangjsac2023}
H.~Du, J.~Wang, D.~Niyato, J.~Kang, Z.~Xiong, J.~Zhang, and X.~Shen, ``Semantic
  communications for wireless sensing: {RIS}-aided encoding and self-supervised
  decoding,'' \emph{{IEEE} J. Sel. Areas Commun.}, vol.~41, no.~8, pp.
  2547--2562, 2023.

\bibitem{xujieWCL2024}
X.~Hu, Y.~Tian, Q.~Li, Y.~H. Kho, X.~Wang, B.~Xiao, Z.~Yang, and W.~Li, ``A
  novel {RIS}-aided optimization strategy for semantic communication system,''
  \emph{IEEE Wireless Commun. Lett.}, pp. 1--1, to appear, 2024.

\bibitem{shuyi2024ris}
S.~Chen, Y.~Hui, Y.~Qin, Y.~Yuan, W.~Meng, X.~Luo, and H.-H. Chen,
  ``{RIS}-based on-the-air semantic communications -- a diffractional deep
  neural network approach,'' \emph{IEEE Wireless Commun.}, pp. 1--8, 2024.

\bibitem{Lan2023OFDM}
L.~Lin, W.~Xu, F.~Wang, Y.~Zhang, W.~Zhang, and P.~Zhang,
  ``Channel-transferable semantic communications for multi-user {OFDM-NOMA}
  systems,'' \emph{IEEE Wireless Commun. Lett.}, Early Access, 2023.

\bibitem{9714510}
M.~Yang, C.~Bian, and H.-S. Kim, ``{OFDM}-guided deep joint source channel
  coding for wireless multipath fading channels,'' \emph{IEEE Trans. Cogn.
  Commun. Netw.}, vol.~8, no.~2, pp. 584--599, Feb. 2022.

\bibitem{Yulin_WCL}
Y.~Shao and D.~G{\"u}nd{\"u}z, ``Semantic communications with discrete-time
  analog transmission: A {PAPR} perspective,'' \emph{IEEE Wireless Commun.
  Lett.}, vol.~12, no.~3, pp. 510--514, Mar. 2023.

\bibitem{OFDM_Chuanhong2024}
C.~Liu, C.~Guo, Y.~Yang, W.~Ni, and T.~Q.~S. Quek, ``{OFDM}-based digital
  semantic communication with importance awareness,'' \emph{arXiv preprint
  arXiv:2401.02178}, 2024.

\bibitem{TungDeepJSCCQ}
T.-Y. Tung, D.~B. Kurka, M.~Jankowski, and D.~G{\"u}nd{\"u}z, ``{DeepJSCC-Q}:
  Constellation constrained deep joint source-channel coding,'' \emph{IEEE J.
  Sel. Areas Info. Theory}, vol.~3, no.~4, pp. 720--731, Dec. 2022.

\bibitem{Mengyang2022Q}
M.~Wang, J.~Li, M.~Ma, and X.~Fan, ``Constellation design for deep joint
  source-channel coding,'' \emph{IEEE Signal Process. Lett.}, vol.~29, pp.
  1442--1446, Jun. 2022.

\bibitem{Kristy2019}
K.~Choi, K.~Tatwawadi, A.~Grover, T.~Weissman, and S.~Ermon, ``Neural joint
  source-channel coding,'' in \emph{Proc. Int. Conf. Mach. Learn. (ICML)},
  vol.~97, Jun. 2019, pp. 1182--1192.

\bibitem{guo2024digitalsc}
L.~Guo, W.~Chen, Y.~Sun, and B.~Ai, ``Digital-{SC}: Digital semantic
  communication with adaptive network split and learned non-linear
  quantization,'' \emph{arXiv preprint arXiv:2305.13553}, 2023.

\bibitem{wang2023spiking}
M.~Wang, J.~Li, M.~Ma, and X.~Fan, ``Spiking semantic communication for feature
  transmission with {HARQ},'' \emph{arXiv preprint arXiv:2310.08804}, 2023.

\bibitem{PeiwenTCOM}
P.~Jiang, C.-K. Wen, S.~Jin, and G.~Y. Li, ``Deep source-channel coding for
  sentence semantic transmission with {HARQ},'' \emph{IEEE Trans. Commun.},
  vol.~70, no.~8, pp. 5225--5240, Aug. 2022.

\bibitem{mu2022semi_NOMA}
X.~Mu, Y.~Liu, L.~Guo, and N.~Al-Dhahir, ``Heterogeneous semantic and bit
  communications: A semi-{NOMA} scheme,'' \emph{IEEE J. Sel. Areas Commun.},
  vol.~41, no.~1, pp. 155--169, Jan. 2023.

\bibitem{mu2022mul_sem}
X.~Mu and Y.~Liu, ``Semantic communications in multi-user wireless networks,''
  \emph{arXiv preprint arXiv:2211.08932}, 2022.

\bibitem{liu2023semrelay}
T.~Liu, C.~You, Z.~Hu, C.~Wu, Y.~Gong, and K.~Huang, ``Semantic-relay-aided
  text transmission: Placement optimization and bandwidth allocation,''
  \emph{arXiv preprint arXiv:2311.09850}, 2023.

\bibitem{hu2023semrelay}
Z.~Hu, T.~Liu, C.~You, Z.~Yang, and M.~Chen, ``Multiuser resource allocation
  for semantic-relay-aided text transmissions,'' \emph{arXiv preprint
  arXiv:2311.06854}, 2023.

\bibitem{yang2023energy}
Z.~Yang, M.~Chen, Z.~Zhang, and C.~Huang, ``Energy efficient semantic
  communication over wireless networks with rate splitting,'' \emph{arXiv
  preprint arXiv:2301.01987}, 2023.

\bibitem{yan2022resource}
L.~Yan, Z.~Qin, R.~Zhang, Y.~Li, and G.~Y. Li, ``Resource allocation for text
  semantic communications,'' \emph{IEEE Wireless Commun. Lett.}, vol.~11,
  no.~7, pp. 1394--1398, Jul. 2022.

\bibitem{Jiajia2023ICSPCC}
J.~Shi, T.-T. Chan, H.~Pan, and T.-M. Lok, ``Reconfigurable intelligent surface
  assisted semantic communication systems,'' in \emph{IEEE Int. Conf. Signal
  Process., Commun. Comput.(ICSPCC)}, 2023, pp. 1--6.

\bibitem{you2016energy}
C.~You, K.~Huang, H.~Chae, and B.-H. Kim, ``Energy-efficient resource
  allocation for mobile-edge computation offloading,'' \emph{IEEE Trans.
  Wireless Commun.}, vol.~16, no.~3, pp. 1397--1411, Mar. 2016.

\bibitem{you2018asynchronous}
C.~You, Y.~Zeng, R.~Zhang, and K.~Huang, ``Asynchronous mobile-edge computation
  offloading: Energy-efficient resource management,'' \emph{IEEE Trans.
  Wireless Commun.}, vol.~17, no.~11, pp. 7590--7605, Nov. 2018.

\bibitem{arda2024semantic}
E.~Arda, E.~Kutay, and A.~Yener, ``Semantic forwarding for next generation
  relay networks,'' in \emph{Proc. Ann. Conf. Info. Sci. Syst. (CISS)}, Mar.
  2024.

\bibitem{LAI2023}
F.~Jiang, Y.~Peng, L.~Dong, K.~Wang, and X.~Y. Kun~Yang, Cunhua~Pan, ``Large
  {AI} model empowered multimodal semantic communications,'' \emph{arXiv
  preprint arXiv.2309.01249}, 2023.

\bibitem{Shuaishuai2023SI}
S.~Guo, Y.~Wang, S.~Li, and N.~Saeed, ``Semantic importance-aware
  communications using pre-trained language models,'' \emph{IEEE Commun.
  Lett.}, vol.~27, no.~9, pp. 2328--2332, Jul. 2023.

\bibitem{he2021checkerboard}
D.~He, Y.~Zheng, B.~Sun, Y.~Wang, and H.~Qin, ``Checkerboard context model for
  efficient learned image compression,'' in \emph{Proc. IEEE Conf. Comput. Vis.
  Pattern Recognit. (CVPR)}, 2021, pp. 14\,771--14\,780.

\bibitem{Hanju2023Access}
H.~Yoo, L.~Dai, S.~Kim, and C.-B. Chae, ``On the role of {ViT} and {CNN} in
  semantic communications: Analysis and prototype validation,'' \emph{IEEE
  Access}, vol.~11, pp. 71\,528--71\,541, Jul. 2023.

\bibitem{Dong2022WCSP}
H.~Dong, W.~Yue, and K.~Niu, ``A demo of semantic communication: Rosefinch,''
  in \emph{Proc. Int. Conf. Wireless Commun. Signal Process. (WCSP)}, Nov.
  2022, pp. 373--377.

\bibitem{Utkovski_IOTJ}
Z.~Utkovski, A.~Munari, G.~Caire, J.~Dommel, P.-H. Lin, M.~Franke, A.~C.
  Drummond, and S.~Stańczak, ``Semantic communication for edge intelligence:
  Theoretical foundations and implications on protocols,'' \emph{IEEE Internet
  Things Mag.}, vol.~6, no.~4, pp. 48--53, 2023.

\bibitem{10024901}
X.~Mu, Z.~Wang, and Y.~Liu, ``{NOMA} for integrating sensing and communications
  towards {6G}: A multiple access perspective,'' \emph{{IEEE} Wireless
  Commun.}, Early Access, 2023.

\bibitem{Zuo}
J.~Zuo, X.~Mu, and Y.~Liu, ``Non-orthogonal multiple access for near-field
  communications,'' \emph{arXiv preprint arXiv:2304.13185}, 2023.

\bibitem{10129111}
Z.~Ding, R.~Schober, and H.~V. Poor, ``{NOMA}-based coexistence of near-field
  and far-field massive {MIMO} communications,'' \emph{{IEEE} Wireless Commun.
  Lett.}, vol.~12, no.~8, pp. 1429--1433, Aug. 2023.

\bibitem{10315058}
Z.~Ding, ``Resolution of near-field beamforming and its impact on {NOMA},''
  \emph{{IEEE} Wireless Commun. Lett.}, vol.~13, no.~2, pp. 456--460, Feb.
  2024.

\bibitem{9801736}
Y.~Liu, X.~Mu, X.~Liu, M.~Di~Renzo, Z.~Ding, and R.~Schober, ``Reconfigurable
  intelligent surface-aided multi-user networks: Interplay between {NOMA} and
  {RIS},'' \emph{{IEEE} Wireless Commun.}, vol.~29, no.~2, pp. 169--176, April
  2022.

\bibitem{9139273}
X.~Mu, Y.~Liu, L.~Guo, J.~Lin, and N.~Al-Dhahir, ``Exploiting intelligent
  reflecting surfaces in {NOMA} networks: Joint beamforming optimization,''
  \emph{{IEEE} Trans. Wireless Commun.}, vol.~19, no.~10, pp. 6884--6898, Oct.
  2020.

\bibitem{9240028}
J.~Zhu, Y.~Huang, J.~Wang, K.~Navaie, and Z.~Ding, ``Power efficient
  {IRS}-assisted {NOMA},'' \emph{{IEEE} Trans. Commun.}, vol.~69, no.~2, pp.
  900--913, Feb. 2021.

\bibitem{Zheng}
J.~Zheng, T.~Wu, X.~Lai, C.~Pan, M.~Elkashlan, and K.-K. Wong, ``{FAS}-assisted
  {NOMA} short-packet communication systems,'' \emph{arXiv preprint
  arXiv:2310.14251}, 2023.

\bibitem{10158994}
X.~Mu and Y.~Liu, ``Exploiting semantic communication for non-orthogonal
  multiple access,'' \emph{{IEEE} J. Sel. Areas Commun.}, vol.~41, no.~8, pp.
  2563--2576, Aug. 2023.

\bibitem{li2023non}
W.~Li, H.~Liang, C.~Dong, X.~Xu, P.~Zhang, and K.~Liu, ``Non-orthogonal
  multiple access enhanced multi-user semantic communication,'' \emph{IEEE
  Trans. Cogn. Commun. Netw.}, vol.~9, no.~6, pp. 1438--1453, Dec. 2023.

\bibitem{ITU_NOMA}
\BIBentryALTinterwordspacing
``framework and overall objectives of the future development of {IMT} for 2030
  and beyond''. [Online]. Available:
  \url{https://www.itu.int/dms_pubrec/itu-r/rec/m/R-REC-M.2160-0-202311-I%21%21PDF-E.pdf}
\BIBentrySTDinterwordspacing

\bibitem{liu2022integrated}
F.~Liu, Y.~Cui, C.~Masouros, J.~Xu, T.~X. Han, Y.~C. Eldar, and S.~Buzzi,
  ``Integrated sensing and communications: Toward dual-functional wireless
  networks for {6G} and beyond,'' \emph{IEEE J. Sel. Areas Commun.}, vol.~40,
  no.~6, pp. 1728--1767, Mar. 2022.

\bibitem{lu2024integrated}
S.~Lu, F.~Liu, Y.~Li, K.~Zhang, H.~Huang, J.~Zou, X.~Li, Y.~Dong, F.~Dong,
  J.~Zhu \emph{et~al.}, ``Integrated sensing and communications: Recent
  advances and ten open challenges,'' \emph{IEEE Internet Things J.}, Early
  Access, 2024.

\bibitem{chen2023cramer}
A.~Chen, L.~Chen, Y.~Chen, C.~You, G.~Wei, and F.~R. Yu, ``{Cram{\'e}r-Rao}
  bounds of near-field positioning based on electromagnetic propagation
  model,'' \emph{IEEE Trans. Veh. Technol.}, vol.~72, no.~11, pp.
  13\,808--13\,825, Nov. 2023.

\bibitem{chen2023near}
A.~Chen, L.~Chen, Y.~Chen, N.~Zhao, and C.~You, ``Near-field positioning and
  attitude sensing based on electromagnetic propagation modeling,'' \emph{arXiv
  preprint arXiv:2310.17327}, 2023.

\bibitem{8736783}
B.~Friedlander, ``Localization of signals in the near-field of an antenna
  array,'' \emph{{IEEE} Trans. Signal Process.}, vol.~67, no.~15, pp.
  3885--3893, Aug. 2019.

\bibitem{wang2023nearisac}
Z.~Wang, X.~Mu, and Y.~Liu, ``Near-field integrated sensing and
  communications,'' \emph{{IEEE} Commun. Lett.}, vol.~27, no.~8, pp.
  2048--2052, Aug. 2023.

\bibitem{wang2024cramer}
H.~Wang, Z.~Xiao, and Y.~Zeng, ``{Cram{\'e}r-Rao} bounds for near-field sensing
  with extremely large-scale {MIMO},'' \emph{{IEEE} Trans. Signal Process.},
  vol.~72, pp. 701--717, Jan. 2024.

\bibitem{wang2023near_v}
Z.~Wang, X.~Mu, and Y.~Liu, ``Near-field velocity sensing and predictive
  beamforming,'' \emph{arXiv preprint arXiv:2311.09888}, 2023.

\bibitem{wang2023rethinking}
------, ``Rethinking integrated sensing and communication: When near field
  meets wideband,'' \emph{{IEEE} Commun. Mag.}, accepted to appear, 2024.

\bibitem{shao2022target}
X.~Shao, C.~You, W.~Ma, X.~Chen, and R.~Zhang, ``Target sensing with
  intelligent reflecting surface: Architecture and performance,'' \emph{{IEEE}
  J. Sel. Areas Commun.}, vol.~40, no.~7, pp. 2070--2084, Jul. 2022.

\bibitem{shao2024intelligent}
X.~Shao, C.~You, and R.~Zhang, ``Intelligent reflecting surface aided wireless
  sensing: {Applications} and design issues,'' \emph{IEEE Wireless Commun.},
  Early Access, 2024.

\bibitem{hua2023intelligent}
M.~Hua, Q.~Wu, W.~Chen, Z.~Fei, H.~C. So, and C.~Yuen, ``Intelligent reflecting
  surface assisted localization: Performance analysis and algorithm design,''
  \emph{IEEE Wireless Commun. Lett.}, vol.~13, no.~1, pp. 84--88, Jan. 2024.

\bibitem{hua2023secure}
M.~Hua, Q.~Wu, W.~Chen, O.~A. Dobre, and A.~L. Swindlehurst, ``Secure
  intelligent reflecting surface aided integrated sensing and communication,''
  \emph{{IEEE} Trans. Wireless Commun.}, vol.~23, no.~1, pp. 575--591, Jan.
  2024.

\bibitem{10243495}
S.~P. Chepuri, N.~Shlezinger, F.~Liu, G.~C. Alexandropoulos, S.~Buzzi, and
  Y.~C. Eldar, ``Integrated sensing and communications with reconfigurable
  intelligent surfaces: From signal modeling to processing,'' \emph{{IEEE}
  Signal Process. Mag.}, vol.~40, no.~6, pp. 41--62, Sep. 2023.

\bibitem{wang2023stars}
Z.~Wang, X.~Mu, and Y.~Liu, ``{STARS} enabled integrated sensing and
  communications,'' \emph{{IEEE} Trans. Wireless Commun.}, vol.~22, no.~10, pp.
  6750--6765, Oct. 2023.

\bibitem{zhang2023stars}
Z.~Zhang, Y.~Liu, Z.~Wang, and J.~Chen, ``{STARS-ISAC}: How many sensors do we
  need?'' \emph{{IEEE} Trans. Wireless Commun.}, vol.~23, no.~2, pp.
  1085--1099, Feb. 2024.

\bibitem{rihan2023passive}
M.~Rihan, A.~Zappone, S.~Buzzi, G.~Fodor, and M.~Debbah, ``Passive vs. active
  reconfigurable intelligent surfaces for integrated sensing and communication:
  Challenges and opportunities,'' \emph{IEEE Netw.}, Early Access, 2023.

\bibitem{zhang2022integrated}
R.~Zhang, B.~Shim, W.~Yuan, M.~Di~Renzo, X.~Dang, and W.~Wu, ``Integrated
  sensing and communication waveform design with sparse vector coding: Low
  sidelobes and ultra reliability,'' \emph{{IEEE} Trans. Veh. Technol.},
  vol.~71, no.~4, pp. 4489--4494, Apr. 2022.

\bibitem{xiong2023fundamental}
Y.~Xiong, F.~Liu, Y.~Cui, W.~Yuan, T.~X. Han, and G.~Caire, ``On the
  fundamental tradeoff of integrated sensing and communications under
  {Gaussian} channels,'' \emph{IEEE Trans. Inf. Theory}, vol.~69, no.~9, pp.
  5723--5751, Jun. 2023.

\bibitem{abrardo2023design}
A.~Abrardo, A.~Toccafondi, and M.~Di~Renzo, ``Design of reconfigurable
  intelligent surfaces by using {S}-parameter multiport network
  theory--optimization and full-wave validation,'' \emph{arXiv preprint
  arXiv:2311.06648}, 2023.

\bibitem{ivrlavc2010toward}
M.~T. Ivrla{\v{c}} and J.~A. Nossek, ``Toward a circuit theory of
  communication,'' \emph{IEEE Trans. Circuits Syst. I, Reg. Papers}, vol.~57,
  no.~7, pp. 1663--1683, Jul. 2010.

\bibitem{gradoni2021end}
G.~Gradoni and M.~Di~Renzo, ``End-to-end mutual coupling aware communication
  model for reconfigurable intelligent surfaces: {An} electromagnetic-compliant
  approach based on mutual impedances,'' \emph{IEEE Wireless Commun. Lett.},
  vol.~10, no.~5, pp. 938--942, May 2021.

\bibitem{nerini2023universal}
M.~Nerini, S.~Shen, H.~Li, M.~Di~Renzo, and B.~Clerckx, ``A universal framework
  for multiport network analysis of reconfigurable intelligent surfaces,''
  \emph{arXiv preprint arXiv:2311.10561}, 2023.

\bibitem{akrout2022achievable}
M.~Akrout, V.~Shyianov, F.~Bellili, A.~Mezghani, and R.~W. Heath, ``Achievable
  rate of near-field communications based on physically consistent models,''
  \emph{IEEE Trans. Wireless Commun.}, vol.~22, no.~2, pp. 1266--1280, Sept.
  2022.

\end{thebibliography}
\end{document}